# Anisotropic polymer nanoparticles with controlled dimensions from the morphological transformation of isotropic seeds

*Zan Hua[1, 2], Joseph R. Jones[2], Marjolaine Thomas[2], Maria C. Arno[2], Anton Souslov[3], Thomas R. Wilks[2]* and Rachel K. O'Reilly[2]**

[1]Department of Chemistry, University of Warwick, Gibbet Hill Road, Coventry, CV4 7AL, UK.

[2]School of Chemistry, University of Birmingham, Edgbaston, Birmingham, B15 2TT, UK.

[3]Department of Physics, University of Bath, Claverton Down, Bath BA2 7AY, UK.

*Corresponding authors: Thomas R. Wilks (t.r.wilks@bham.ac.uk) and Rachel K. O'Reilly (r.oreilly@bham.ac.uk)

## ABSTRACT

Understanding and controlling self-assembly processes at multiple length scales is vital if we are to design and create advanced materials. In particular, our ability to organise matter on the nanoscale has advanced considerably, but still lags far behind our skill in manipulating individual molecules. New tools allowing controlled nanoscale assembly are sorely needed, as well as the physical understanding of how they work. Here, we report a new method for the production of highly anisotropic nanoparticles with controlled dimensions based on a morphological transformation process (MORPH for short) driven by the formation of supramolecular bonds. We present a minimal physical model for MORPH which suggests it will be generalisable to a large number of polymer/nanoparticle systems. We envision MORPH becoming a valuable tool for controlling nanoscale self-assembly, and for the production of functional nanostructures for diverse applications.

## MAIN TEXT

Complex structures are a hallmark of natural systems, achieved through hierarchical self-assembly across multiple length scales – bone, with its remarkable combination of stiffness and toughness, exhibits ordering across at least nine distinct levels, from the molecule upwards.[1] Scientists have long been interested in mimicking biological organisation to create artificial materials with similarly exceptional properties. However, while chemists have become adept at manipulating molecules, it has been more challenging to achieve the same degree of control at the nanoscale, just one level up. In particular, highly anisotropic structures are a natural target for nanoscale self-assembly because they are ubiquitous in biology (e.g. microtubules, muscle filaments) and have been shown to possess unique properties in applications as diverse as photonics[2–4] and drug delivery.[5–10] The ideal building blocks for the bottom-up self-assembly of anisotropic nanostructures would (1) be readily accessible (i.e. cheap and scalable syntheses), (2) allow straightforward tuning of chemical composition (so materials can be tailored for different applications) and (3) allow fine control over nanoparticle dimen-sions (enabling controlled higher-order assembly at still larger length scales). However, marrying all three of these requirements has proven difficult.

DNA nanotechnology[11–13] allows well-defined anisotropic nanostructures to be built with a high degree of precision, but concerns remain about the scalability and cost-effectiveness of this approach (despite recent advances[14]). Inorganic nanoparticles are more accessible, and high aspect ratio structures can be produced, but only limited variation of the chemical composition is possible.[3,4] In this context, synthetic polymers are highly promising building blocks because their synthesis is cheap and scalable, and the development of controlled polymerisation techniques has allowed straightforward modulation of the length, architecture and chemical composition of polymer chains.[15]

However, while the self-assembly of polymers in solution affords bottom-up access to nanoscale objects,[16] controlling this process to make highly anisotropic nanoparticles with well-defined dimensions has proven to be very challenging. This is because conventional methods rely on exploiting differences in the stabilities of particles with different shapes, which are not significant between high aspect ratio structures of different lengths.[17] While a large body of work exists demonstrating the formation of anisotropic polymer nanostructures,[8,10,16,18,19] and even triggered switching between different shapes,[20,21] it remains challenging to create stable systems with fine control of nanoparticle dimensions. For example, it is possible to generate pure phases of wormlike nanoparticles using polymerisation-induced self-assembly,[18] but the products are highly disperse with little control over length and width.

The "seeded growth" method (Figure 1) circumvents these problems, but is somewhat limited in scope. In the seeded growth approach, anisotropic nanoparticle seeds are fed with polymer unimers in solution and 1-dimensional (1D) growth is driven by the formation of bonding interactions between the unimers and the exposed ends of the seed. Under the right conditions, this enables the growth of long, cylindrical particles with lengths determined simply by the amount of added unimer. Crystallisation-driven self-assembly (CDSA)[22–25] is the most well-known example, but metal–metal interactions,[26,27] hydrogen bonding (H-bonding)[28] and π–π stacking[29] have also been exploited. However, seeded growth has so far proven limited to a narrow range of polymers, and depends on the generation of a uniform population of anisotropic seed particles, which can be a major challenge.

Here, we report a more general method for the production of high aspect ratio anisotropic polymer nanoparticles with controlled lengths and widths, which we term morphological transformation (MORPH, Figure 1). Unlike 1D seeded growth, MORPH uses readily-accessed *isotropic* nanoparticles as a starting point, which are then driven to transform and grow into anisotropic structures by the formation of supramolecular bonds with an added polymer. We show that the amount of growth is determined by the amount of added polymer, allowing fine control over length. We further demonstrate that multi-functional nanoparticles of defined dimensions can be built using MORPH by simple stepwise growth using appropriately modified polymers. Finally, we develop a physical model for the process which suggests it is a generic approach applicable to a broad range of polymer systems and supramolecular interactions, and with many potential applications across materials science.

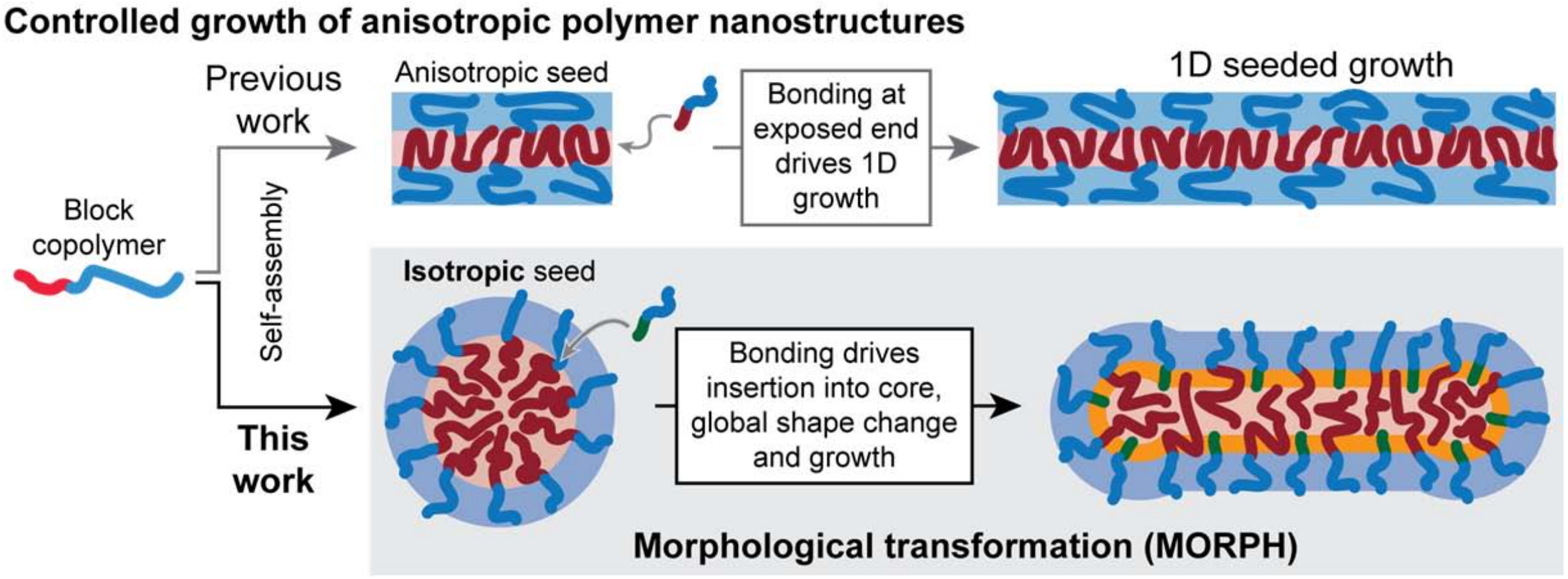

**Figure 1.** Methods for the growth of anisotropic polymer nanostructures with controlled dimensions. Previous efforts have been based on 1-dimensional growth from an anisotropic seed particle (top). We report a qualitatively different process based on the global morphological transformation of an isotropic seed particle (lower).

## RESULTS

We first encountered the possibility for morphological transformation while investigating the self-assembly behaviour of nucleobase-containing polymers,[30–36] in a nanoparticle system which exhibited poorly controlled shape-changing behaviour in response to the addition of a polymer in solution. A handful of other examples of this behaviour were subsequently found in the literature,[37,38] which had similarly failed to demonstrate control of product dimensions. We reasoned that achieving this control would result in a useful new approach to the generation of well-defined anisotropic nanoparticles, and hypothesised that these previous efforts had not succeeded because of nanoparticle disassembly/reassembly during the transformation process, which led to the formation of a range of structures of different sizes. To retain control it would be necessary to eliminate this pathway, and we hypothesised that this could be achieved by building the seed nanoparticle out of a polymer with low water solubility and a high glass transition temperature ($T_g$). This would inhibit disassembly by making extraction of polymer chains into the bulk solvent highly unfavourable and introducing a significant barrier to rearrangement of the nanoparticle core.

To test these hypotheses, we synthesised the amphiphilic block copolymer **PT** with a long, hydrophobic thymine-containing block and a short hydrophilic block (see supporting information (SI), section S3). This design was intended to minimise water-solubility and maximise the polymer's $T_g$, which was measured to be 73 °C. **PT** was self-assembled in water to give well-defined nanoparticle **NT** (Figure 2a, SI section S4) with a diameter of ~60 nm according to transmission electron microscopy (TEM) (Figure 2b). A second polymer, **PA** (SI section S3), was designed with the same length of hydrophilic block as **PT** and an adenine-containing block that was sufficiently short to allow **PA** to freely dissolve in water at moderate concentrations. **PA** was added at different molar equivalents to separate solutions of **NT** (SI section S5), and the mixtures stirred for 2 h at 24 °C. Stain-free TEM analysis revealed that A:T H-bonding between **NT** and **PA** drove a morphological transformation from spheres to dumbbells or worms depending on the A:T molar ratio (Figure 2b and SI section S5).

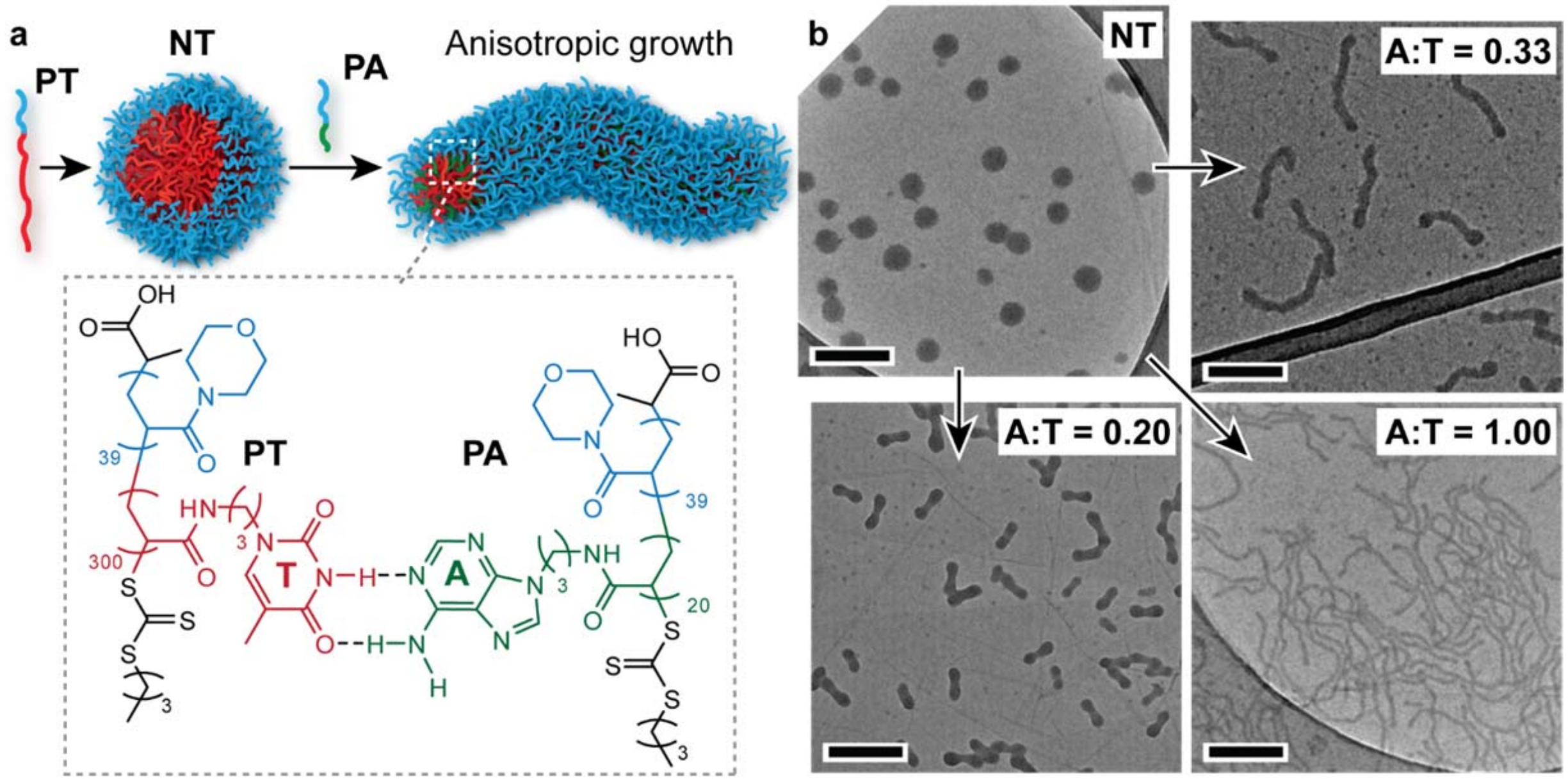

**Figure 2.** Single step transformation of spherical nanoparticles into different anisotropic morphologies by MORPH. (a) Schematic representation of MORPH using nucleobase-containing polymers. Spherical nanoparticle **NT** with a thymine-containing core was formed by self-assembly of the polymer **PT** using a solvent switch method from DMF to water. Introduction of the adenine-containing polymer **PA** induced MORPH at A:T molar ratios above 0.20. (b) TEM images of the nanoparticles formed before and after addition of **PA** at different A:T molar ratios to separate solutions of **NT**. Particles were imaged stain-free on graphene oxide (GO).[39] Scale bars = 200 nm.

We next explored whether *sequential* addition of **PA** could achieve the same morphological transformation, by feeding **NT** with small aliquots (0.07 molar equivalents) of **PA**, leaving 2 hours between additions. Surprisingly, this did not lead to the same morphological transformation process. Instead, the particles remained spherical, swelling in size before disassembling into much smaller spherical particles as the A:T ratio approached 1:1 (SI section S6). We speculated that a threshold concentration of **PA** was required to induce anisotropy, after which stepwise growth of the worms might be possible. To explore this idea, short seed worms approximately 300 nm long were fabricated by adding **PA** to **NT** at an A:T molar ratio of 0.33 (Figure 3a), followed by stepwise addition of further **PA**. Figure 3b-e shows that longer and thinner worms were obtained with each addition of **PA**. The average contour length gradually extended to over 1000 nm, with an approximately linear relationship between worm length and A:T molar ratio, and narrow length distributions (Figure 3f and SI section S6). TEM images suggested the average width of the worms decreased from approximately 22 nm to 14 nm, with a slight increase in the volumes of individual worms (Figure 3g-h and SI section S6). The decrease in width was verified for the bulk sample by small angle X-ray scattering (SAXS) analyses (Figure 3g). These results confirmed that the growth process could be controlled, and that uniformly-sized wormlike nanoparticles could be produced using this method.

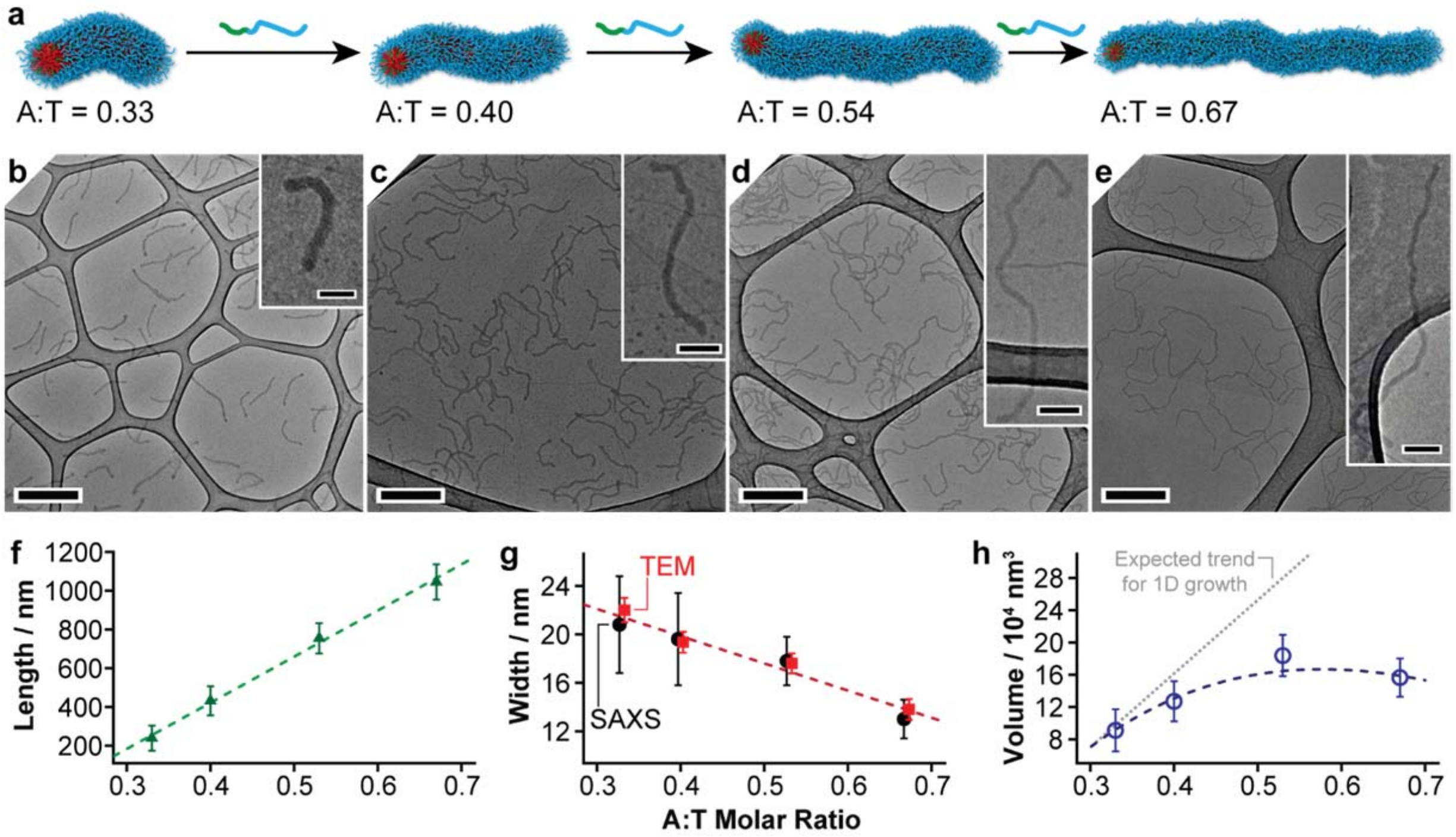

**Figure 3.** Stepwise growth of well-defined anisotropic nanoparticles by MORPH. (a) Schematic representation and (b-h) characterisation of the wormlike nanoparticles. (b-e) Stain-free TEM images of the nanoparticles at different A:T molar ratios; the scale bars in (b-e) and their insets are 500 and 100 nm, respectively. Average lengths (f), widths (g) and volumes (h) of the worms from analyses of TEM images (b-e) and SAXS. Note that the TEM and SAXS data in (g) have been offset for clarity.

We performed a series of control experiments in which H-bonding for one of the polymers was blocked or removed altogether (SI section S7). In no case was a morphological transformation observed, confirming that the MORPH process was driven by the formation of strong H-bonding interactions as opposed to weaker hydrophobic effects. We also investigated whether the process was indeed a single-particle transformation, rather than particle–particle fusion mediated by **PA**, by obtaining further information about the mass average molar masses ($\overline{M}_w$) of **NT** and the dumbbells formed on addition of **PA** (A:T = 0.20) using light scattering (LS). The dumbbells were found to have a $\overline{M}_w$ of 69 ± 1 × $10^6$ Da, indicating an average mass increase of only 8% compared with **NT** ($\overline{M}_w$ = 63 ± 1 × $10^6$ Da) (SI section S8), consistent only with a single-particle transformation process.

To illustrate the potential to use MORPH to produce well-defined, functional nanoparticles **PA** was derivatised with two different dyes – BODIPY-FL (**PA**$^{G}$) and -TR (**PA**$^{R}$) (SI section S9) – and used in worm growth experiments. Starting with unfunctional worms at an A:T ratio of 0.33 we added 0.07 equivalents of **PA**$^{G}$ followed by 0.07 equivalents of **PA**$^{R}$ and inspected the resulting worms using TEM and confocal fluorescence microscopy (Figure 4). As expected, feeding the unfunctional worms with **PA**$^{G}$ gave green fluorescent worms (Figure 4, middle column). Further feeding with **PA**$^{R}$ gave yellow fluorescent worms as a result of colocalisation of the dyes (Figure 4, right hand column – see SI section S9 for controls). Control over worm dimensions was retained, as evidenced by stain-free TEM images (Figure 4, bottom row).

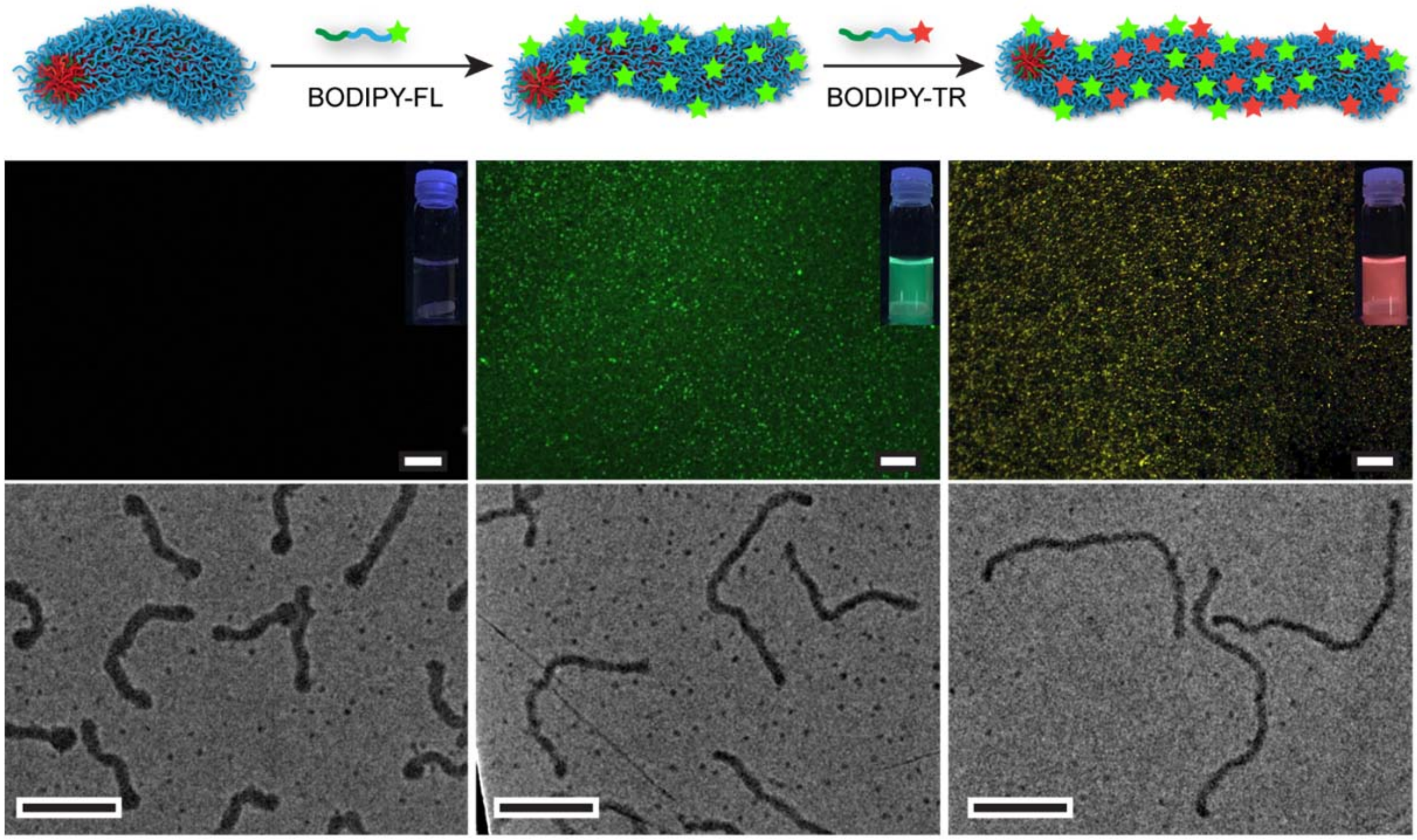

**Figure 4.** Controlled fabrication of fluorescent anisotropic nanoparticles by MORPH. Non-fluorescent seed worms with A:T molar ratio 0.33 were converted into green and then yellow fluorescent nanoparticles by adding BODIPY-FL (**PA**$^{G}$) and -TR (**PA**$^{R}$) tagged polymers sequentially. The top row shows confocal fluorescence images (overlay of green and red channels, scale bars = 10 μm) with photographs of the nanoparticle

solutions under UV light as insets. The bottom row shows stain-free TEM images confirming the controlled morphological transformation (scale bars = 200 nm).

## DISCUSSION

MORPH operates differently from 1D seeded growth – as illustrated in Figure 3f-h, an increase in worm length is accompanied by a corresponding *decrease* in width. LS and TEM data confirm that the growth is uniform along the length of the worms, suggesting that polymers can insert at any point with equal probability. In other words, addition of further polymer triggers a *global* morphological transformation, as opposed to local growth at the nanoparticle ends. MORPH's other unique feature is *induced* anisotropy, which allows readily-accessed isotropic particles to be used as seeds. In the system we describe, the MORPH process is highly controlled, reproducibly giving wormlike nanoparticles with well-defined dimensions determined simply by the amount of added complementary copolymer. The growing particles retain their reactivity and remain capable of incorporating more of the complementary polymer as they grow. As a result, worm length scales approximately linearly with the amount of added copolymer, allowing particular particle dimensions to be targeted.

To confirm which features of our system are necessary for MORPH to occur, we begin by considering the behaviour of the polymer chains in the cores of the nanoparticles. The glass transition temperature ($T_g$) of **PT** was found to be 73 °C, far above the experimental temperature (24 °C), indicating that the core of **NT** is most likely in a 'frozen', glassy state, with low chain mobility.[40] Bulk $T_g$ does not always give a reliable indication of the behaviour of a polymer in a solvated nanoparticle,[41] but other evidence also points to glassy core dynamics: when we mixed **PA** and **PT** in the appropriate ratios in DMF (a good solvent for all blocks) and performed a slow solvent switch to water, only spheres were observed (SI section S7), which indicates that these are the thermodynamically most stable structures.[40] This observation implies that in the worm-like nanoparticles formed there is a significant barrier to core chain mobility, and these glassy dynamics are essential to preserve the anisotropic shape and prevent disassembly into spheres. Because core chain mobility is restricted in **NT**, there must be a driving force behind MORPH – in the system we consider, this is the formation of energetically-favourable H-bonds between thymine and adenine. The absence of any rearrangement driven solely by the hydrophobic effect (see SI section S7) further implies that this driving force must exceed a threshold value. Below, we show that MORPH can be explained on the basis of these properties (glassy core dynamics and the presence of a driving force for polymer insertion), without the need to invoke more specific details of the chemical bonding involved, such as directionality.

To develop a physical model for the process, we consider the response of a nanoparticle to the insertion of a polymer at the core–corona interface (Figure 5). We assume that the polymer will diffuse to the interface and insert into the nanoparticle core with an associated timescale, $\tau_I$, but that because of the short length of the core block in the added polymer this insertion will occur only in a thin shell region. It is predicted that $\tau_I$ will be inversely proportional to the concentration of the polymer (i.e., more polymer will lead to faster insertion). Insertion of the polymer will introduce steric crowding which the system can relieve by two possible routes. The first, which is thought to be operational in many copolymer systems capable of undergoing morphological transitions,[42,43] is removal of material from the nanoparticle by extraction of polymer chains into solution. However, in MORPH this process is strongly disfavoured because of a combination of the glassy core dynamics and strong bonding between the nanoparticle and inserted polymer. With chain extraction suppressed, the only way that the additional mass can be accommodated is by an increase in the surface area of the core–corona interface, requiring rearrangement of the nanoparticle core chains, with timescale $\tau_R$. We assume that $\tau_R$ will be determined principally by the bulk properties of the core chains (i.e. the bulk modulus and viscosity) and remain more or less independent of the polymer concentration.

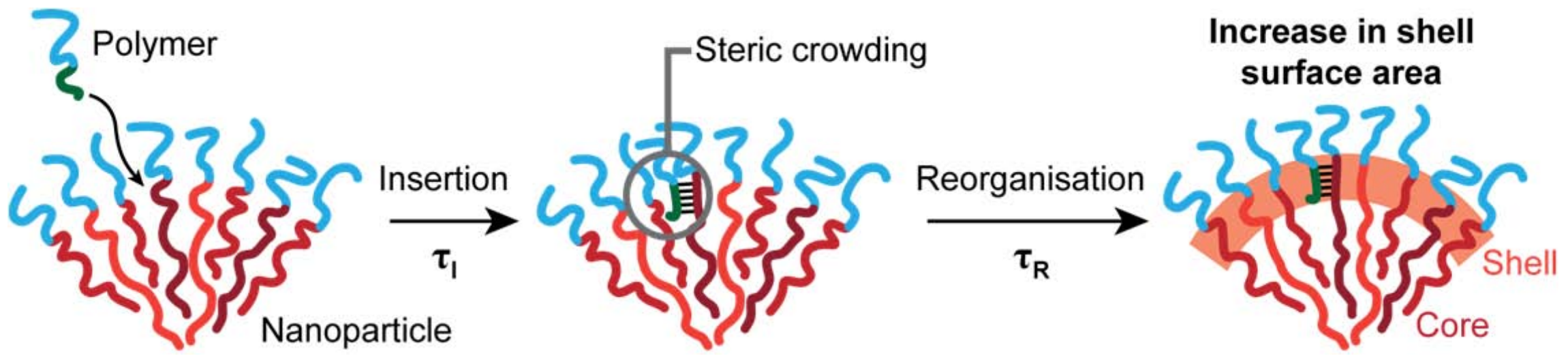

**Figure 5.** Illustration of the polymer insertion process in MORPH. The polymer inserts with timescale $\tau_I$ into a thin shell region at the core–corona interface. Insertion causes steric crowding that must be relieved by reorganisation to increase the shell's surface area, which proceeds with timescale $\tau_R$.

Since the sphere is the shape with the least surface area per unit volume, an increase in the surface area of the nanoparticle's core–corona interface can be achieved in only two ways: swelling or shape change. At sufficiently low polymer concentrations $\tau_I^{-1} < \tau_R^{-1}$ and insertion of the polymer is slow enough that the core chains have time to rearrange and allow expansion of the interface by simple isotropic swelling, minimising the core surface-tension energy (Figure 6, upper pathway).[42] Swelling requires stretching of the core polymer chains, so it cannot continue indefinitely – at some point no further stretching is possible and the only way for the system to continue increasing the shell surface area to accommodate more of the added polymer is by disassembly into small particles, in agreement with the experimental observations (SI section S6). At higher polymer concentrations, $\tau_I^{-1} > \tau_R^{-1}$ and rapid insertion of the polymer requires an equally fast increase in the surface area of the interface, which can only be achieved by *both* core chain rearrangement *and* shape change (Figure 6, lower pathway). Induction of anisotropy is only possible if the sphere cannot isotropically swell on the timescale of polymer insertion, and proceeds by overcoming the core surface-tension energy barrier that favours a spherical shape. Once this barrier has been overcome and anisotropy induced, elongation is expected to become the favoured pathway because, unlike swelling, it does not require the energetically unfavourable stretching of core chains. Regardless of the polymer concentration, further insertion is therefore anticipated to cause the dumbbell to increase in length and shrink in width as more material is drawn from the core to form bonds with the added polymer in the shell region. Each nanoparticle core can be thought of as a reservoir of material capable of driving elongation by forming bonds with the added polymer. When the core is saturated with bonds to the added polymer, this reservoir is exhausted, and elongation is no longer possible. Addition of further polymer is then expected to drive disassembly, as for the low concentration swelling pathway, and this was indeed observed in the experimental system at A:T ratios around 1:1 (see SI section S6).

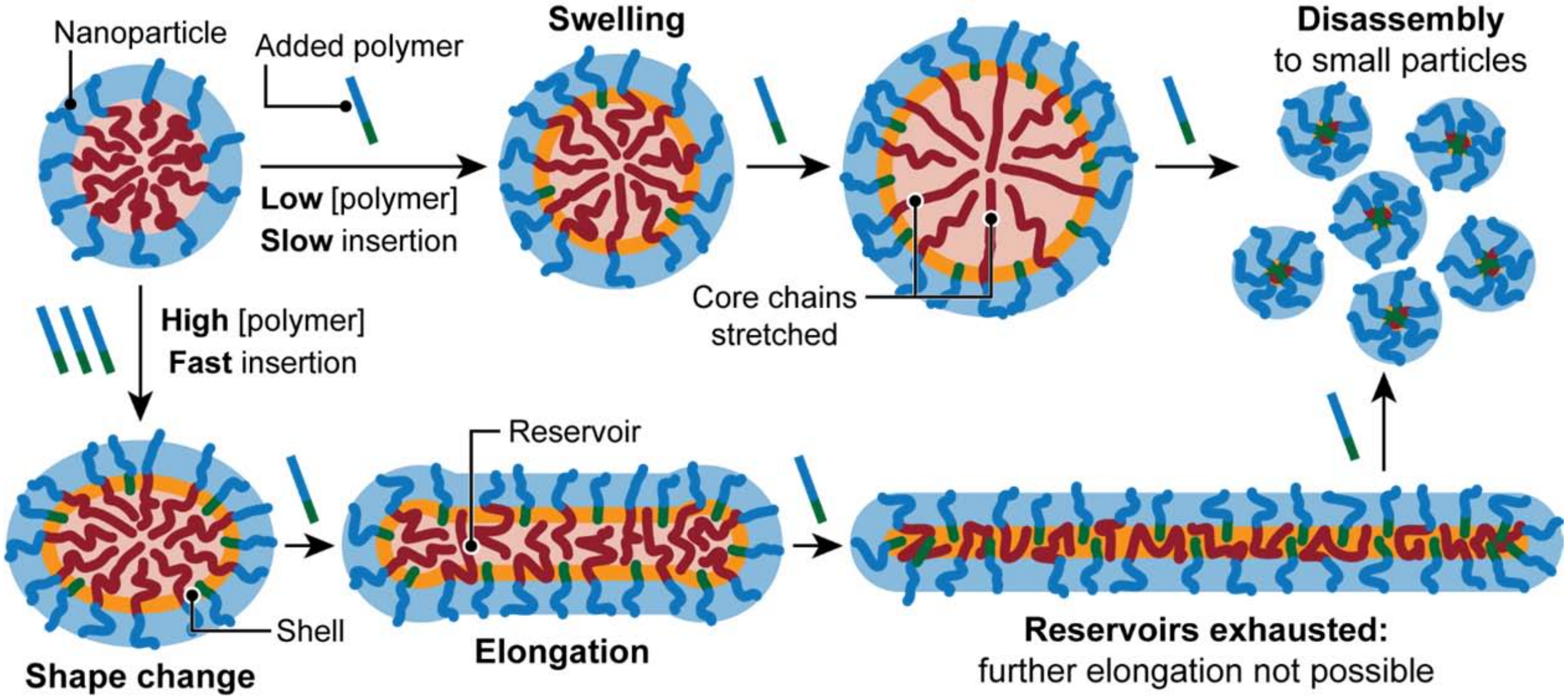

**Figure 6.** Schematic illustration of the proposed MORPH pathways at low (upper pathway) and high (lower pathway) concentrations of the added polymer.

We can formalise this physical argument by writing down the following equation for the evolution of small shape eccentricity, $\epsilon$ (i.e. a parameter to represent particle anisotropy): $\partial_t \epsilon^2 = (\tau_I^{-1} - \tau_R^{-1})$ [See SI section S10 for a phenomenological derivation]. When $\tau_I^{-1} < \tau_R^{-1}$, $\epsilon = 0$ is a stable solution of the equation, indicating isotropic growth due to minimisation of core surface-tension; when $\tau_I^{-1} > \tau_R^{-1}$, $\epsilon \neq 0$ and eccentricity grows in time, indicating the induction of anisotropy. Note that unlike many instabilities in equilibrium polymer micelles,[44–46] the glassy core prevents the anisotropic nanoparticles from disassembling in the absence of an external source of added polymer. This model predicts that MORPH proceeds through an ellipsoidal intermediate, so we attempted to access this in the experimental system by forming particles at an A:T ratio intermediate between iso-tropic swelling (A:T = 0.07) and dumbbell formation (A:T = 0.20). As predicted, LS and SAXS analyses confirmed the formation of prolate ellipsoids at an A:T molar ratio of 0.14 (SI section S5).

This minimal complexity model therefore succeeds in describing the main features of the MORPH process. It assumes only that (1) the seed nanoparticle has a glassy core, (2) the added polymer is sufficiently short to penetrate only a thin shell around the core, and (3) its insertion results in the formation of reasonably strong bonds. A wide range of polymers assemble into nanoparticles with glassy core dynamics, controlled polymerisation techniques allow polymer length to be easily tuned, and supramolecular chemistry provides us with a broad palette of strong, non-covalent bonding interactions. We therefore suggest that MORPH is a general mechanism for the generation of anisotropy applicable to a large number of polymer/nanoparticle systems beyond the specific case reported here.

## CONCLUSIONS

We report a new method, MORPH, for the production of anisotropic polymer nanoparticles with well-defined aspect ratios, based on the morphological transformation of isotropic seeds driven by the formation of supramolecular bonds. MORPH exhibits a number of useful features: the isotropic seed nanoparticles are straightforward to assemble because they are the thermodynamic product of a simple self-assembly process; the shape change can be controlled with precision simply by mixing aqueous solutions of nanoparticles and complementary polymer; and the global transformation process makes it straightforward to incorporate additional functionality across the nanostructure, as demonstrated by our stepwise incorporation of fluorescent tags. One can imagine incorporating functional ligands, therapeutics and reporter groups to build up multifaceted delivery platforms with a high degree of control. In future work, particles like these could be used to explore the precise effects of aspect ratio on their interactions with cells. Most significantly, the physical model we have developed implies that it will be possible to exploit the huge diversity of supramolecular bonding motifs to perform controlled nanoparticle growth using MORPH – we therefore predict that this approach will become an invaluable addition to the self-assembly toolkit.

## METHODS

### Polymer Synthesis

The adenine- and thymine-functionalised monomers (**AAm** and **TAm**) were synthesised as reported in our previous work[33] and full details are given in the SI. The poly(4-acryloylmorpholine) (**PNAM$_{39}$**) macro-chain transfer agent (macro-CTA) was synthesised by RAFT polymerisation as follows. A 10 mL ampoule was charged with NAM (500 µL, 4.0 mmol), 2-(((butylthio)carbonothiolyl)thio)-propanoic acid (23.8 mg, 0.1 mmol), VA-044 (1.3 mg, 0.004 mmol) and a mixture of 1,4-dioxane and water (2.0 mL, v:v 1:4). The mixture was thoroughly degassed *via* 4 freeze-pump-thaw cycles, filled with nitrogen and then immersed in an oil bath at 70 °C for 2 h. The polymerisation solution was precipitated three times from cold $CH_3OH$. The light yellow polymer was dried in a vacuum oven overnight at room temperature and analysed by $^1H$ NMR spectroscopy and DMF SEC. The degree of polymerisation (DP) was calculated using $^1H$ NMR spectroscopy by comparing the integrated signals corresponding to the backbone signals ($\delta$ = 1.62 ppm) with those of the methyl group from the CTA ($\delta$ = 0.87 ppm). **PNAM$_{39}$** was then chain extended to give the diblock copolymers used in this report – an example procedure for **PT** follows. **PNAM$_{39}$** (14 mg, 0.0025 mmol), **TAm** (178 mg, 0.75 mmol), and the initiator VA-044 (0.08 mg, 0.00025 mmol) were dissolved in a mixture of DMF and water (0.5 mL, v:v 1:1). The mixture was thoroughly degassed *via* 4 freeze-pump-thaw cycles, filled with nitrogen and then immersed in an oil bath at 70 °C overnight. An aliquot of the crude product was taken and analyzed by $^1H$ NMR spectroscopy to calculate the conversion and degree of polymerization. The residual solution was then precipitated three times from cold $CH_3OH$. The light yellow polymer was dried in a vacuum oven overnight at room temperature and analysed by $^1H$ NMR spectroscopy and DMF SEC.

### Self-Assembly of PT to form NT

The seed nanoparticles **NT** were assembled as follows. **PT** was dissolved in DMF (at 8 mg mL$^{-1}$) and stirred for 2 h at 70 °C. Then an excess of 18.2 MΩ·cm water was added *via* a syringe pump at a rate of 1 mL h$^{-1}$. The final volume ratio between water and organic solvent was about 8:1. The solution was then dialysed against 18.2 MΩ·cm water, incorporating at least 6 water changes, to afford nanoparticles **NT** at a concentration of *ca.* 1 mg mL$^{-1}$.

### Morphological Transformation of NT

For the single step transformations shown in Figure 2, **PA** was dispersed in $H_2O$ at 5 mg mL$^{-1}$. This was then added to separate solutions of the nanoparticle **NT** (0.5 mg mL$^{-1}$) with stirring at A:T molar ratios of 0.07, 0.20, 0.33, 0.67, 1.0, 1.33. The molar ratios were calculated according to the $M_n$ determined from $^1$H NMR spectroscopic analyses and the polymers' mass concentrations. The mixtures were sealed and allowed to stir at room temperature for 2 h, and then characterised by LS and TEM. For the stepwise growth experiments shown in Figure 3, **PA** (0.33 molar ratio of A relative to T) was added to a solution of **NT** (0.5 mg mL$^{-1}$) to give short "seed" worms. After 2 h stirring at 24 °C, further **PA** solution (0.07 molar ratio of A relative to T) was added. This process was repeated until A:T molar ratios of 0.33, 0.40, 0.53 and 0.67 were achieved. Each stage was characterised by TEM and SAXS analyses.

### Controlled Production of Fluorescent Wormlike Nanoparticles

**PA** was modified with BODIPY-FL or BODIPY-TR as described in the SI to give **PA$^{G}$** and **PA$^{R}$** respectively. A solution of **NT** (0.5 mg mL$^{-1}$) was then fed with an aliquot of **PA** solution to give non-fluorescent wormlike nanoparticles with an A:T molar ratio of 0.33. This solution was then fed with an aliquot of **PA$^{G}$** followed by an aliquot of **PA$^{R}$** (0.07 molar equivalents of A relative to T in both cases), with 2 h stirring at 24 °C in between each addition. Each stage was characterised by TEM and confocal fluorescence microscopy. In addition, control experiments were performed in which the seed worms were fed with two successive additions of the same dye-functionalised polymer to give pure red or green worms.

### Light (LS) and Small Angle X-Ray Scattering (SAXS) Experiments

Scattering experiments were conducted to determine the uniformity of particles within the bulk sample. Initial estimates for hydrodynamic diameters and the particle size distribution were made using a Malvern Zetasizer Nano S instrument fitted with a (He-Ne) 633 nm laser module and the refractive index increment was measured using a DnDc1260 differential refractometer supplied by PSS GmbH. The mass average molar mass and particle size information were then determined using an ALV-CGS3 goniometer-based system supplied by ALV GmbH, also operating at $\lambda = 633$ nm.

SAXS measurements were made using a Xenocs Xeuss 2.0 equipped with a micro-focus Cu Kα source collimated with Scatterless slits. The scattering was measured using a Pilatus 300k detector with a pixel size of 0.172 mm × 0.172 mm. A radial integration as a function of scattering length, $q$, was performed on the 2-dimensional scattering profile and the resulting data corrected for absorption and background from the sample holder.

## SUPPORTING INFORMATION

Full details of materials, instrumentation and analysis techniques, synthetic procedures for all monomers and polymers, detailed assembly and characterisation data for the seed nanoparticles and morphological transformation products, control experiments and phenomenological derivation of the physical model.

## AUTHOR CONTRIBUTIONS

T.R.W. and R.K.O. designed the study, supervised research and wrote the paper. Z.H. and M.T. developed the monomer and polymer designs and performed the synthetic work. J.R.J. designed and performed light scattering experiments and analysed all data associated with these experiments. M.C.A. designed and performed confocal microscopy measurements. A.S., J.R.J., T.R.W. and R.K.O. developed the theory for the growth mechanism. All authors analysed the data, provided figures and commented on the manuscript.

## NOTES

The authors declare no competing financial interests.

## ACKNOWLEDGMENTS

The authors thank the University of Warwick, China Scholarship Council (Z. H.), EPSRC, and the ERC (grant number 615142) for research funding. The University of Warwick Advanced BioImaging Research Technology Platform, BBSRC ALERT14 award BB/M01228X/1 are thanked for confocal fluorescence microscopy analysis. X-ray Diffraction Research Technology Platform in the University of Warwick is thanked for conducting the SAXS analysis.

# Supporting Information

# for

# Anisotropic polymer nanoparticles with controlled dimensions from the morphological transformation of isotropic seeds

*Zan Hua[1, 2], Joseph R. Jones[2], Marjolaine Thomas[2], Maria C. Arno[2], Anton Souslov[3], Thomas R. Wilks[2]* and Rachel K. O'Reilly[2]**

[1]Department of Chemistry, University of Warwick, Gibbet Hill Road, Coventry, CV4 7AL, UK.

[2]School of Chemistry, University of Birmingham, Edgbaston, Birmingham, B15 2TT, UK.

[3]Department of Physics, University of Bath, Claverton Down, Bath BA2 7AY, UK.

*Corresponding authors: Thomas R. Wilks (t.r.wilks@bham.ac.uk) and Rachel K. O'Reilly (r.oreilly@bham.ac.uk)

## TABLE OF CONTENTS

## S1 MATERIALS

2,2′-Azo-bis(isobutyronitrile) (AIBN) was obtained from Molekula and recrystallized from methanol. 2,2′-Azobis[2-(2-imidazolin-2-yl)propane]dihydrochloride (VA-044, Wako) was used without further purification. 4-Acryloylmorpholine (NAM) was bought from Aldrich and was purified by vacuum distillation. 2-(((Butylthio)carbonothiolyl)thio)propanoic acid was synthesized as described previously and stored at 4 °C.[1] Wafers of p-silicon (100) were purchased from Sigma-Aldrich and cut into plates with a size of 10 mm × 10 mm for AFM imaging. Dialysis membranes (molecular weight cut-off = 3.5 kDa) were purchased from Spectra/Por. DMF, DMSO and other chemicals were obtained from Fisher Chemicals and used without further purification. Dry solvents were obtained by passing over a column of activated alumina using an Innovative Technologies solvent purification system.

## S2 INSTRUMENTATION & ANALYSIS

### S2.1 NMR Spectroscopy

$^1H$ NMR spectra were recorded on a Bruker DPX-400 or HD500 spectrometer with DMSO-$d_6$ as the solvent. The chemical shifts of protons were relative to solvent residues (DMSO 2.50 ppm, $CDCl_3$ 7.26 ppm).

### S2.2 Size Exclusion Chromatography (SEC)

SEC data were obtained in HPLC grade DMF containing 5 mM $NH_4BF_4$ at 50 °C, with a flow rate of 1.0 mL $min^{-1}$, on a set of two PLgel 5 µm Mixed-D columns, and a guard column. SEC data were analyzed with Cirrus SEC software calibrated using poly(methyl methacrylate) (PMMA) standards.

### S2.3 Refractive Index (RI) Measurements

Values for the refractive index increment (dn/dc) of polymers listed in Table S1 were determined using a PSS DnDc1260 differential refractometer fitted with a 620 nm laser.

**Table S1.** Refractive index increments (d*n*/d*c*) for the polymers and nanoparticle formulations used in this study.

| **Sample** | **d*n*/d*c*** / mL $g^{-1}$ |
|---|---|
| **PA** | 0.162 |
| **PT** | 0.175 |
| **NT** + **PT** (0.14 A:T) | 0.174* |
| **NT** + **PT** (0.20 A:T) | 0.172* |

* Refractive index increments for mixed nanoparticle systems were calculated using a weighted sum of the d*n*/d*c* values of the individual copolymers.[1]

### S2.4 Dynamic Light Scattering (DLS) Analysis

Initial estimates for hydrodynamic diameters ($D_H$) and size distributions of particles were determined using a Malvern Zetasizer Nano S instrument fitted with a 4 mW He-Ne 633 nm laser module, which records measurements at a single detection angle, 173°. The proprietary software was used to calculate $D_H$ according to the Stokes-Einstein equation for diffusion of particles through a liquid with low Reynolds number (see S2.5).

In Figures S7, S8, S20 and S22, `PD' refers to (poly)dispersity index, a measure of the particle size distribution provided by the proprietary software, where $\mathrm{PD} = \log_{10}\left(\frac{\bar{M}_W}{\bar{M}_N}\right)$.

### S2.5 Static and Dynamic Light Scattering (LS) Analysis

LS experiments were conducted using an ALV-CGS3 goniometer-based system operating a $\lambda$ = 633 nm wavelength laser, with the sample maintained at 25 °C. Samples were contained in 5 mm borosilicate glass tubes. Aliquots of the particles in solution at concentration c = 0.7 g $L^{-1}$ were passed through 1.2 μm cellulose syringe filters and a dilution series prepared over the concentration range $0.2 \leq c \leq 0.7$ g $L^{-1}$. Intensity of light scattering, *I(t)*, was recorded over the angular range $30° \leq \theta \leq 130°$ at 5° intervals for each of the standard (toluene), the solvent ($H_2O$) and the solution. At each datum (*q, c*), where the magnitude of the scattering vector is $q = \left(\frac{4\pi}{\lambda}\right) \cdot n_D \sin\left(\frac{\theta}{2}\right)$ and $n_D$ is the refractive index of the solvent, the proprietary software records the following measurements:

1. $\frac{R(q,c)}{Kc}$, which is the Rayleigh ratio, $= \left(\frac{I_{solution} - I_{solvent}}{I_{standard}}\right) \cdot I_{std(abs)}$ , normalized with respect to sample concentration, *c*, and an instrument constant, $K = \left(\frac{4\pi^2}{\lambda^4 N_A}\right) \cdot \left(n_D \cdot \frac{dn}{dc}\right)^2$ , where $N_A$ is the Avogadro number and $\frac{dn}{dc}$ is the refractive index increment (S2.3).

2. The normalized scattering intensity autocorrelation function, $g_2(q,\tau) = \frac{\langle I(q,t)I(q,t+\tau)\rangle}{\langle I(q,\tau)^2\rangle}$, and from this the amplitude correlation function, $g_1(q,\tau)$, according to the Siegert relation, $g_2(q,\tau) = 1 + g_1(q,\tau)^2$, both of these calculated by the ALV LSE-5004 correlator module.

Zimm plots were constructed using the Berry transformation, as recommended by Andersson,[2] using a first order polynomial fit for both variables to perform the double extrapolation

$$\frac{R}{Kc}(q \to 0, c \to 0)$$

and thereby estimate the intensity weighted radius of gyration, $\langle R_G\rangle_Z$, mass average molar mass, $\overline{M}_W$, and a virial coefficient to represent pairwise interactions amongst particles, $A_2$.

The REPES algorithm was used to determine relaxation rates, $\tau^{-1}(\theta, c)$, that were consistent with a diffusion process from the amplitude correlation function. The intensity weighted mean translational diffusion coefficient, $D$, was then estimated according to the relation $\tau^{-1} = Dq^2$ and thereby the hydrodynamic radius, $\langle R_H\rangle_Z$, calculated according to the Stokes-Einstein equation, $\langle R_H\rangle_Z = \frac{k_B T}{6\pi\eta D}$, where $k_B$ is the Boltzmann constant, $T$ is solution temperature and $\eta(T)$ is the kinematic viscosity of the solvent. Statistical analysis and parameter fitting were conducted using R statistical software and the library 'FME'.[3]

## S2.6 Small-Angle X-Ray Scattering (SAXS) Analysis

Small-angle X-ray scattering (SAXS) measurements were made using a Xenocs Xeuss 2.0 equipped with a micro-focus Cu $K_\alpha$ source collimated with Scatterless slits. The scattering was measured using a Pilatus 300k detector with a pixel size of 0.172 mm × 0.172 mm. The distance between the detector and the sample was calibrated using silver behenate ($AgC_{22}H_{43}O_2$), giving a value of 2.481(5) m. Samples were mounted in 1 mm borosilicate glass capillaries.

## S2.7 Transmission Electron Microscopy (TEM)

TEM observations were performed on a JEOL 2100 electron microscope at an acceleration voltage of 200 kV. All TEM samples were prepared on graphene-oxide (GO)-coated lacey carbon grids (400 Mesh, Cu, Agar Scientific), to enable high contrast TEM images without any staining.[4] Generally, a drop of sample (10 μL) was pipetted onto a grid and left for several minutes, then blotted away. TEM images were analyzed using the ImageJ software, and over 100 particles were counted for each sample to obtain number-average diameter $D_n$ (for spheres), length $L_n$ and width $W_n$ (for worms). Volumes of worms were calculated according to volume, $V = \pi W_n^2 L_n/4$.

## S2.8 Atomic Force Microscopy (AFM)

AFM imaging and analysis were performed on an Asylum Research MFP3D-SA atomic force microscope in tapping mode. Samples for AFM analysis were prepared by drop casting 5 μL of solution (0.1 mg $mL^{-1}$) onto a silicon wafer that had been freshly cleaned with water and ethanol, then activated using plasma treatment to generate a hydrophilic surface.

## S2.9 Confocal Microscopy

Confocal microscopy images were taken using a Zeiss LSM 880 confocal fluorescent microscope. The solution of the assembly being studied (5 μL of a 0.1 mg $mL^{-1}$ in $H_2O$) was dropped onto a plasma-cleaned microscope slide and left to dry overnight. Assemblies tagged with BODIPY-FL dye (green) were excited using a 488 nm laser, while assemblies tagged with BODIPY-TR dye (red) were excited using a 633 nm laser. Both channels were used at the same time to detect the presence of green or red assemblies and overlays were produced.

# S3 SYNTHETIC METHODS

## S3.1 Monomer Syntheses

### S3.1.i Synthesis of N-(3-bromopropyl)acrylamide

*N*-(3-Bromopropyl) acrylamide was synthesized using procedures similar to the previous literature.[1] To a solution of 3-bromopropylamine (10.1 g, 45 mmol), triethylamine (TEA) (14 mL, 100 mmol) and 4-(dimethylamino)pyridine (DMAP) (288 mg, 2.3 mmol) in $CH_2Cl_2$ (150 mL), acryloyl chloride (4.2 mL, 50 mmol) was added dropwise in an ice bath and then left at room temperature for another 4.5 h. The reaction solution was washed with saturated $NaHCO_3$ aqueous solution (100 mL) and water (2 × 100 mL). The organic layer was collected and dried with anhydrous $MgSO_4$ and filtered. Then 2,6-bis(1,1-dimethylethyl)-4-methylphenol (6.3 mg, 1.5 mmol) was added to the filtrate followed by concentration under vacuum to give a brown oil. The brown oil (6.3 g, 73%) was used for the following reaction immediately without further purification. $^1$H NMR (400 MHz, $CDCl_3$) δ = 6.27 (d, *J* = 16.8 Hz, 1H, C*H*=CH-CO), 6.11 (d, *J* = 10.0 Hz, 1H, C*H*=CH-CO), 6.06 (s, 1H, N*H*CO), 5.63 (dd, *J* = 16.8 Hz, 10.0 Hz, 1H, $CH_2$=C*H*-CO), 3.42-3.50 (m, 4H, $CH_2$-$CH_2$-$CH_2$-Br), 2.08-2.15 (m, 2H, $CH_2$-$CH_2$-$CH_2$-Br) ppm; $^{13}$C NMR (400 MHz, $CDCl_3$) δ = 166.0, 130.8, 126.7, 38.2, 32.2, 31.0 ppm.

### S3.1.ii Synthesis of 3-(adenine-9-yl)propyl acrylamide (AAm)

The **AAm** monomer was synthesised according to our previous work.[1] To a suspension of adenine (3.0 g, 24.2 mmol) in dry DMF (100 mL), 60% NaH dispersed in mineral oil (1.0 g, 25.4 mmol NaH) was slowly added in small portions under a nitrogen atmosphere. The mixture was stirred for 1 h until no gas was produced. The viscous mixture was immersed into an ice bath and *N*-(3-bromopropyl) acrylamide freshly synthesized (5.4 g, 28.2 mmol) was added dropwise. The ice bath was left in place and the yellow viscous mixture was stirred overnight. The resulting suspension was concentrated under high vacuum at 50 °C to give a highly viscous oil, to which $CH_2Cl_2$ was added and the contents mixed by gentle swirling. The $CH_2Cl_2$ was then poured off and the process repeated several times, followed by concentration under vacuum. The crude residue was further purified by column chromatography using a mixture of $CH_2Cl_2$ and $CH_3OH$ as eluent and a gradient from 1:0 to 9:1* to give a white solid, **AAm** (3.18 g, 52%). $^1$H NMR (400 MHz, DMSO-*d*6) δ = 8.19 (t, *J* = 5.2 Hz, CON*H*), 8.15 (s, 1H, purine *H*-2), 8.14 (s, 1H, purine *H*-8), 7.20 (s, 2H, N$H_2$), 6.21 (dd, J = 16.8 Hz, 10.0 Hz, 1H, $CH_2$=C*H*-CO), 6.08 (dd, J = 16.8 Hz, 2.0 Hz, 1H, C*H*=CH-CO), 5.59 (dd, J = 10.0 Hz, 2.0 Hz, 1H, C*H*=CH-CO), 4.15 (t, 2H, J =6.8 Hz, C$H_2$-purine) 3.13 (m, 2H, OC-NH-C$H_2$), 1.97 (m, 2H, OC-NH-$CH_2$-C$H_2$-$CH_2$-purine) ppm; 13C NMR (400 MHz, DMSO-*d*6) δ = 165.6, 153.3, 150.4, 148.3, 141.8, 132.6, 126.1, 119.7, 41.8, 36.8, 30.4 ppm; HR-MS (m/z) found 269.1119, calc. 269.1127 $[M+Na]^+$ .

### S3.1.iii Synthesis of 3-benzoylthymine

Following the procedures in a previous report,[1] benzoyl chloride (11.24 mL, 96.8 mmol) and thymine (3.0 g, 24.2 mmol) were suspended in a mixture of acetonitrile (30 mL) and pyridine (12 mL) under nitrogen. The reaction was stirred under a nitrogen atmosphere supplied by a balloon at room temperature overnight. The reaction solution was then concentrated under vacuum. The viscous liquid was partitioned between $CH_2Cl_2$ and water. The aqueous layer was extracted three times with $CH_2Cl_2$ and the combined organic layers were dried over anhydrous $K_2CO_3$. The solvent was removed under vacuum. The residue was dissolved in dioxane (30 mL) and $K_2CO_3$ (4.1 g) in 30 mL of water was added and the reaction mixture was stirred until TLC showed complete conversion to the mono-protected thymine (around 2 h). The crude product was concentrated and colourless crystals crystallised from the solution upon standing at room temperature and further cooling to 4 °C. The crystals were isolated by filtration and washed on the filter with water to remove salt impurities, then dried to yield the final product (4.5 g, 80%). $^1$H NMR (400 MHz, DMSO-*d*6) δ = 11.4 (br, 1H,

* Approximately 500 mL of eluent was used for 1:0, 99:1, 95:5, 93:7 and 91:9 $CHCl_3$:MeOH during the gradient column.

pyrimidine-*H*1), 7.94 (d, *J* = 10.0 Hz, 2H, benzene-*H*1,*H*5), 7.77 (t, *J* = 10.0 Hz, 1H, benzene-*H*3), 7.59 (d, *J* = 10.0 Hz, 2H, benzene-*H*2,*H*4), 7.53 (s, 1H, pyrimidine-*H*6), 1.82 (d, 3H, J = 6.8 Hz, $CH_3$-pyrimidine) ppm; $^{13}C$ NMR (400 MHz, DMSO-$d_6$) δ = 170.7, 164.1, 150.5, 139.3, 135.8, 131.9, 130.7, 129.9, 108.4, 12.2 ppm.

**S3.1.iv Synthesis of 3-(3-benzoylthymin-1-yl)propyl acrylamide**

To a solution of 3-benzoylthymine (2.3 g, 10.0 mmol) in dry DMF (50 mL), 60% NaH (0.42 g, 10.5 mmol NaH) was slowly added. The mixture was stirred for 1 h until no gas was produced. The viscous mixture was immersed in an ice bath and *N*-(3-bromopropyl) acrylamide freshly synthesized (2.3 g, 12.0 mmol) was added dropwise. The ice bath was left in place and the yellow, viscous mixture was stirred overnight. The resulting solution was concentrated under high vacuum at 50 °C. The residue was partitioned between EtOAc and water. The aqueous layer was extracted three times with EtOAc and the combined organic layers were dried over anhydrous $MgSO_4$. The solvent was removed under vacuum. The mixture was further purified by column chromatography using EtOAc as eluent to give a viscous liquid (2.0 g, 58%). $^1H$ NMR (400 MHz, DMSO-$d_6$) δ = 8.17 (t, *J* = 5.2 Hz, CON*H*), 7.96 (d, *J* = 6.0 Hz, 2H, benzene-*H*1,*H*5), 7.79 (s, 1H, pyrimidine-*H*6), 7.77 (t, *J* = 6.0 Hz, 1H, benzene-*H*3), 7.59 (d, *J* = 6.0 Hz, 2H, benzene-*H*2, *H*4), 6.20 (dd, *J* = 17.0 Hz, 10.0 Hz, 1H, $CH_2$=C*H*-CO), 6.10 (dd, *J* = 17.0 Hz, 2.0 Hz, 1H, C*H*=CH-CO), 5.58 (dd, *J* = 10.0 Hz, 2.0 Hz, 1H, C*H*=CH-CO), 3.73 (t, 2H, *J* = 7.0 Hz, $CH_2$-pyrimidine), 3.20 (m, 2H, OC-NH-$CH_2$), 1.84 (d, 3H, *J* = 6.8 Hz, $CH_3$-pyrimidine), 1.82 (m, 2H, OC-NH-$CH_2$-$CH_2$-$CH_2$-pyrimidine) ppm; $^{13}C$ NMR (400 MHz, DMSO-$d_6$) δ = 170.3, 165.2, 163.4, 149.9, 143.0, 135.9, 132.1, 131.7, 130.8, 130.0, 125.6, 109.0, 46.6, 36.3, 28.9, 12.3 ppm.

**S3.1.v Synthesis of 3-(thymin-1-yl)propyl acrylamide (TAm)**

The **TAm** was synthesised as reported previously.[1] (3-Benzoylthymin-1-yl)propyl acrylamide (2.0 g, 5.9 mmol) was dissolved in a mixture of TFA/$CH_2Cl_2$ (3:1) (20 mL). The reaction solution was stirred at room temperature overnight. After completion of the reaction, solvent was removed under vacuum. The residue was purified by column chromatography with a gradient of $CHCl_3/CH_3OH$ from 1:0 to 93:7[†] to give a viscous liquid. Ethanol (20 mL) was then added and the solution cooled to −20 °C to precipitate a white solid[‡], **TAm** (1.0 g, 70%). $^1$H NMR (500 MHz, DMSO-*d*$_6$) δ = 11.23 (s, 1H, pyrimidine-*H*3), 8.12 (t, *J* = 5.2 Hz, CON*H*), 7.51 (s, 1H, pyrimidine-*H*6), 6.18 (dd, *J* = 16.8 Hz, 10.0 Hz, 1H, $CH_2$=C*H*-CO), 6.07 (dd, *J* = 16.8 Hz, 2.0 Hz, 1H, C*H*=CH-CO), 5.58 (dd, *J* = 10.0 Hz, 2.0 Hz, 1H, C*H*=CH-CO), 3.63 (t, 2H, *J* = 6.8 Hz, C*H*$_2$-pyrimidine), 3.14 (m, 2H, OC-NH-C*H*$_2$), 1.74 (d, 3H, *J* = 1.0 Hz, C*H*$_3$-pyrimidine), 1.74 (m, 2H, OC-NH-$CH_2$-C*H*$_2$-$CH_2$-pyrimidine) ppm; $^{13}$C NMR (500 MHz, DMSO-*d*$_6$) δ = 165.1, 164.8, 151.3, 142.0, 132.2, 125.6, 108.9, 45.9, 36.4, 29.1, 12,4 ppm; HR-MS (m/z) found 260.1004, calc. 260.1011 $[M+Na]^+$.

### S3.1.vi Synthesis of 3-(3-methylthymin-1-yl)-propylacrylamide (T$^{Me}$Am)

A mixture of 3-(thymin-1-yl)- propylacrylamide (**TAm**) (71 mg, 0.3 mmol), dry $K_2CO_3$ (66 mg, 0.48 mmol), and iodomethane (75 µL) in anhydrous DMF (0.4 mL) was stirred at room temperature for 24 h and then diluted with ethyl acetate (20 mL), washed with water (2 × 20 mL), and dried with anhydrous $Na_2SO_4$. The solvent was removed under vacuum. The mixture was further purified by column chromatography with a mixture of $CH_2Cl_2/CH_3OH$ (95:5) to give a white solid, **T$^{Me}$Am** (73 mg, 0.29 mmol, 97%). $^1$H NMR (500 MHz, DMSO-*d*$_6$) δ 8.12 (t, *J* = 5.0 Hz, CON*H*), 7.59 (s, 1H, pyrimidine-*H*6), 6.18 (dd, *J* = 17.5, 10.5 Hz, 1H, $CH_2$-C*H*−CO), 6.07 (dd, *J* = 17.5, 2.0 Hz, 1H, C*H*-CH−CO), 5.58 (dd, *J* = 10.5, 2.0 Hz, 1H, C*H*-CH−CO), 3.70 (t, 2H, *J* = 7.5 Hz, C*H*$_2$-pyrimidine), 3.17 (s, 3H, OC−NC*H*$_3$), 3.14 (m, 2H, OC−HN−C*H*$_2$), 1.80 (s, 3H, C*H*$_3$-pyrimidine), 1.76 (m, 2H, OC−NH−$CH_2$−C*H*$_2$−$CH_2$-pyrimidine) ppm. $^{13}$C NMR (125 MHz, DMSO-*d*$_6$) δ 165.1, 163.8, 151.5, 140.4,

† Approximately 500 mL of eluent was used for 1:0, 99:1, 95:5 $CHCl_3$:MeOH during the gradient column.
‡ In some cases further concentration and/or addition of $Et_2O$ was required to trigger the crystallisation.

132.2, 125.5, 107.9, 47.1, 36.3, 29.0, 28.0, 13.1 ppm; HR-MS (m/z) found 274.1165, calcd 274.1162 $[M + Na]^+$.

### S3.1.vii Synthesis of 3-(*N*6,*N*6-dimethyladenine-9-yl)propyl acrylamide ($A^{Me}Am$)

To a suspension of *N*6,*N*6-dimethyladenine (0.16 g, 1.0 mmol) in dry DMF (5 mL), NaH (0.025 g, 1.05 mmol) was slowly added (Scheme S1). The mixture was stirred for 1 h until no gas was produced. The viscous mixture was immersed into an ice bath and 3-bromopropyl acrylamide freshly synthesized (0.23 g, 1.2 mmol) was added dropwise. The yellow viscous mixture was stirred overnight and the resulting suspension was concentrated under vacuum. The obtained mixture was purified by column chromatography using a mixture of $CH_2Cl_2$ and $CH_3OH$ as eluent and a gradient from 1:0 to 95:5 to give a white solid, **$A^{Me}Am$** (0.22 g, 80%). Assigned $^1H$, $^{13}C$ NMR spectra are shown in Figure S1.

$^1H$ NMR (500 MHz, DMSO-$d_6$) $\delta$ = 8.21 (s, 1H, purine *H*-2), 8.17 (s, 1H, purine *H*-8), 8.19 (t, $J$ = 4.5 Hz, 1H, CON*H*), 6.20 (dd, $J$ = 17.0 Hz, 10.0 Hz, 1H, $CH_2$=C*H*-CO), 6.07 (dd, $J$ = 17.0 Hz, 2.0 Hz, 1H, C$H_2$=CH-CO), 5.59 (d, $J$ = 10.0 Hz, 2.0 Hz, 1H, C$H_2$=CH-CO), 4.17 (t, 2H, $J$ = 6.5 Hz, C$H_2$-purine) 3.45 (s, 6H, purine N-(C$H_3$)$_2$), 3.12 (q, 2H, $J$ = 6.5 Hz, OC-NH-C$H_2$), 1.96 (m, 2H, $J$ = 6.5 Hz, OC-NH-$CH_2$-C$H_2$-$CH_2$-purine) ppm. $^{13}C$ NMR (125 MHz, DMSO-$d_6$) $\delta$ = 165.1, 154.7, 152.2, 150.7, 140.2, 132.2, 125.6, 119.7, 41.4, 40.2, 36.3, 29.9 ppm. HR-MS (m/z) found 275.1616, calc. 275.1615 $[M+H]^+$.

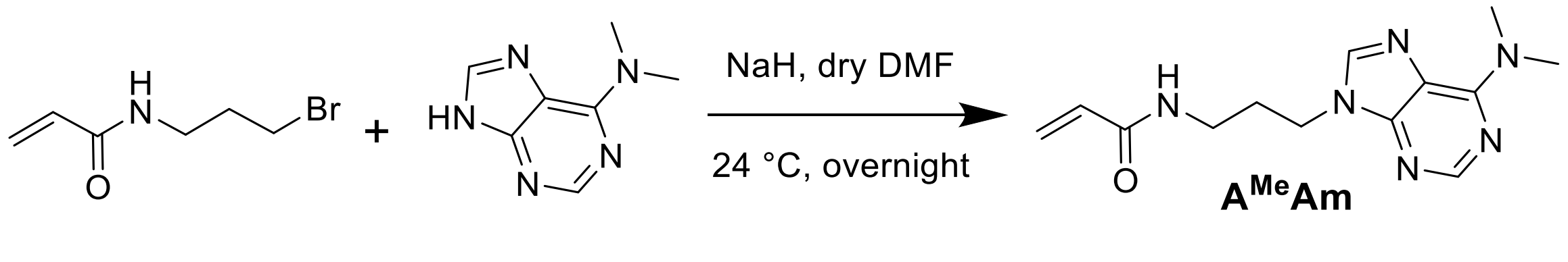

**Scheme S1.** The synthesis of 3-(*N*6,*N*6-dimethyladenine-9-yl)propyl acrylamide (**$A^{Me}Am$**).

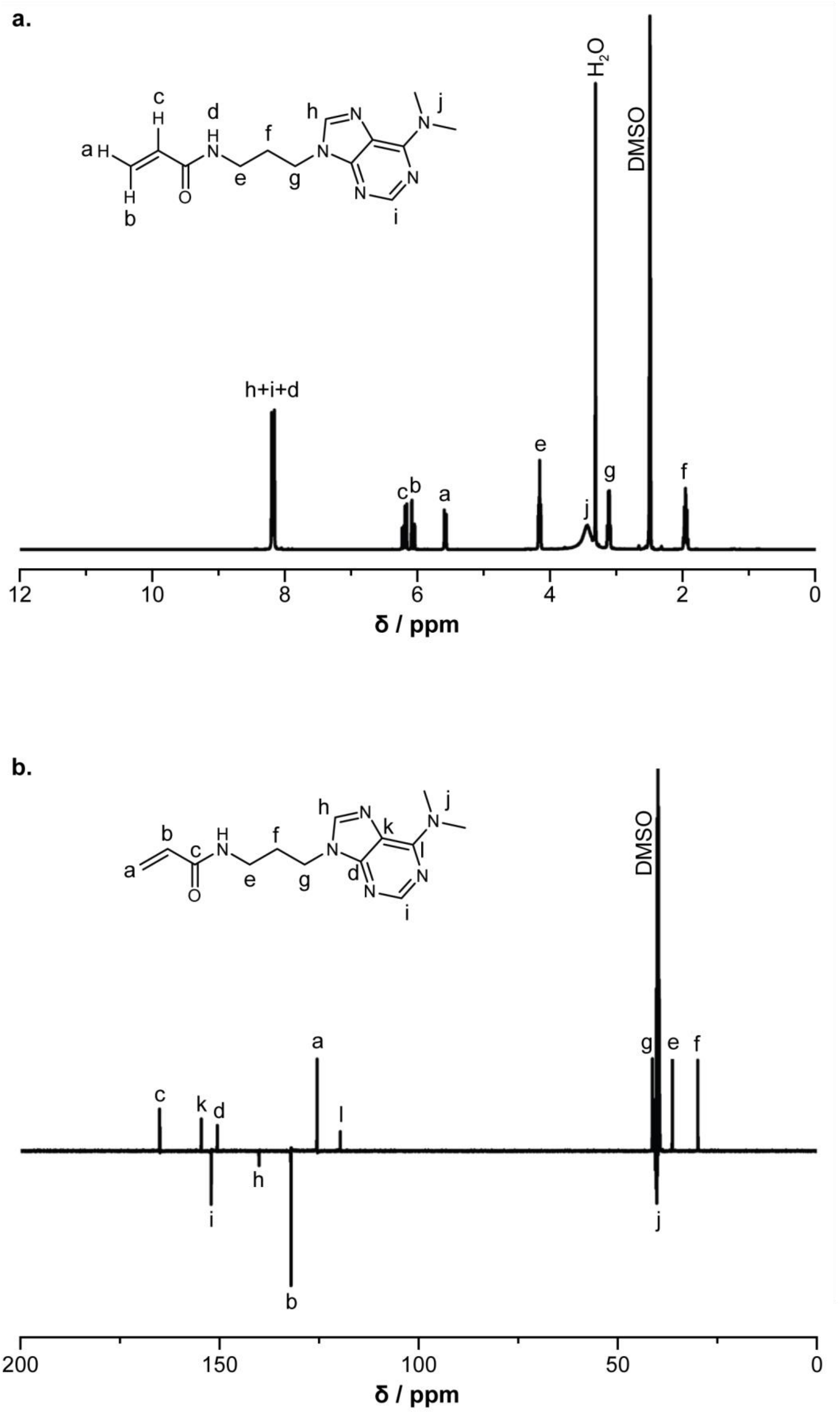

**Figure S1.** Assigned [1]H, [13]C NMR spectra of 3-(*N*6,*N*6-dimethyladenine-9-yl)propyl acrylamide (**A$^{Me}$Am**).

## S3.2 Polymer Syntheses

The synthetic strategies for the macroCTA, **PA** and **PT** are shown in Scheme S2 – the other polymers used were synthesised using similar procedures. Characterization data for all polymers are shown in Table S2, with $^{1}$H NMR spectra presented in Figures S2-3 and S6a, and size exclusion chromatograms in Figures S4-5 and S6b.

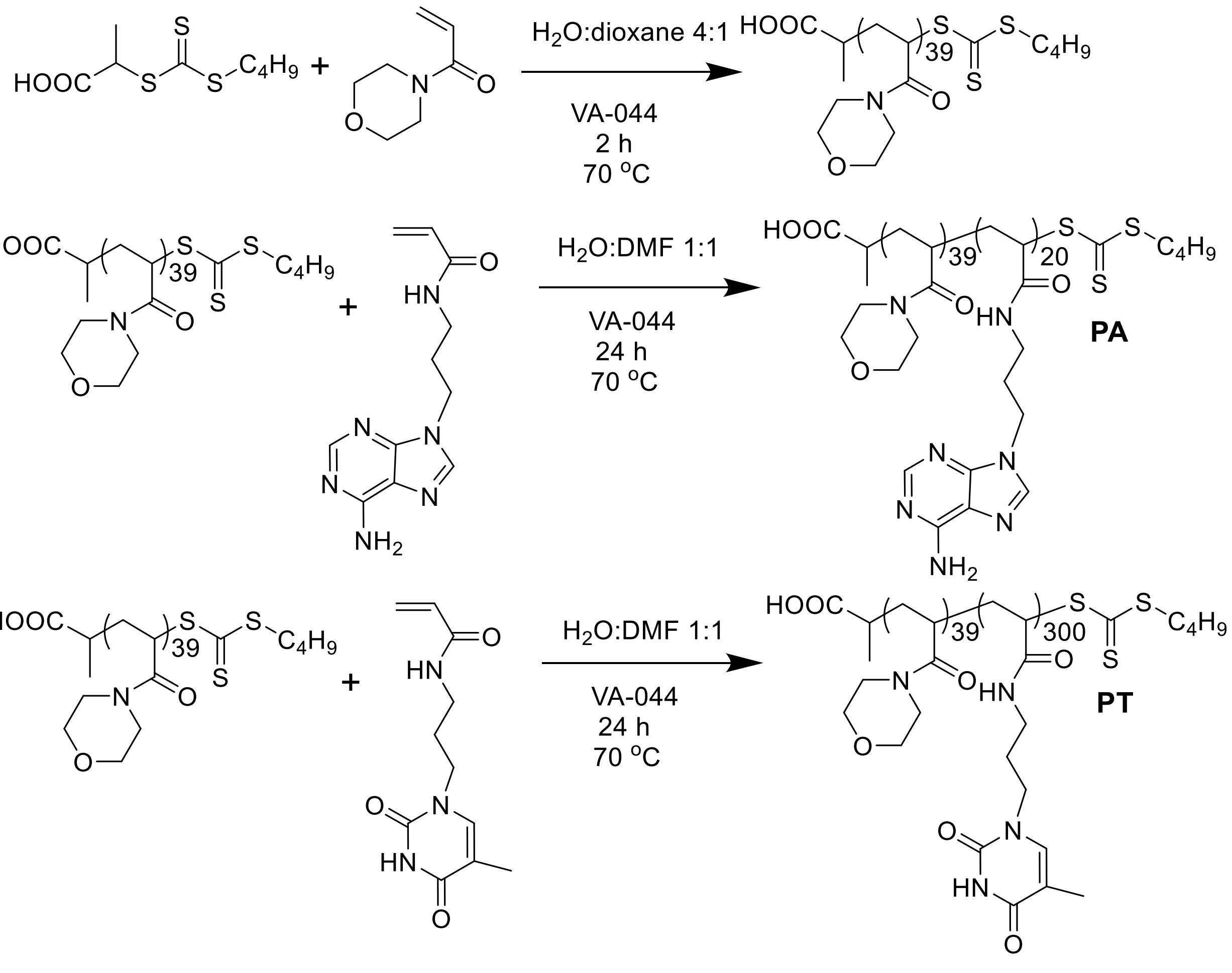

**Scheme S2.** Synthetic routes for PNAM$_{39}$, **PA** (PNAM$_{39}$-*b*-PAAm$_{20}$) and **PT** (PNAM$_{39}$-*b*-PTAm$_{300}$).

### S3.2.i Synthesis of Poly(4-acryloylmorpholine) (PNAM$_{39}$) Macro-CTA *via* RAFT Polymerization

The procedures were similar to our previous work.[1] The typical procedure was as follows. For PNAM$_{39}$, a 10 mL ampoule was charged with NAM (500 μL, 4.0 mmol), 2-(((butylthio)carbonothiolyl)thio)- propanoic acid (23.8 mg, 0.1 mmol), VA-044 (1.3 mg, 0.004 mmol) and a mixture of 1,4-dioxane and water (2.0 mL, v:v 1:4). The mixture was thoroughly degassed *via* 4 freeze-pump-thaw cycles, filled with nitrogen and then immersed in an oil bath at 70 °C for 2 h. The polymerization solution was precipitated three times from cold $CH_3OH$. The light yellow polymer was dried in a vacuum oven overnight at room temperature and analyzed by $^1H$ NMR spectroscopy and DMF SEC (Figures S2 and S4). The degree of polymerization (DP) of this PNAM macro-CTA was calculated to be 39 using $^1H$ NMR spectroscopy by comparing the integrated signals corresponding to the backbone signals ($\delta$ = 1.62 ppm) with those of the methyl group from the CTA ($\delta$ = 0.87 ppm).

**S3.2.ii Syntheses of Diblock Copolymers**

The typical procedure was as follows. For $PNAM_{39}$-*b*-$PTAm_{300}$, $PNAM_{39}$ (14 mg, 0.0025 mmol), **TAm** (178 mg, 0.75 mmol), and VA-044 (0.08 mg, 0.00025 mmol) were dissolved in a mixture of DMF and water (0.5 mL, v:v 1:1). The mixture was thoroughly degassed *via* 4 freeze-pump-thaw cycles, filled with nitrogen and then immersed in an oil bath at 70 °C overnight. An aliquot of the crude product was taken and analyzed by $^1H$ NMR spectroscopy to calculate the conversion. The degree of polymerization (DP) of obtained diblock copolymers was calculated using the conversion from $^1H$ NMR spectroscopy. The residual solution was then precipitated three times from cold $CH_3OH$. The light yellow polymer was dried in a vacuum oven overnight at room temperature and analysed by $^1H$ NMR spectroscopy and DMF SEC. See Table S2 for NMR and SEC characterization of polymers used.

**Table S2.** Characterization data for the macroCTA and nucleobase-containing diblock copolymers.

| Polymer | $M_{n,NMR}$* / kDa | $M_{n,SEC}$† / kDa | $Đ_M$† |
|---|---|---|---|
| $PNAM_{39}$ | 5.7 | 5.9 | 1.07 |
| $PNAM_{39}$-*b*-$PTAm_{300}$ **PT** | 76.9 | 58.5 | 1.33 |
| $PNAM_{39}$-*b*-$PAAm_{20}$ **PA** | 10.7 | 12.6 | 1.07 |
| $PNAM_{39}$-*b*-$PMAAm_{20}$ **$PA^{Me}$** | 11.2 | 9.7 | 1.12 |
| $PNAM_{39}$-*b*-$PTAm_{20}$ **PT1** | 10.5 | 13.1 | 1.07 |
| $PNAM_{39}$-*b*-$PSt_{20}$ **PS** | 7.8 | 8.0 | 1.09 |
| $PNAM_{39}$-*b*-$PMTAm_{300}$ **$PT^{Me}$** | 81.1 | 53.7 | 1.35 |

* Determined by $^{1}H$ NMR spectroscopy (400 MHz) in deuterated DMSO. † Determined by DMF SEC, with poly(methyl methacrylate) (PMMA) standards.

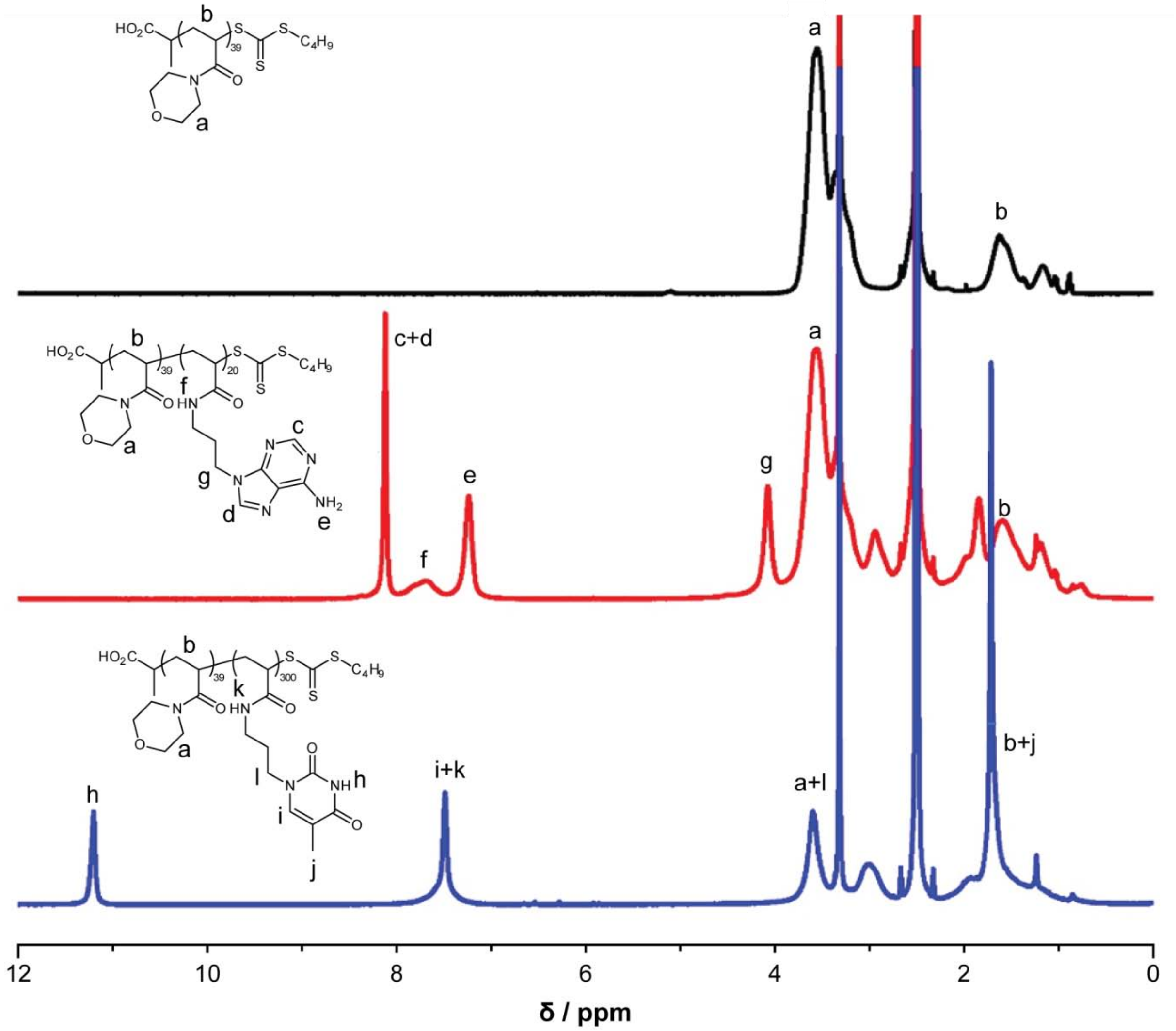

**Figure S2.** [1]H NMR spectra of $PNAM_{39}$, **PA** ($PNAM_{39}$-*b*-$PAAm_{20}$) and **PT** ($PNAM_{39}$-*b*-$PTAm_{300}$) (400 MHz, $d_6$-DMSO).

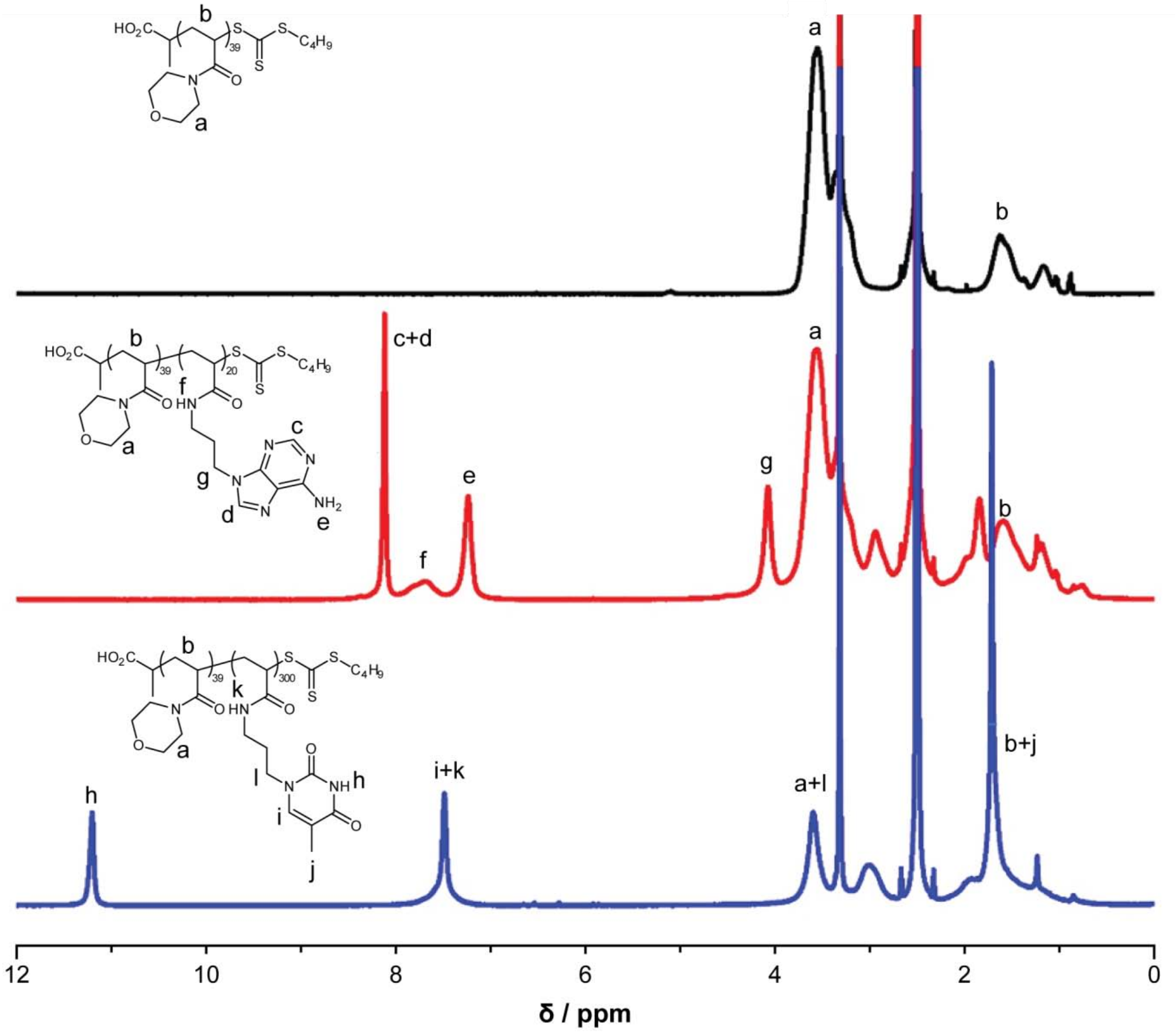

**Figure S3.** $^1$H NMR spectra of **PS** (PNAM$_{39}$-*b*-PSt$_{20}$), **PT1** (PNAM$_{39}$-*b*-PTAm$_{20}$) and **PA$^{Me}$** (PNAM$_{39}$-*b*-PMAAm$_{20}$) (400 MHz, $d_6$-DMSO).

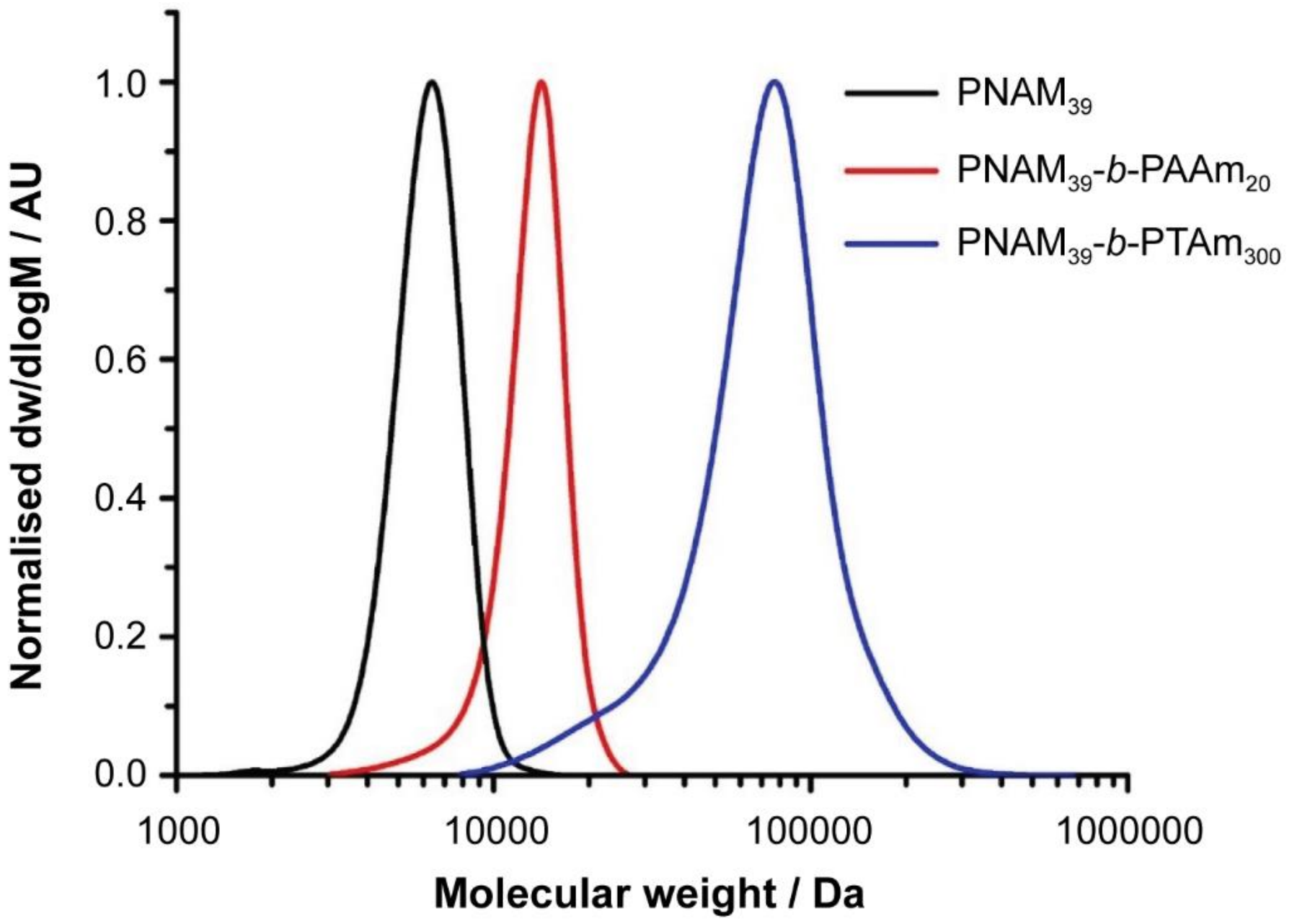

**Figure S4.** Size exclusion chromatograms of $PNAM_{39}$, **PA** ($PNAM_{39}$-*b*-$PAAm_{20}$) and **PT** ($PNAM_{39}$-*b*-$PTAm_{300}$) from DMF SEC using poly(methyl methacrylate) (PMMA) standards.

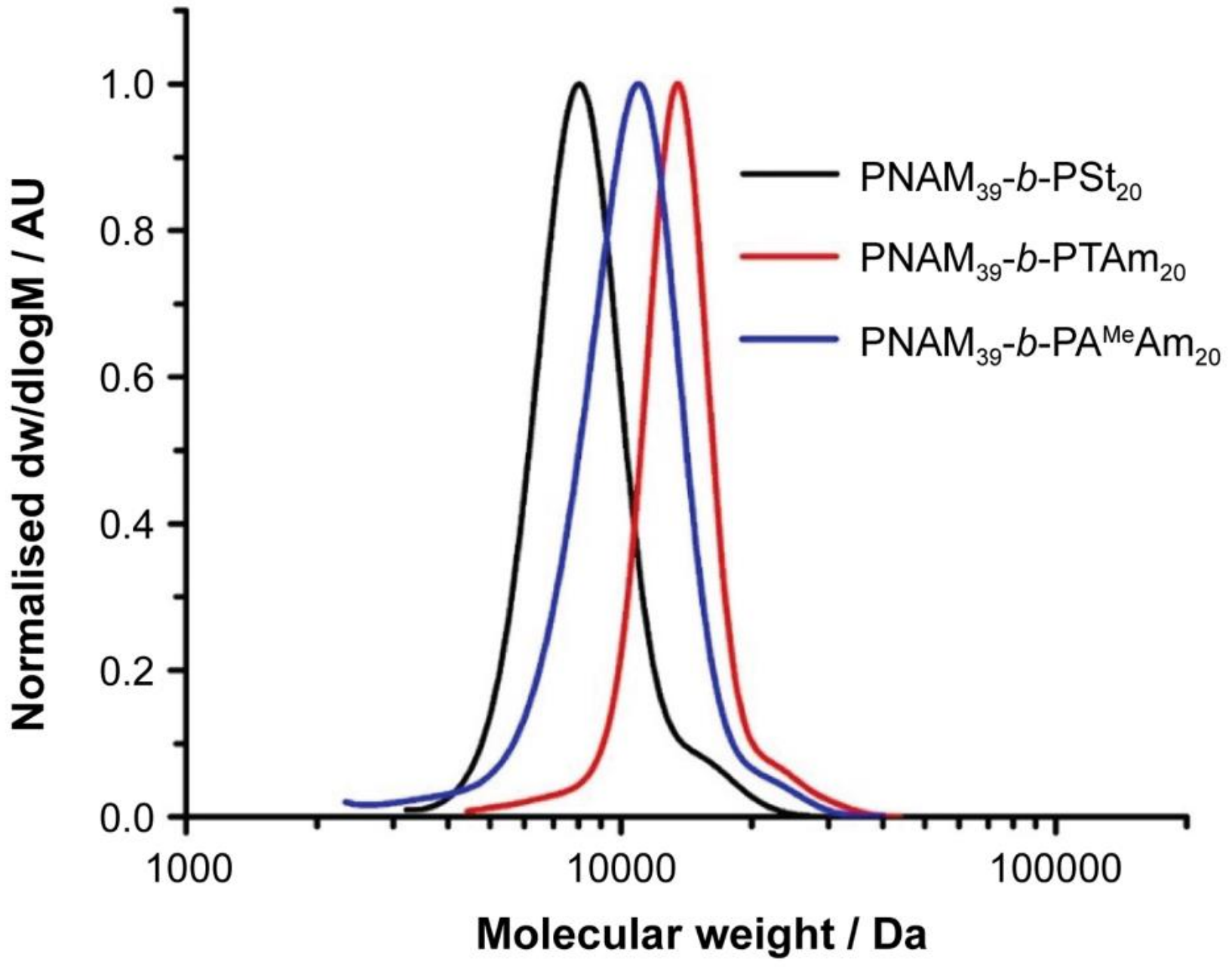

**Figure S5.** Size exclusion chromatograms for **PS** (PNAM39-*b*-$PSt_{20}$), **PT1** ($PNAM_{39}$-*b*-$PTAm_{20}$) and **$PA^{Me}$** ($PNAM_{39}$-*b*-$PMAAm_{20}$) from DMF SEC using poly(methyl methacrylate) (PMMA) standards.

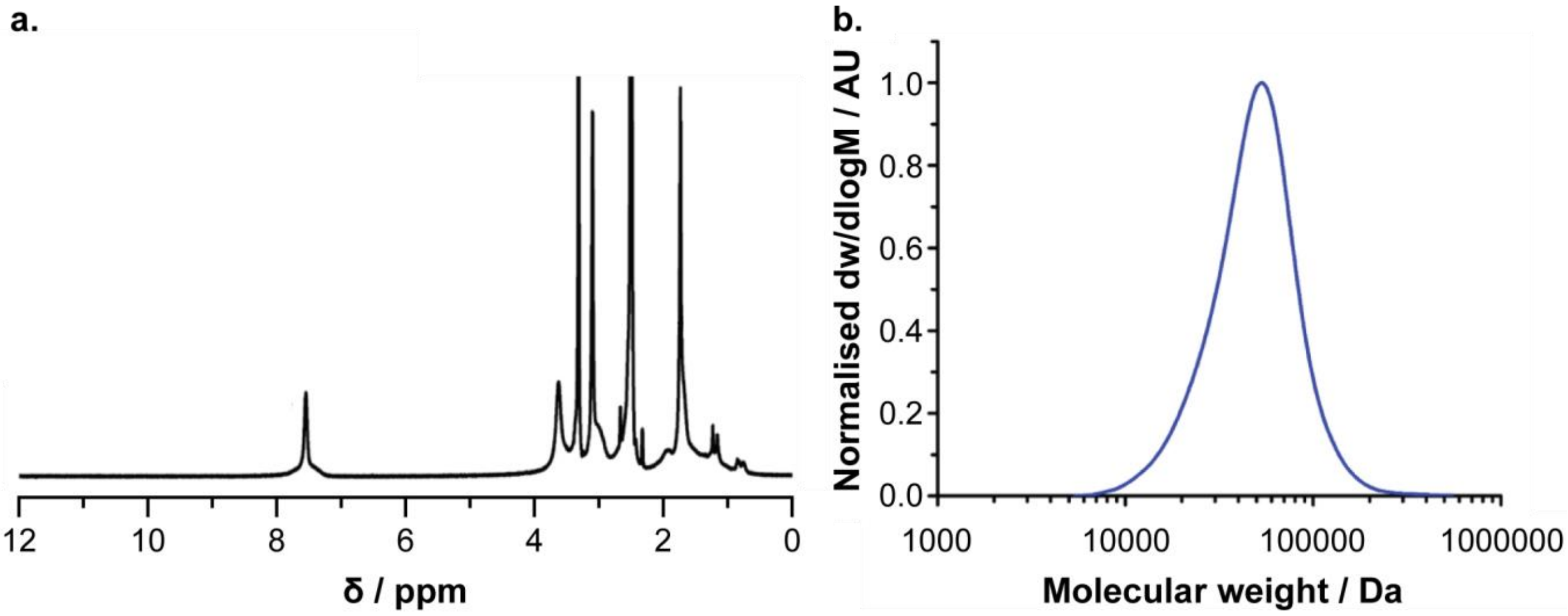

**Figure S6.** (a) $^{1}$H NMR spectrum of **PT$^{\mathbf{Me}}$** (PNAM$_{39}$-*b*-PMTAm$_{300}$) (400 MHz, *d*$_6$-DMSO); (b) Size exclusion chromatogram of **PT$^{\mathbf{Me}}$** (PNAM$_{39}$-*b*-PMTAm$_{300}$) from DMF SEC using poly(methyl methacrylate) (PMMA) standards.

## S4 ASSEMBLY AND CHARACTERISATION OF SEED NANOPARTICLES NT

### S4.1 Self-Assembly of PT in Water

The seed nanoparticles **NT** were assembled as follows. The copolymer was dissolved in DMF (at 8 mg $mL^{-1}$) and stirred for 2 h at 70 °C. Then an excess of 18.2 MΩ·cm water was added *via* a syringe pump at a rate of 1 mL $h^{-1}$. The final volume ratio between water and organic solvent was about 8:1. The solution was then dialyzed against 18.2 MΩ·cm water, incorporating at least 6 water changes, to afford self-assemblies **NT** at a concentration of *ca.* 1 mg $mL^{-1}$.

### S4.2 Analysis of Seed Nanoparticles NT

DLS analysis of **NT** is presented in Figure S7. See the main paper for TEM images and section S8.1 for further SLS characterisation, and section S5.2.iii for cryoTEM images.

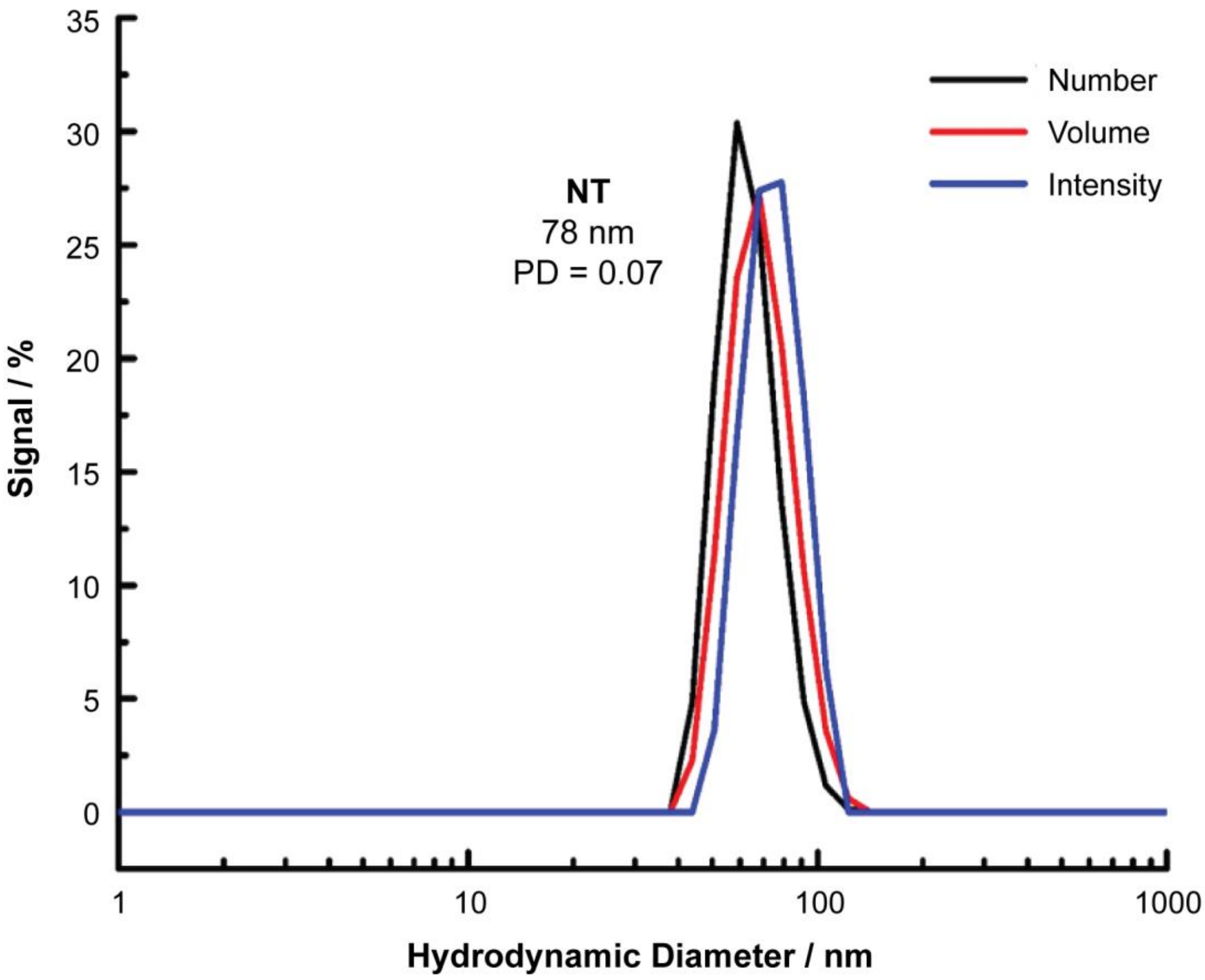

**Figure S7.** DLS analysis of nanoparticles **NT** (0.5 mg mL$^{-1}$) in water.

## S5 MORPHOLOGICAL TRANSFORMATION OF SEED NANOPARTICLES NT

### S5.1 Addition of PA to NT

The typical procedure was as follows. The diblock copolymer $PNAM_{39}$-*b*-$PAAm_{20}$ **PA** was dispersed in $H_2O$ at 5 mg $mL^{-1}$. This was then added to separate solutions of the nanoparticle **NT** (0.5 mg $mL^{-1}$) with stirring at A:T molar ratios of 0.07, 0.20, 0.33, 0.67, 1.0, 1.33. The molar ratios were calculated according to the $M_n$ determined from $^1$H NMR spectroscopic analyses and the polymers' mass concentrations. The mixtures were then sealed and allowed to stir at room temperature for 2 h. The solutions were then characterized by DLS and TEM.

### S5.2 Analysis of the Morphological Transformation Products

#### S5.2.i DLS Analysis of the Transformation Products

The very long worms generated by the morphological transformation process were not amenable to analysis by DLS, since this assumes a spherical particle. However, the dumbbells formed by addition of **PA** to **NT** at an A:T molar ratio of 0.20 were small and compact enough for effective DLS analysis, which is presented in Figure S8.

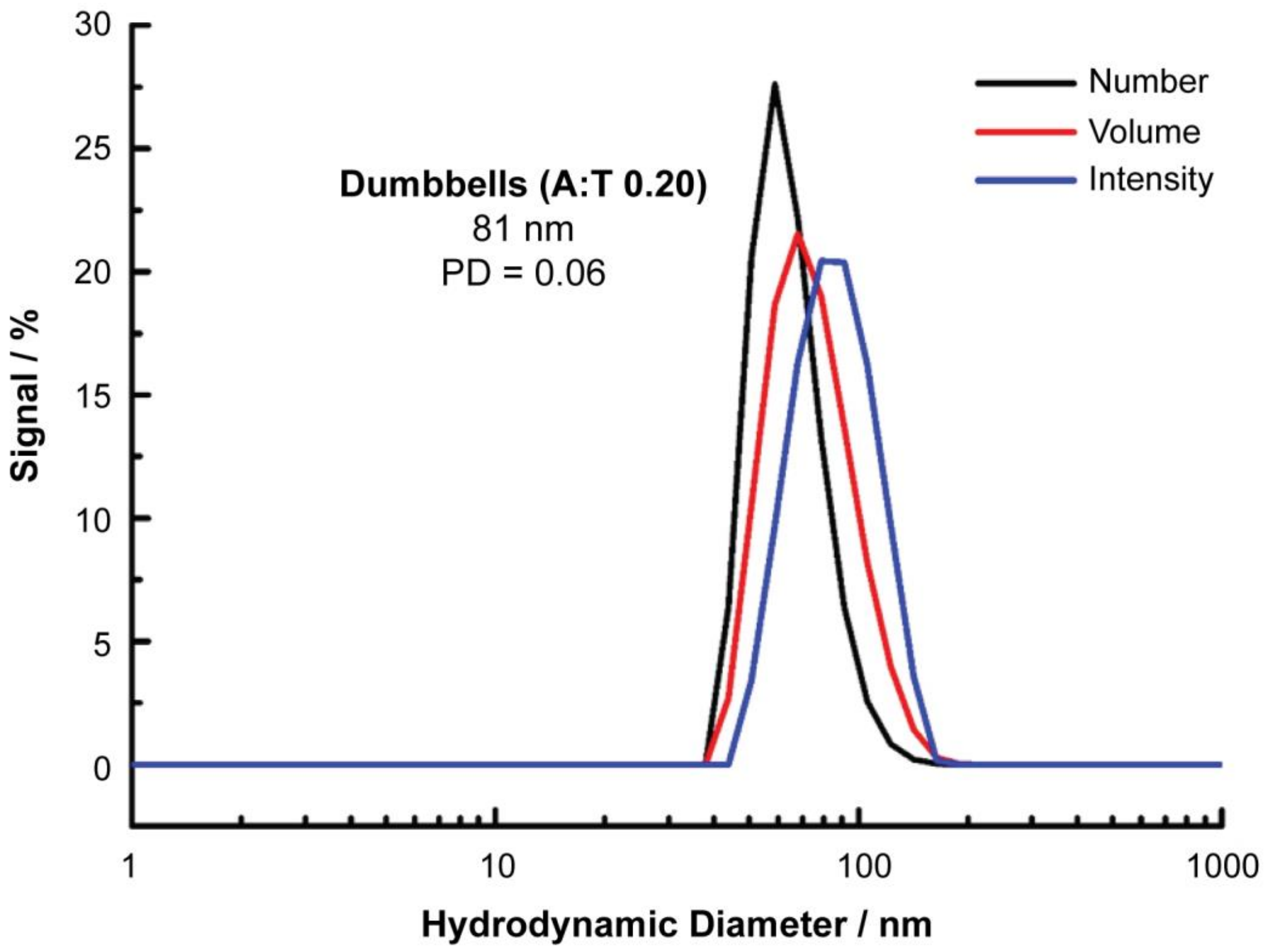

**Figure S8.** DLS analysis of dumbbells (0.5 mg $mL^{-1}$) formed from **NT** after adding **PA** at an **A:T** molar ratio of 0.20.

### S5.2.ii TEM Analysis of the Transformation Products

Further TEM images of the transformation products formed by the addition of a single dose of **PA** at different A:T molar ratios are presented in Figure S9.

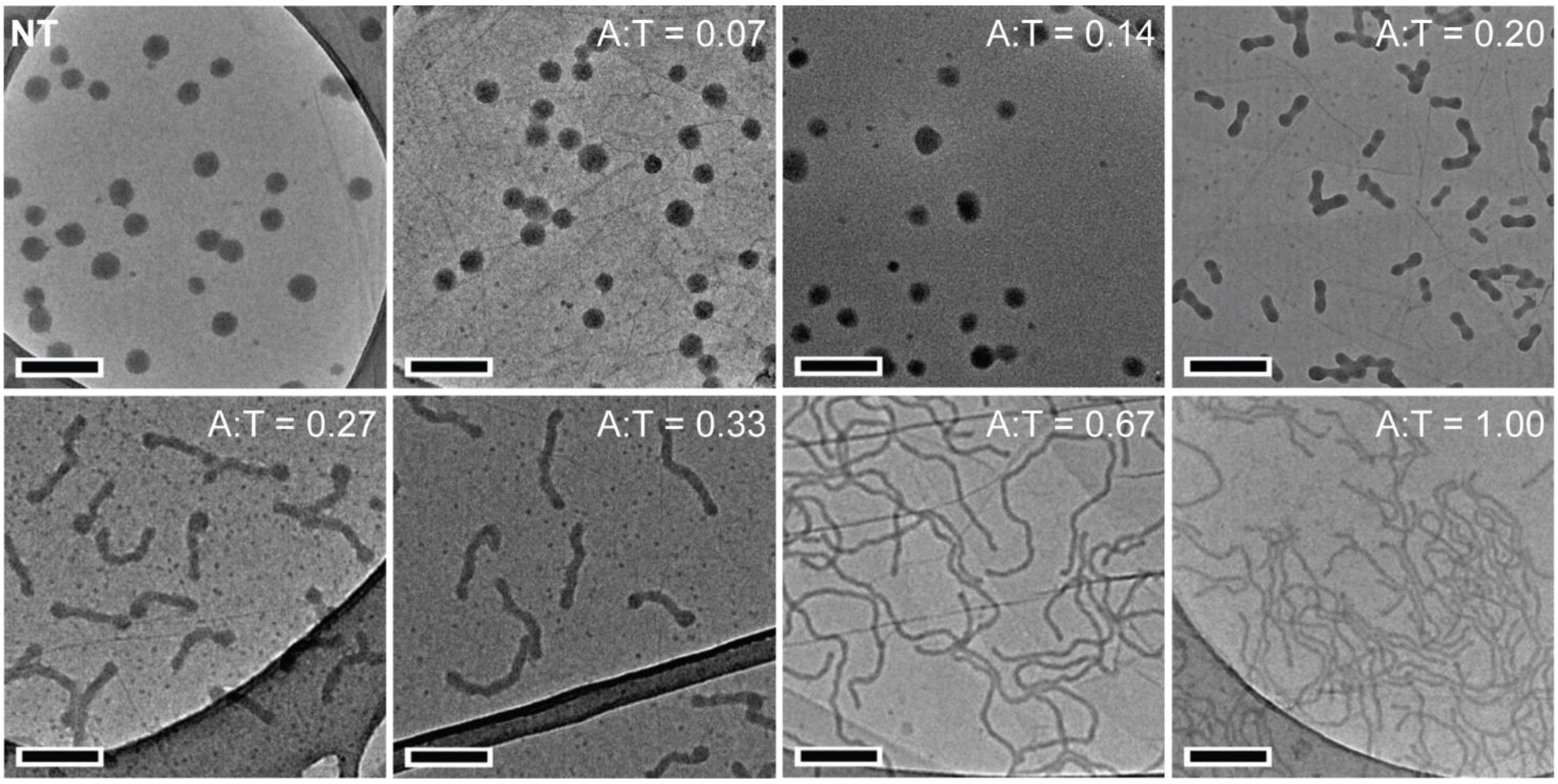

**Figure S9.** Further TEM images of the transformation products formed by the addition of a single dose of **PA** to **NT** at the stated A:T molar ratios.

### S5.2.iii CryoTEM Analysis of the Transformation Products

To rule out the possibility that the morphological transformation products were artefacts observed in dry state TEM, we performed cryoTEM on the same samples. Images are presented in Figure S10, which demonstrate that the structures were not drying artefacts.

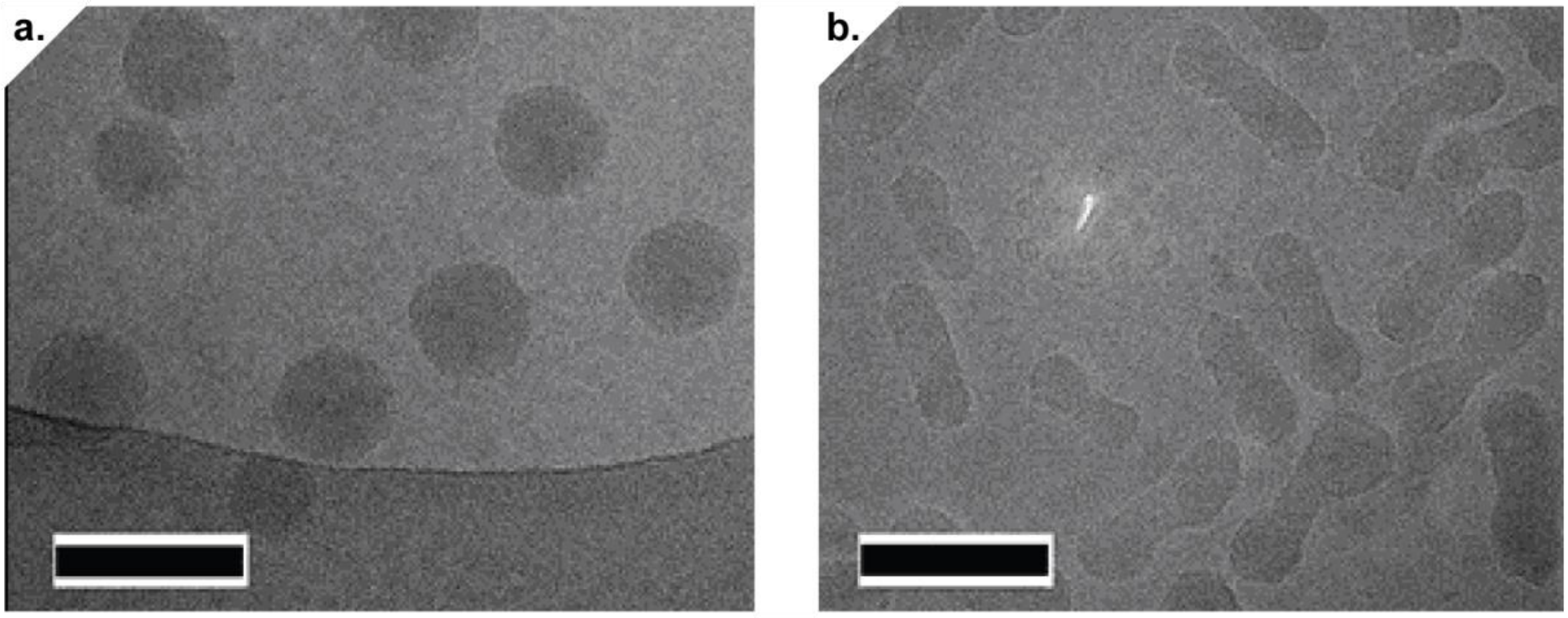

**Figure S10.** Cyro-TEM images of (a) spherical nanoparticles **NT**; (b) dumbbell-like micelles formed by adding **PA** to **NT** at an A:T molar ratio of 0.20. Scale bars = 100 nm.

### S5.2.iv LS and SAXS Analyses of the Transformation Products

The morphologies of the transformation products were investigated using LS and SAXS. The aim of these analyses was to verify that the particles observed by TEM were representative of the bulk sample.

- LS: This was accomplished by comparing the values obtained for $\langle R_G \rangle_Z$ and $\langle R_H \rangle_Z$. The ratio $R_G/R_H$ may be interpreted as measure of a particle's compactness, e.g. the theoretical value for a homogeneous sphere is $R_G/R_H = 0.775$ and the value for a random coil $R_G/R_H \approx 1.5$.[5] As shown in Table S3, the empirical values for this quantity, $\langle R_G \rangle_Z / \langle R_H \rangle_Z$ , increase in correspondence with an increase in A:T molar ratio, in a way that is consistent with the respective particles exhibiting greater anisotropy.

1. SAXS: This was accomplished by fitting an appropriately parameterised model to the empirical form factor. The analytical expressions for (a) a homogeneous sphere and (b) a homogeneous ellipsoid are given by Pedersen[6], i.e.

   a) Homogeneous sphere:

$$P_{SPH}(q,R) = [F]^2\,, \qquad F(q,R) = \frac{3(\sin(qR) - qR \cdot \cos(qR))}{(qR)^3}$$

   b) Homogeneous ellipsoid (semi-axes $R$, $R$, $\varepsilon R$):

$$P_{ELIP}(q,R,\varepsilon) = \int_0^{\frac{\pi}{2}} F^2(q,r) \sin\alpha \, \mathrm{d}\alpha \quad , \qquad r(R,\varepsilon,\alpha) = R(\sin^2\alpha + \varepsilon^2\cos^2\alpha)^{\frac{1}{2}}$$

The SAXS data are shown in relation to these various models in Figure S11, confirming the spherical form of **NT** (Figure S11a) and the ellipsoidal form of particles for **A:T=0.14** (Figure S10b) in keeping with the predictions of the physical model (see discussion section in main paper and section S10). No analytical expression being available, a dumbbell form factor based on TEM measurements was generated using a Monte Carlo method (Figure S10c) and found to fit well to the experimental data at high $q$ (Figure S10d), whilst some discrepancy in the region $2 \leq u \leq 5$ might be explained by variation in the length and thickness of the particle's central region.

**Table S3.** Summary of LS characterization data for spherical nanoparticle, **NT**, ellipsoids (A:T = 0.14) and dumbbells (A:T = 0.20).

| Sample | $\langle R_G \rangle_Z$ , nm | $\langle R_H \rangle_Z$ , nm | $\frac{\langle R_G \rangle_Z}{\langle R_H \rangle_Z}$ |
|---|---|---|---|
| **NT** | 30.8 ± 1.4 | 38.2 ± 0.2 | 0.81 |
| **A:T = 0.14** | 40.8 ± 0.9 | 42.9 ± 0.3 | 0.95 |
| **A:T = 0.20** | 44.9 ± 0.9 | 45.5 ± 0.3 | 0.99 |

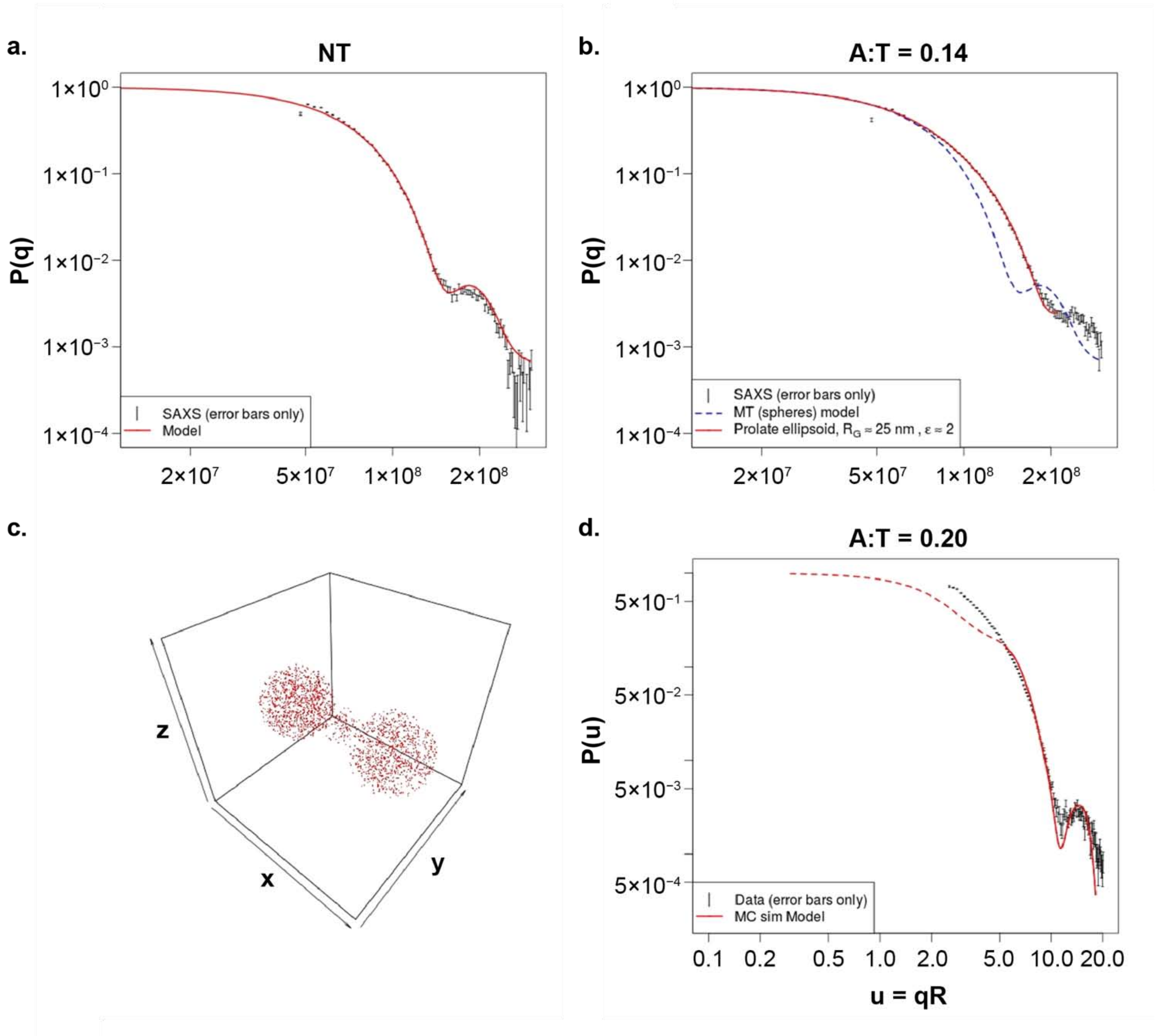

**Figure S11.** SAXS experimental profiles and fittings of: (a) spherical nanoparticles **NT** and (b) nanoparticles with an A:T molar ratio of 0.14, showing the fit to a prolate ellipsoid model. (c) Monte Carlo simulation of the form factor for a dumbbell based on dimensions estimated from TEM images. Pairwise distances were generated either from the enclosed mass to estimate $R_G$ and $P_{DUMB}(q, \boldsymbol{R})$, or from the surface to estimate $R_H$. (d) SAXS experimental profiles and fittings (using Guinier model) of dumbbells at an A:T ratio of 0.20.

# S6 STEPWISE MORPHOLOGICAL TRANSFORMATION OF NT

## S6.1 Morphological Transformation at Low PA Concentrations

We attempted to perform stepwise morphological transformation of the nanoparticles **NT** by adding **PA** in small aliquots (0.07 molar equivalents per addition). However, as shown in Figure S12, this resulted only in swelling and partial disassembly of the nanoparticles, rather than shape change.

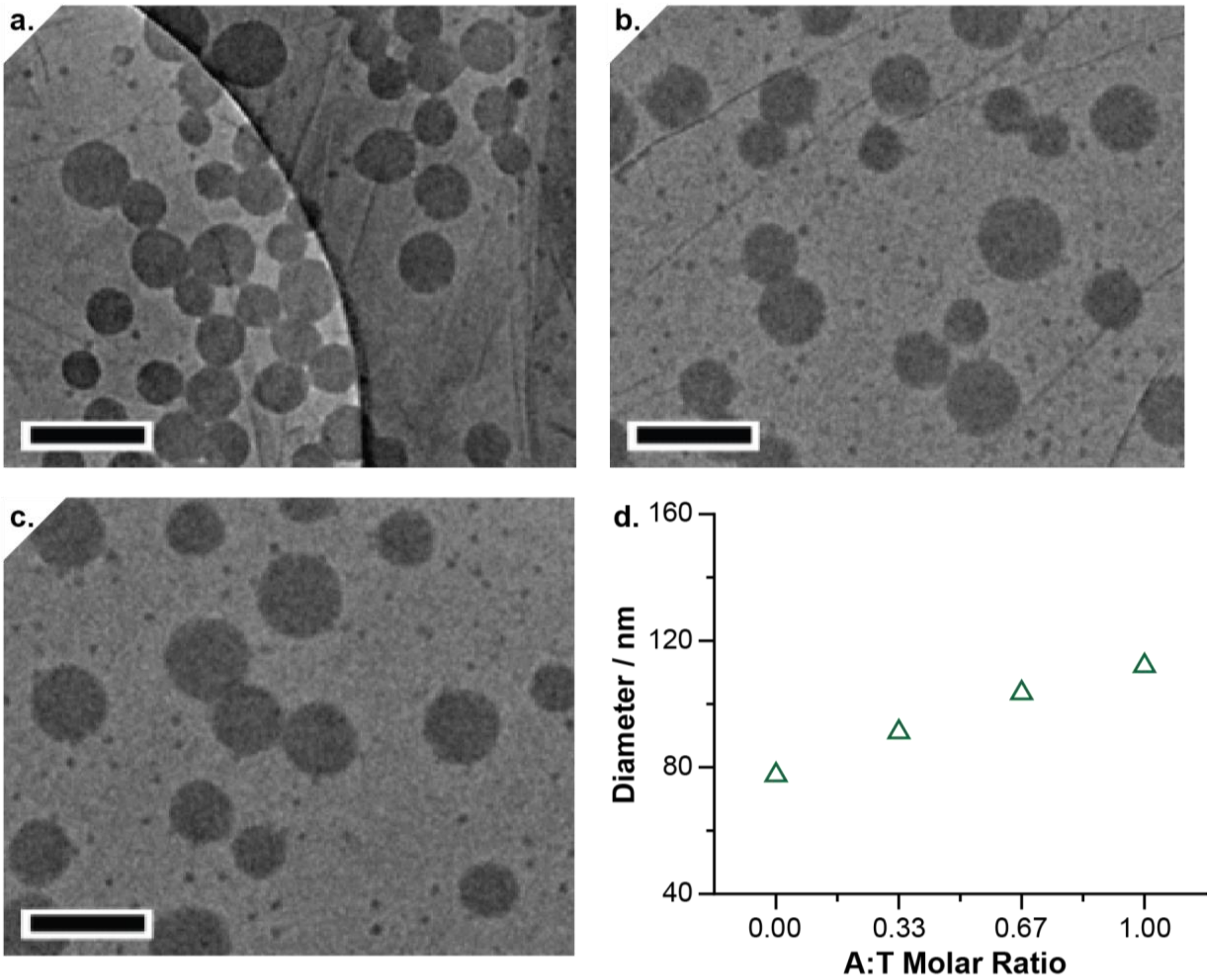

**Figure S12.** TEM images and DLS analyses of nanoparticle **NT** after stepwise addition of low concentrations of **PA** (0.07 molar equivalents relative to thymine per addition) at the following A:T molar ratios: (a) 0.33; (b) 0.67; (c) 1.00; scale bars = 200 nm. (d) Variation of hydrodynamic diameters of the particles shown in (a-c) as measured by DLS.

## S6.2 Stepwise Growth of Anisotropic Nanoparticles

### S6.2.i Stepwise Transformation of Dumbbells

To test the hypothesis that stepwise growth might be possible once anisotropy had been introduced, we first took a sample of the dumbbells (A:T ratio 0.20) and added **PA**, then analysed the resulting particles by TEM (Figure S13). As expected, the dumbbells elongated to form worms of around 300 nm length, suggesting that stepwise growth would be possible.

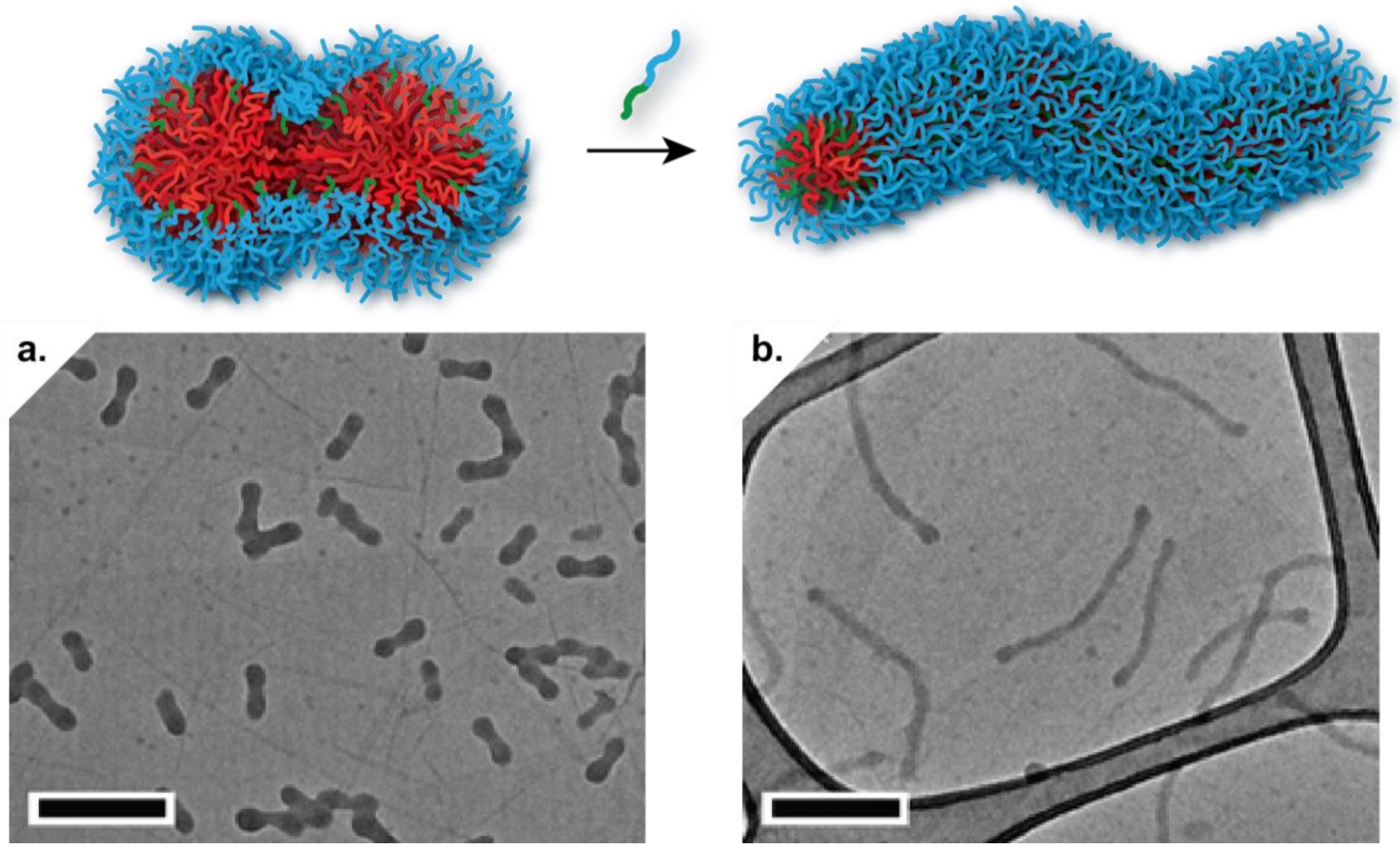

**Figure S13.** Schematic presentation and TEM images of (b) worm-like micelles with lengths over 300 nm formed by further feeding (a) dumbbell-like micelles with **PA** at a total A:T molar ratio of 0.33; scale bars = 200 nm.

### S6.2.ii Stepwise Growth of Long Wormlike Nanoparticles

Having confirmed that stepwise growth was possible, we moved on to the experiments described in Figure 3 of the main paper. The procedure was as follows: a solution of **PA** (0.33 molar ratio of A relative to T) was added to the nanoparticle **NT** solution (0.5 mg $mL^{-1}$) to give short “seed” worms.

After 2 h stirring, further **PA** solution (0.07 molar ratio A relative to T) was added. This process was repeated until A:T ratios of 0.33, 0.40, 0.53 and 0.67 were achieved. Each stage was characterised by TEM and SAXS analyses. Histograms of the TEM particle counting data are presented in Figures S14-15.

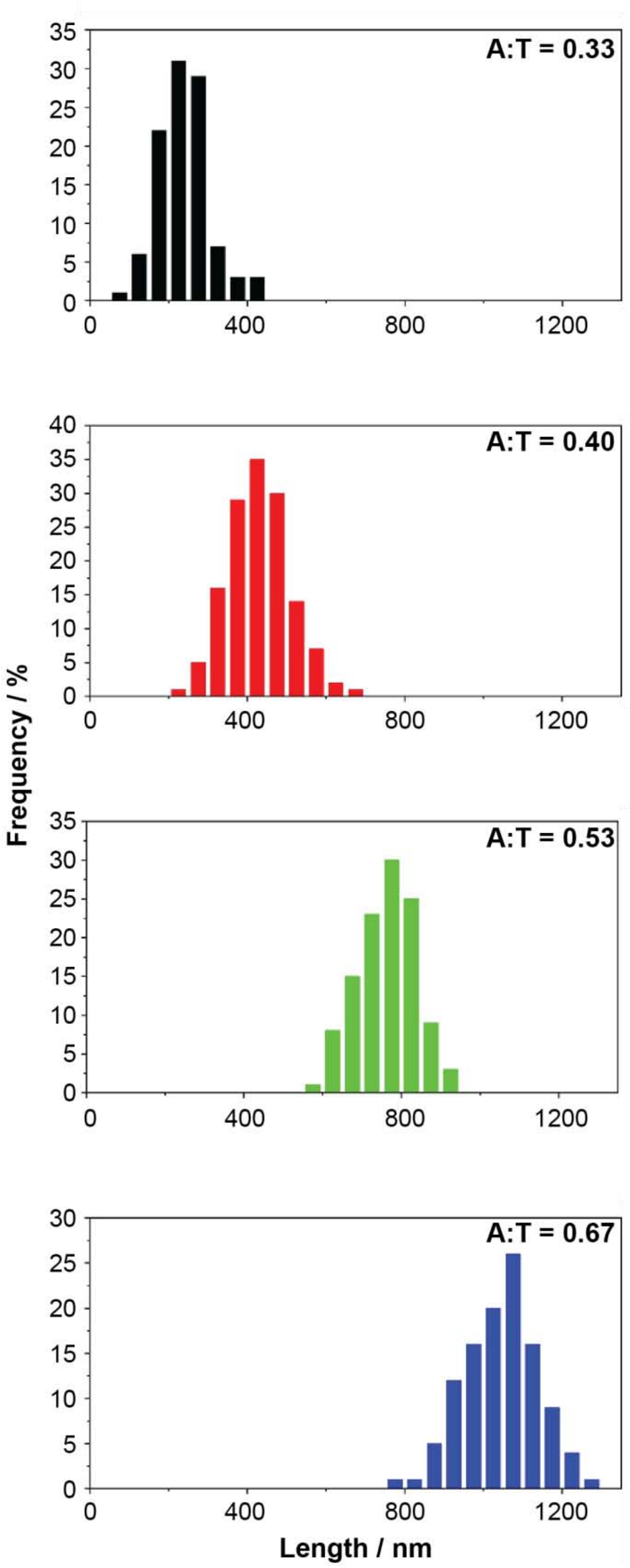

**Figure S14.** Histograms showing the length distributions of the worms grown by stepwise addition of **PA**. Particles were imaged dry on graphene oxide by TEM and measured using ImageJ. The A:T molar ratios in the nanoparticles were: (a) 0.33; (b) 0.40; (c) 0.53; (d) 0.67.

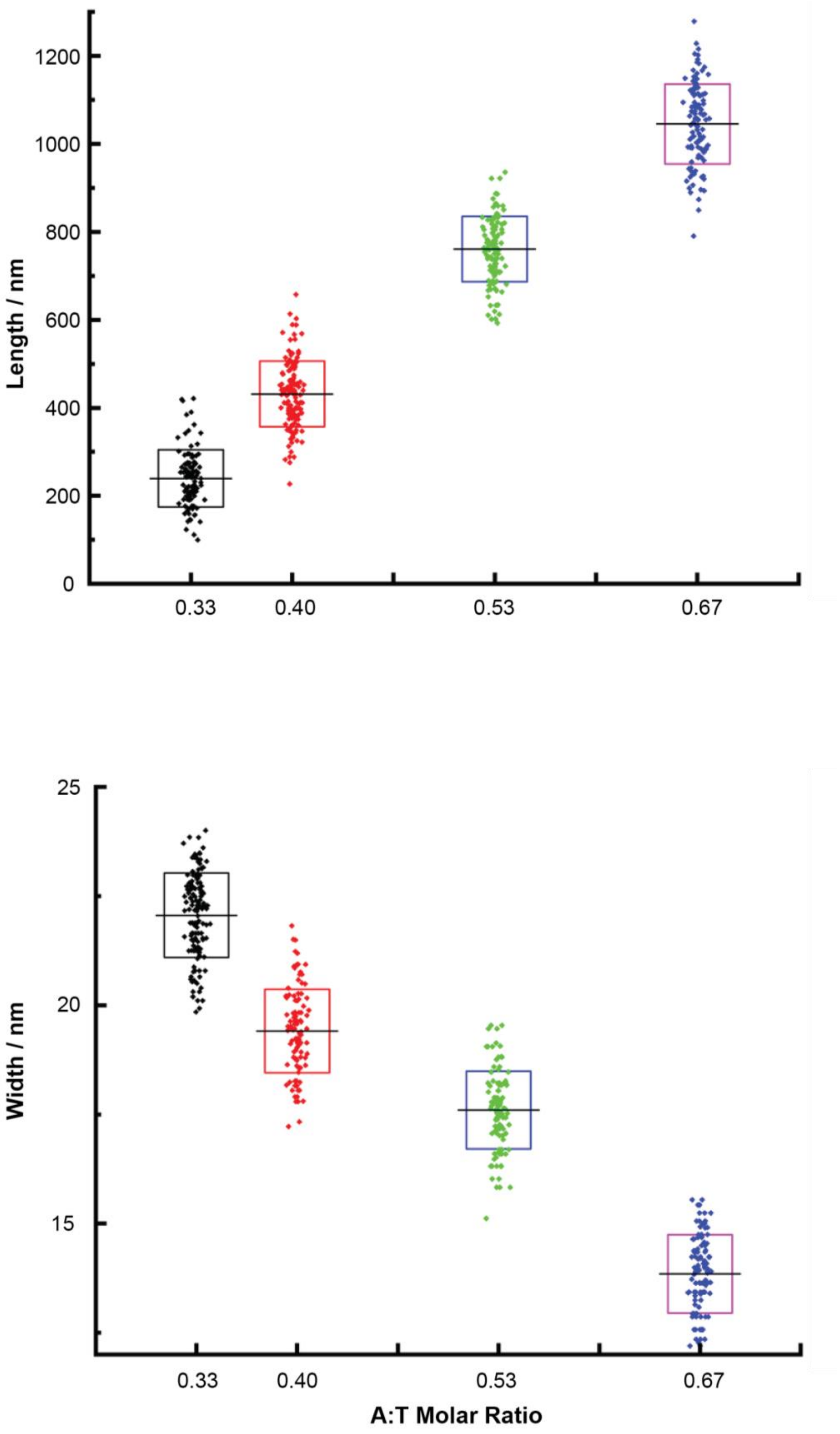

**Figure S15.** Jitter box plot of the (a) lengths and (b) width distributions of the worms fabricated by stepwise growth, as determined by TEM. The central line indicates the mean, and the box encompasses the 95% confidence interval.

### S6.2.iii SAXS Analysis of Wormlike Nanoparticles

An analytical expression for the scattering form factor of a highly anisotropic particle, e.g. a worm-like topology, is given by Pedersen,[6]

$$P_{WORM}(q, \boldsymbol{R}) = P_{CS} P_{SH}$$

where $P_{CS}$ is a component to describe the local shape, e.g. cylindrical, and $P_{SH}$ describes the global shape of the particle. We assumed a circular cross-section radius $r$, i.e.

$$P_{CS}(q, r) = \left(\frac{2J_1(qr)}{qr}\right)^2$$

where $J_1$ is a Bessel function of the first kind. For the global shape component, we assumed a random coil model to be appropriate for $P_{SH}$,

$$P_{SH}(q, \langle R_G^2 \rangle) = \left(\frac{2}{u^2}\right) \cdot [\exp(-u) + u - 1] \ , \qquad u = \langle R_G^2 \rangle q^2 \ , \qquad \langle R_G^2 \rangle = \frac{Lb}{6}$$

But, observing that the relation $\ln(P_{SH}) \sim \ln(q)$ is approximately linear for the combination of worm lengths, L, apparent by TEM and $q$ range accessible to the SAXS experiment, we further simplified the problem in hand to that of only fitting a two parameter distribution for cross-sectional radius, such that

$$r \sim N(\mu, \sigma)$$

The cross-sectional component to the intensity weighted average scattering form factor was then approximated by a sum for values $r_i$ that were sampled from the parameterised distribution:

$$P_{CS}(q, \langle r \rangle_Z) \approx \frac{\sum_i n_i w_i^2 P_{CS}(q, r_i)}{\sum_i n_i w_i^2} \ , \qquad w_i \propto r_i^2$$

Fits to the data are shown in Figure S16. Results are summarised in Figure 3g (main article) and in Table S4.

TEM and SAXS measurements for worm cross-section were analysed by linear regression with the method of measurement, 'TEM' or 'SAXS', incorporated as a factor. According to the fitted regression model, there is very strong evidence to suggest that for each increase in the A:T molar ratio there is a corresponding decrease in worm diameter and this is a statistically significant effect ($F(3,4) = 41.1$ , R-squared = 0.969). On average, an increase in A:T molar ratio of 0.1 is associated with a 2.3 nm decrease in worm diameter ($t = -5.92$, $p = 0.004$). There is no statistical evidence to suggest that the trend in the data is different according to the method of measurement ($t = 0.189$, $p = 0.860$).

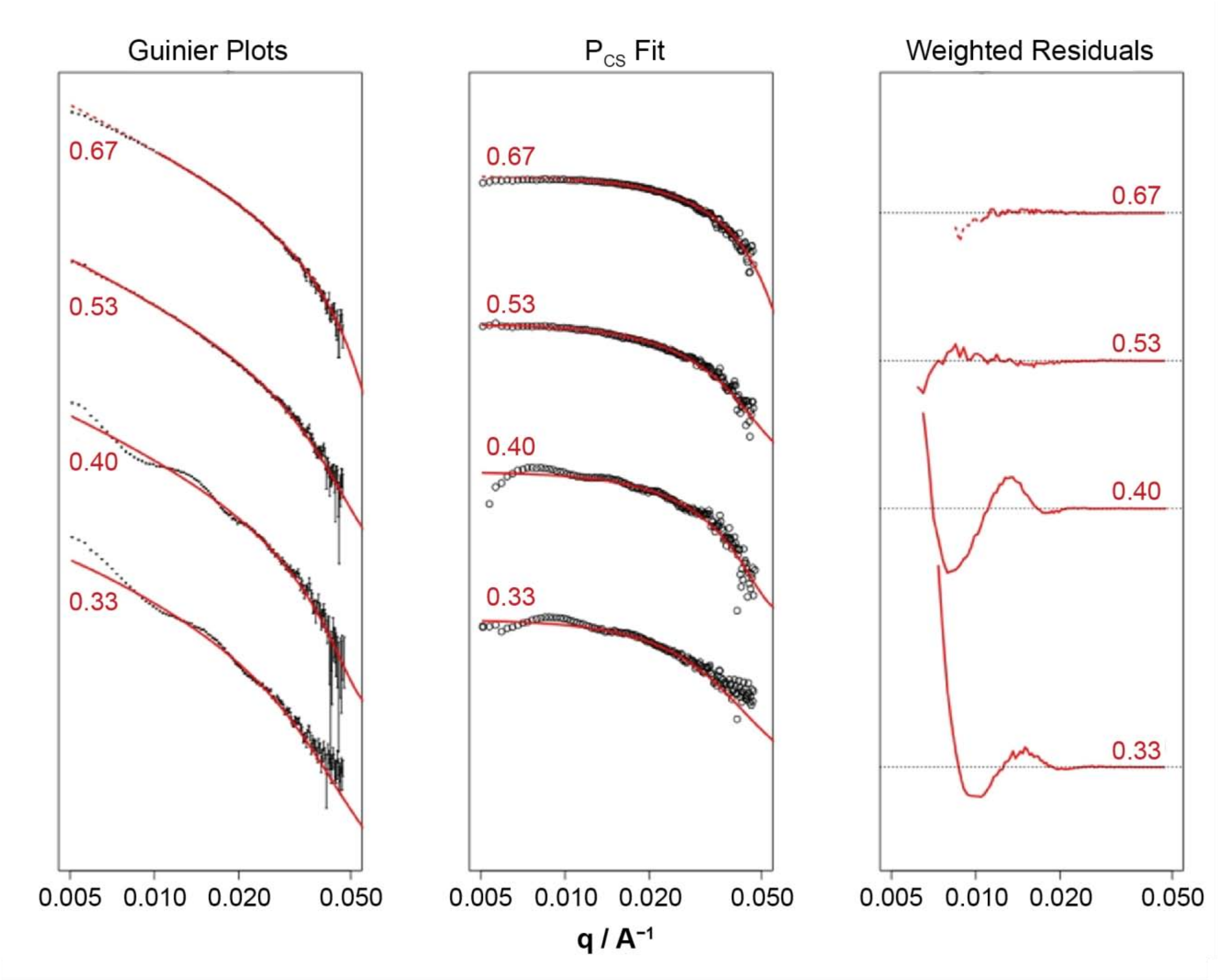

**Figure S16.** Estimation of radius, *R*, from SAXS data, for four samples of worms grown to different mean lengths. Note that in each of the above figures the data are offset for clarity. (a) Guinier plots of the SAXS data and the fit *I*(*q*) / *PRC* · *PCS*. (b) Fit to estimate radius by variation of two parameters, assuming that $R \sim N(\mu R; \sigma R)$. (c) Weighted residuals, normalized to absolute scattering intensity per sample for comparison.

**Table S4.** Summary of SAXS characterization data for cross-sectional dimensions of worms of different lengths. N.B. These results are presented in Figure 3g as diameter, rather than radius, for direct comparison with the measurements from TEM images.

| Sample | $\mu_r = \langle r \rangle_N$ / nm | $\sigma_r$ / nm |
|---|---|---|
| **A:T = 0.33** | 10.4 | 2.0 |
| **A:T = 0.40** | 9.8 | 1.9 |
| **A:T = 0.53** | 8.9 | 1.0 |
| **A:T = 0.67** | 6.5 | 0.8 |

### S6.2.iv Worm Disassembly at High A:T Ratios

Addition of **PA** above A:T ratios of 1.00 was observed to cause disassembly of the nanoparticles into short worms and spheres, as shown in the TEM image in Figure S17.

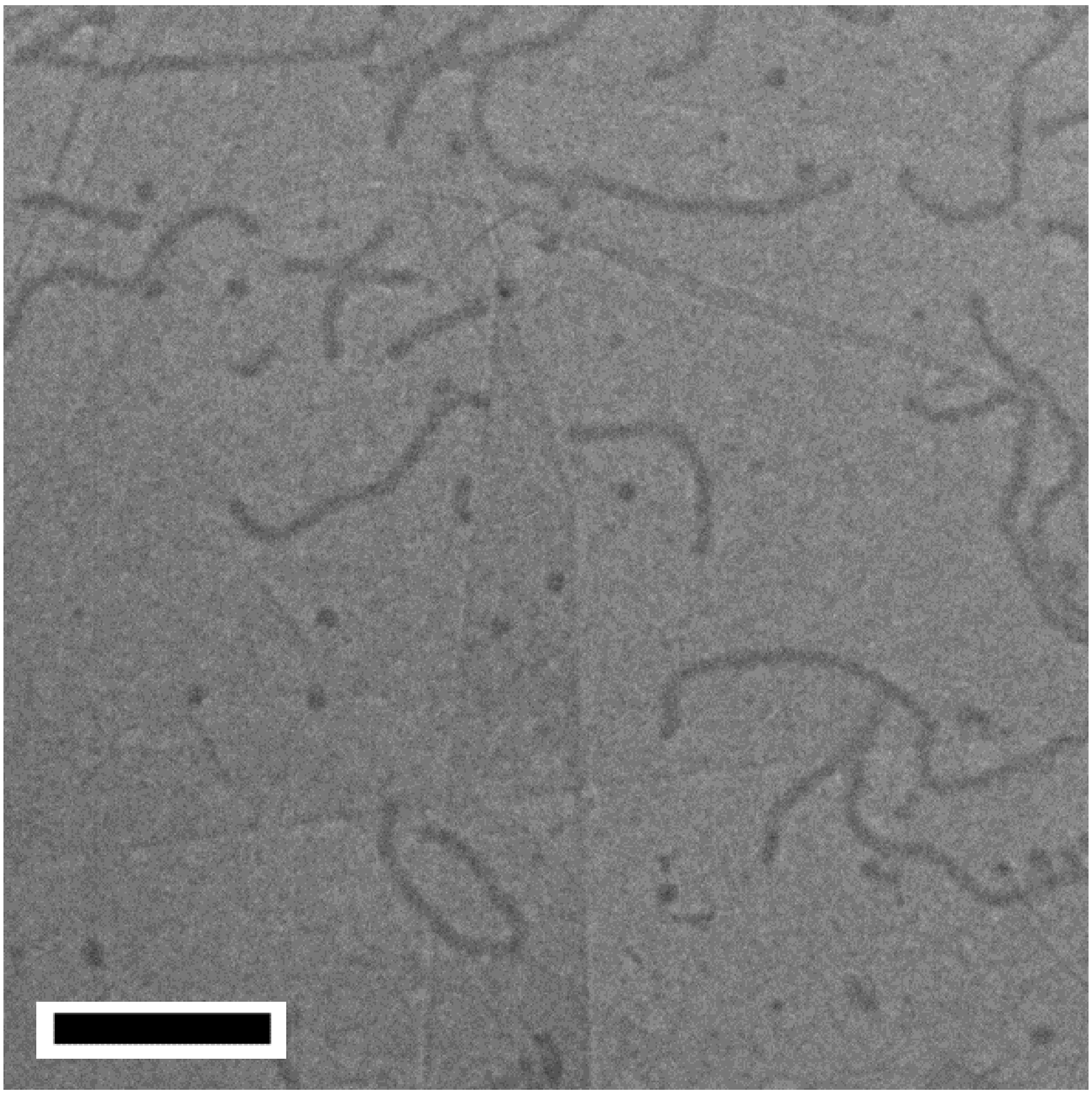

**Figure S17.** TEM image of nanoparticles with an A:T ratio of 1:1 following further addition of **PA**, showing the resulting disassembly into a mixture of short worms and spherical nanoparticle; scale bar = 200 nm.

# S7 MORPHOLOGICAL TRANSFORMATION CONTROL EXPERIMENTS

## S7.1 Blocking of H-Bonding in the Added Polymer

### S7.1.i Analysis of Mixtures of NT with PA$^{Me}$, PT1 and PS

In order to confirm that strong H-bonding between A and T was necessary to drive the morphological transformation process, we performed a series of control experiments using polymers in which H-bonding was either partially blocked or completely removed. We began by synthesising four polymers: PNAM$_{39}$-*b*-PMAAm$_{20}$ (**PA$^{Me}$**), PNAM$_{39}$-*b*-PTAm$_{20}$ (**PT1**), PNAM$_{39}$-*b*-PSt$_{20}$ (**PS**) and PNAM$_{39}$-*b*-PMT$^{Me}$Am$_{300}$ (**PT$^{Me}$**). These were all synthesised from the PNAM$_{39}$ macroCTA whose synthesis is described in section S3.2. Their characterisation data can be found in Table S2. **PA$^{Me}$** was not expected to form strong H-bonds with **NT** because of methylation of the adenine nitrogen; **PT1** was not expected to form strong bonds with **NT** because thymine does not self-dimerise under normal conditions; and **PS** was not expected to form strong bonds with **NT** because of the absence of any H-bond donors or acceptors. **PA$^{Me}$**, **PT1** and **PS** were then mixed with separate solutions of **NT** at different molar ratios, and the resulting nanoparticles analysed by DLS and TEM. DLS analyses (Figure S18) showed no significant changes in the hydrodynamic diameters of the nanoparticles, and TEM analyses showed no noticeable changes in morphology (Figure S19).

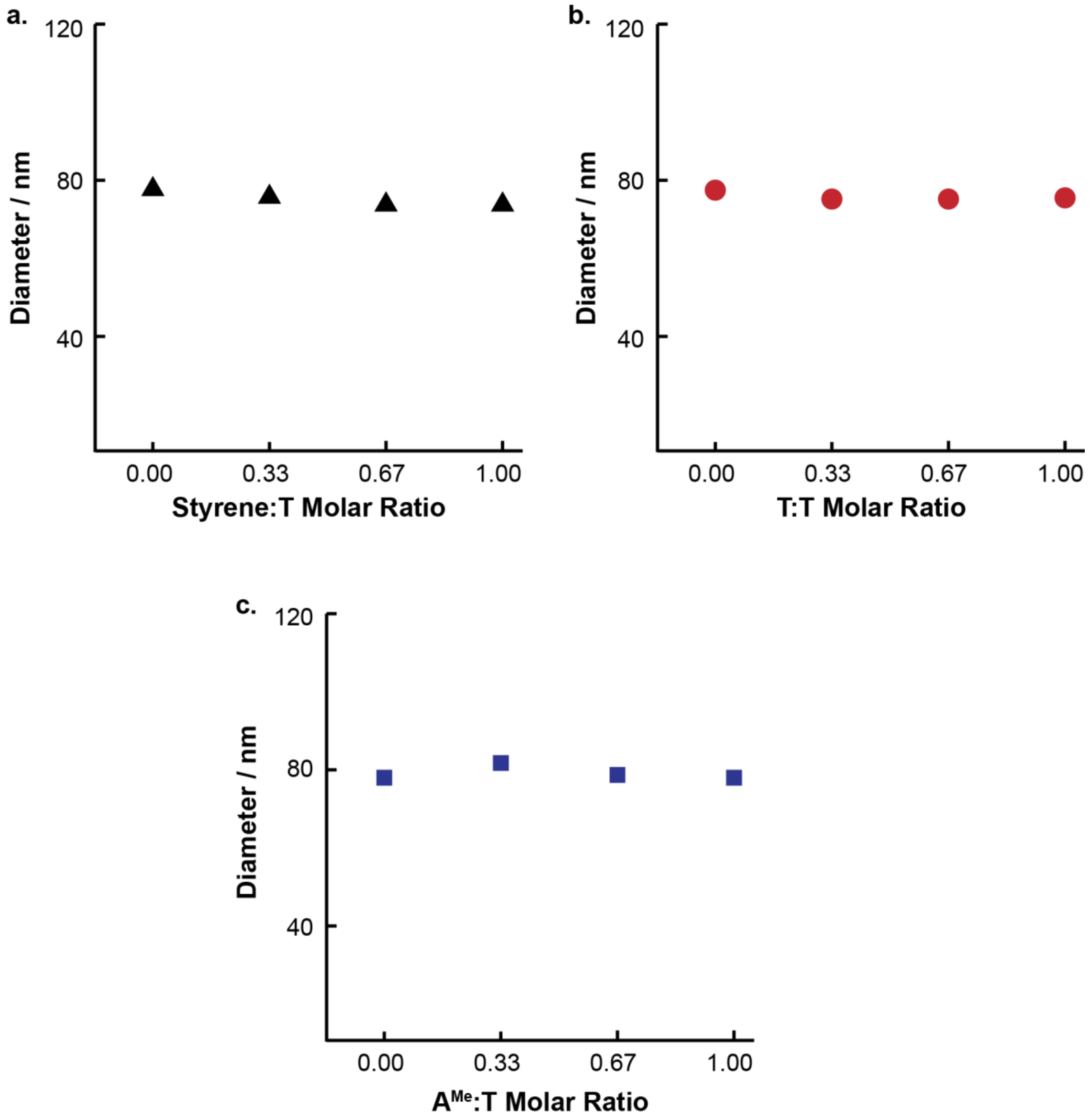

**Figure S18.** Variation of hydrodynamic diameter of the mixture of **NT** with non-complementary copolymers as determined by DLS analyses. (a) Nanoparticle **NT** with **PS**; (b) **NT** with **PT1**; (c) **NT** with **PA**$^{\mathbf{Me}}$; Error bars represent standard deviation of at least three measurements.

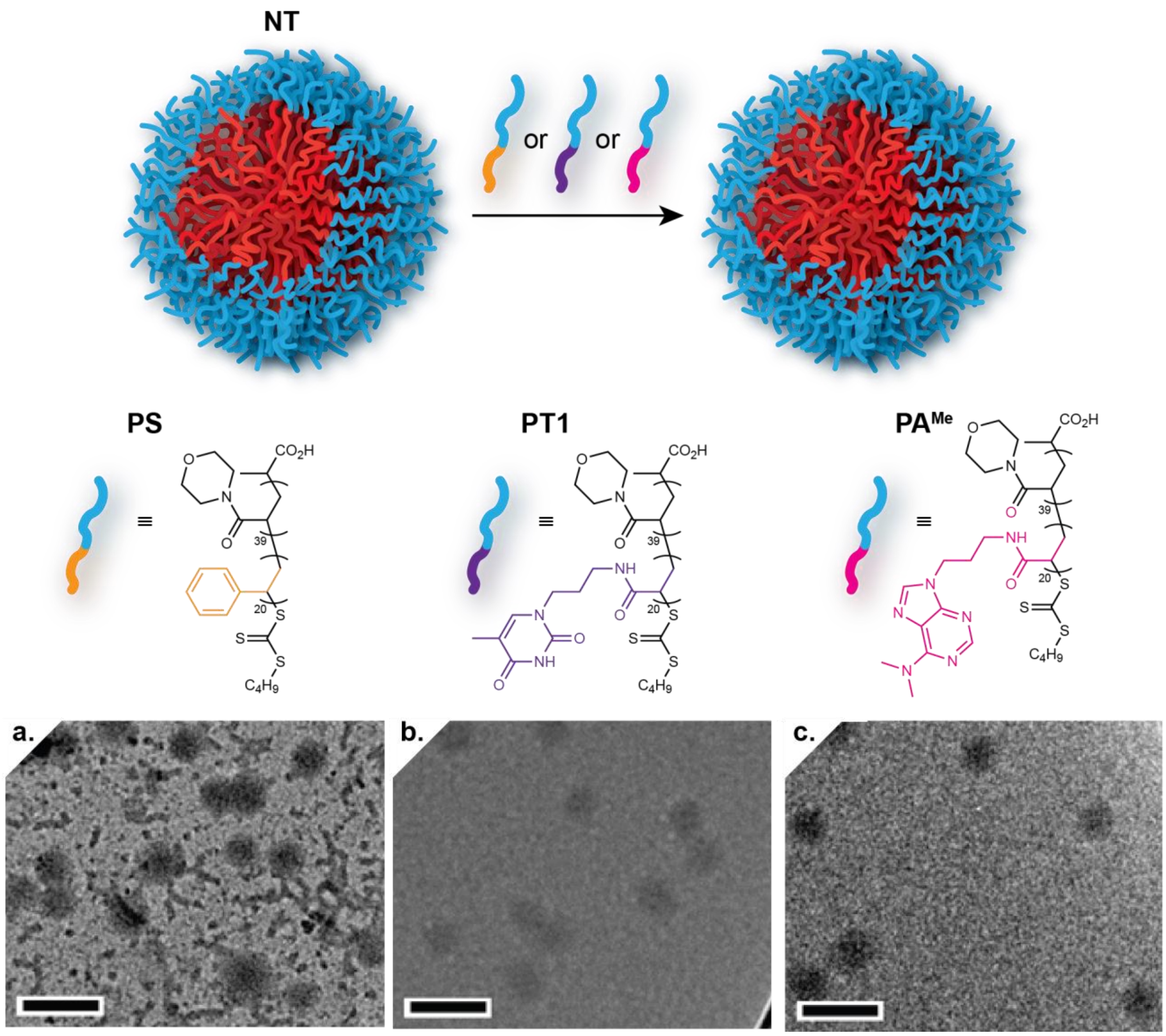

**Figure S19.** Dry-state TEM images on graphene oxide of **NT** following additions of **PS** (a), **PT1** (b) and **PA**$^{Me}$ (c) at the following molar ratios: (a) **St:T** = 1.00; (b) **T:T** = 1.00; (c) **A**$^{Me}$**:T** = 1.00; Small **PS** aggregates were observed when mixing **PS** with **NT**; scale bars = 200 nm.

### S7.1.ii Analysis of the Aggregation Behaviour of PA$^{Me}$, PT1, PS and PA

We wanted to know whether any of the added polymers formed aggregates in solution, which might complicate their interaction with **NT**. We therefore investigated 0.5 mg mL$^{-1}$ solutions of **PA**$^{Me}$, **PT1**, **PS** and **PA** by DLS. DLS analyses (Figure S20) showed that all polymers formed very small aggregates in water at this concentration, but when diluted to the concentrations used in the addition experiments (below 0.1 mg mL$^{-1}$) no particles were observable by light scattering, except in the case of **PS**. We therefore concluded that in the additions experiments the polymers **PA$^{Me}$**, **PT1**, and **PA** were present as unimers. The **PS** aggregates were further investigated by TEM (Figure S21), and observed as small nanoparticles, which were also visible in the TEM images of mixtures of **PS** with **NT** (Figure S19a).

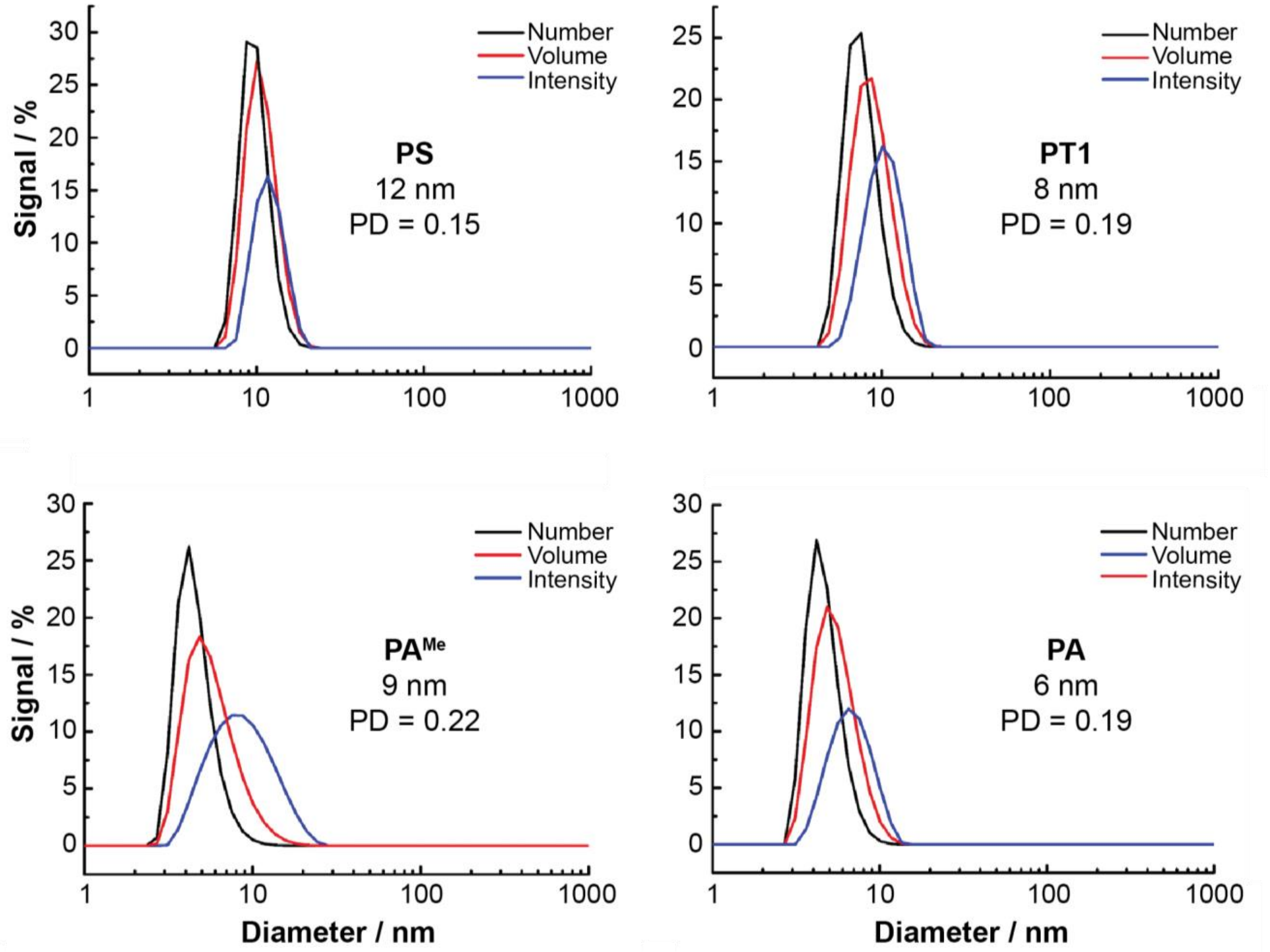

**Figure S20.** DLS analyses of small aggregates (0.5 mg mL$^{-1}$) formed in water by (a) **PS**; (b) **PT1**; (c) **PA$^{Me}$**, (d) **PA**.

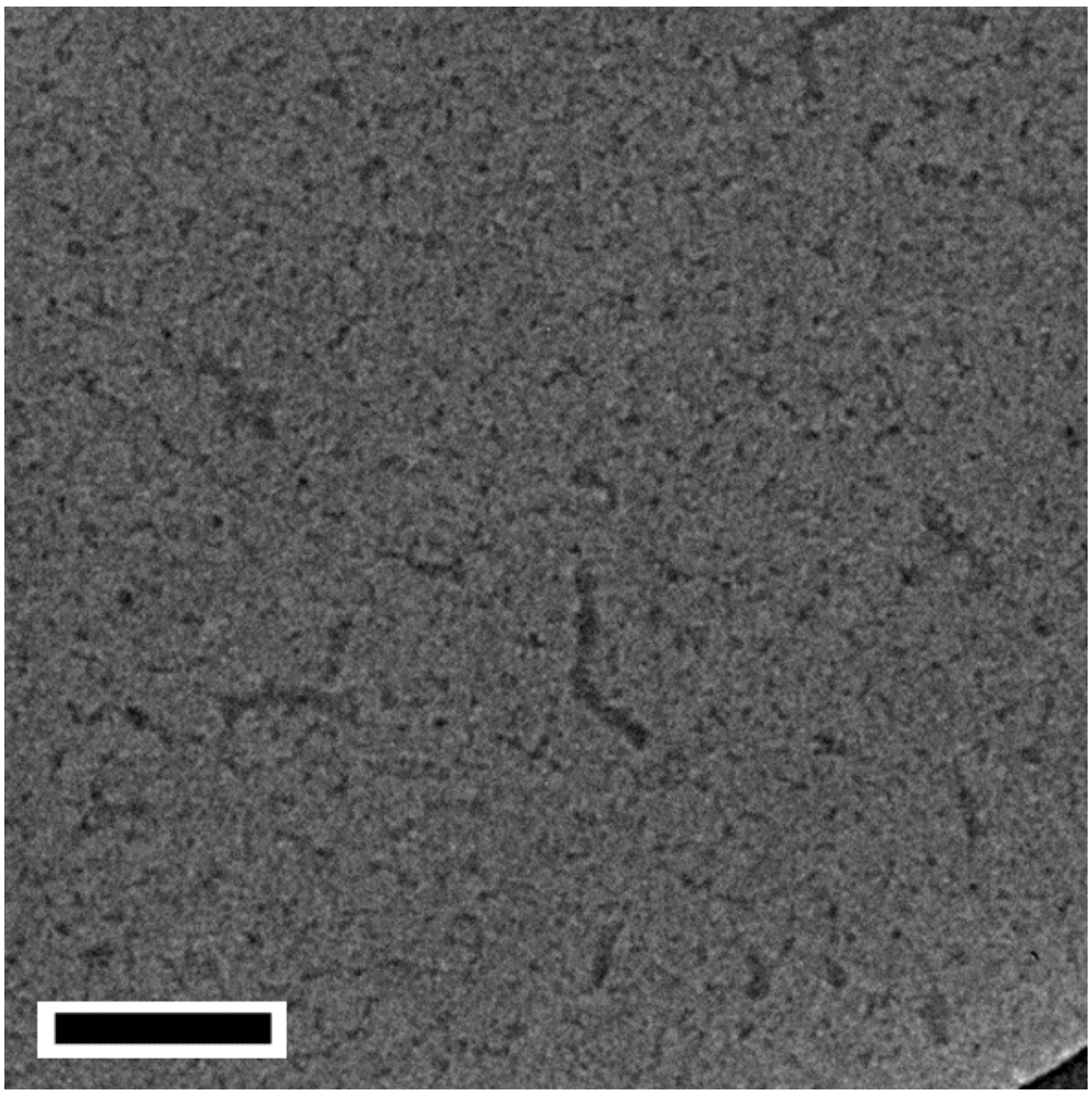

**Figure S21.** TEM images of **PS** aggregates formed in water; scale bar = 200 nm.

## S7.2 Blocking of H-Bonding in the Nanoparticle

We then moved on to investigate the effect of partially blocking H-bonding in the nanoparticle. This was achieved by synthesising a methylated thymine monomer (**T$^{Me}$Am**, see section S3.1.vi for synthesis details) and polymerising it with the $PNAM_{39}$ macroCTA to make a methylated analogue of **PT**: **PT$^{Me}$**. The synthesis of **PT$^{Me}$** was conducted using the general procedure given in section S3.2.ii, and its characterisation data can be found in Table S2. **PT$^{Me}$** was then self-assembled using the protocol in section S4.1 to give spherical nanoparticle **NT$^{Me}$**, which was analysed by DLS and TEM (Figure S22). **PA** was then added to **NT$^{Me}$** at different A:T molar ratios and the resulting particles investigated by DLS and TEM (Figure S23). No change in hydrodynamic diameter or morphology was observed, providing further evidence that strong H-bonding between the nanoparticle core and the added polymer was necessary for morphological transformation to occur.

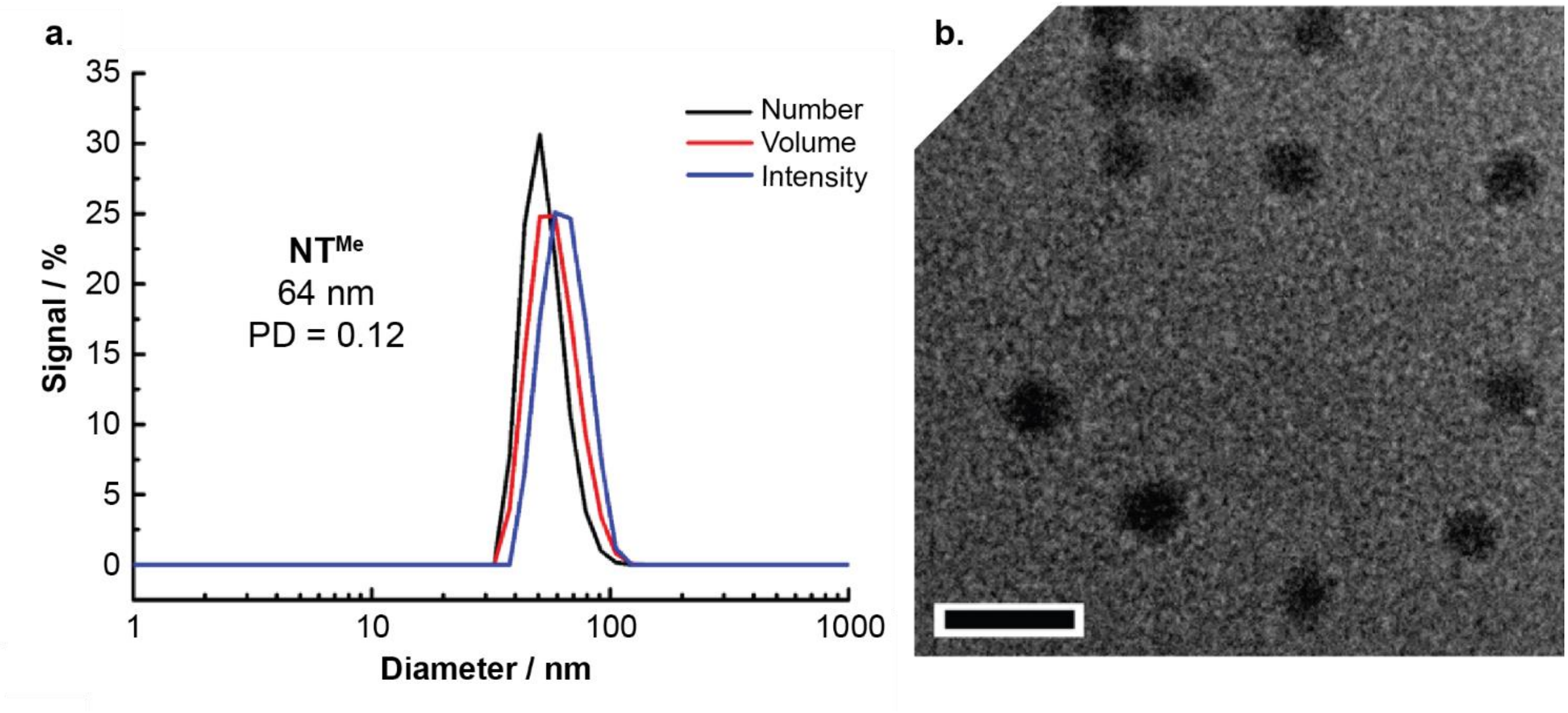

**Figure S22.** (a) DLS analysis of nanoparticle **NT$^{Me}$** (0.5 mg mL$^{-1}$) in water; (b) TEM images of **NT$^{Me}$**; scale bar = 200 nm.

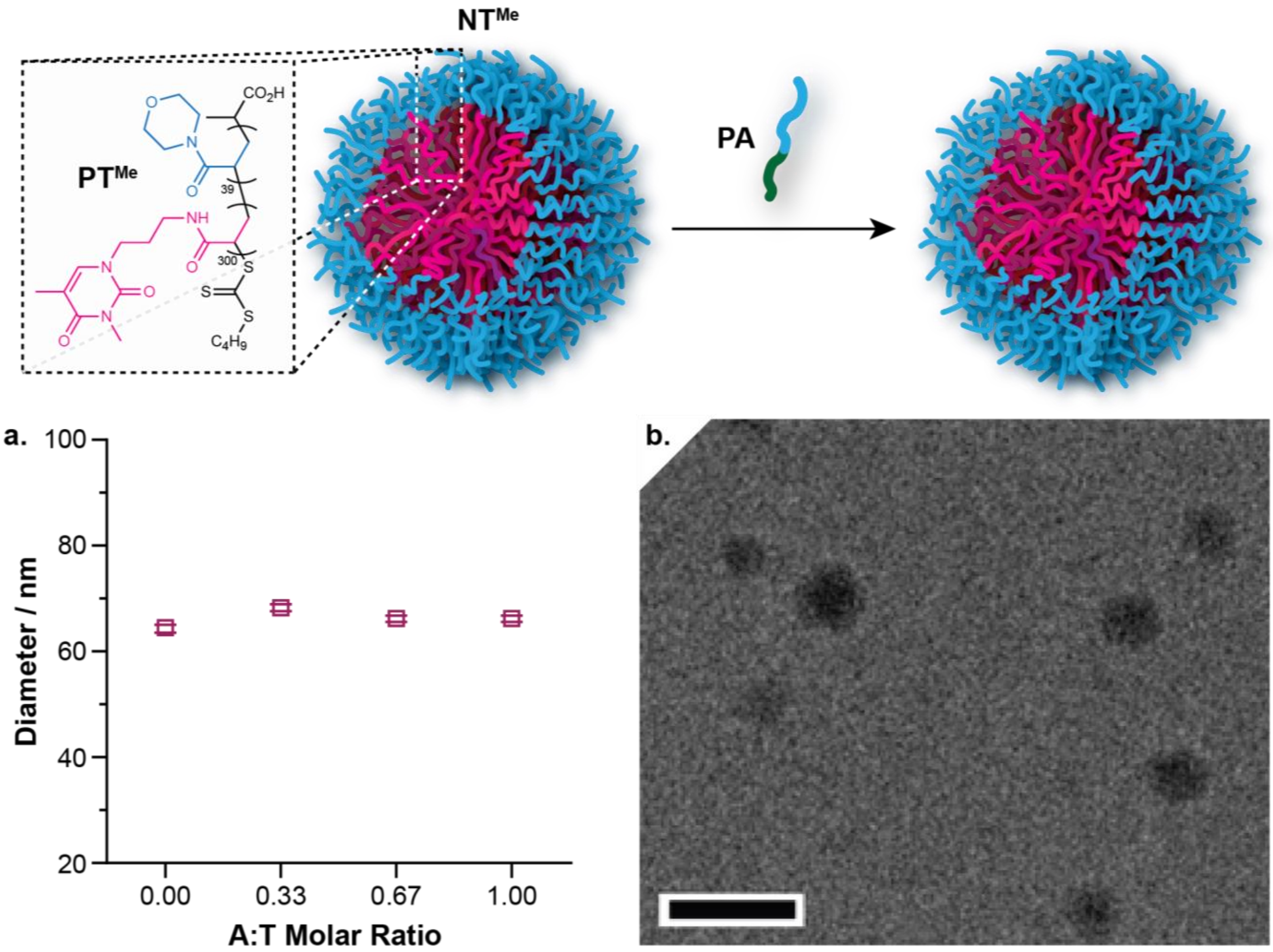

**Figure S23.** (a) Variation of hydrodynamic diameter of the mixture of **NT$^{Me}$** with **PA** as determined by DLS analyses; Error bars represent standard deviation of at least three measurements. (b) dry-state TEM images of nanoparticles **NT$^{Me}$** following addition of **PA** at an A:T$^{Me}$ molar ratio of 1:1; scale bar = 200 nm.

## S7.3 Self-Assembly by Solvent Switch from a Common Solvent

To investigate whether the morphological transformation products represented the thermodynamic assembly products, we mixed **PA** and **PT** in the appropriate ratios and performed a slow solvent switch from DMF (a common solvent for all blocks) to water. As shown in Figure S24, only small spheres were observed, with no apparent formation of anisotropic morphologies.

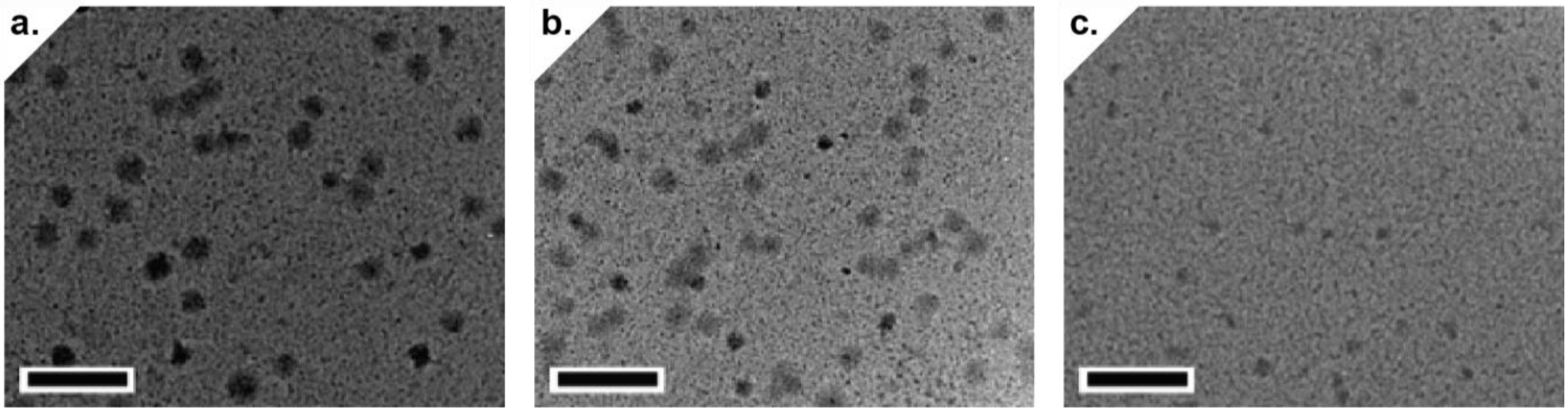

**Figure S24.** TEM images of nanoparticles consisting of **PA** and **PT** prepared through a solvent switch method from DMF to $H_2O$ at the following molar ratios of A:T: (a) 0.33; (b) 0.67; (c) 1.00; scale bars = 200 nm. The sizes of the spherical nanoparticles were 49 ± 5 nm, 42 ± 4 nm and 30 ± 5 nm, respectively.

# S8 EXPERIMENTS CONFIRMING SINGLE PARTICLE TRANSFORMATION PROCESS

## S8.1 SLS Analyses to Determine Nanoparticle Molecular Weights

LS was used to compare estimates for the mass average molar mass of **NT** spherical particles and **A:T = 0.20** dumbbells. Additional filtration of the samples through a 220 nm or 450 nm pore size was found to reduce the sample concentration by up to 30% but allowed estimation of $\langle R_G \rangle_Z$ from data over the full angular range and these results were used to constrain the estimation of $\overline{M}_W$ from samples filtered at 1.2 µm, for which no reduction in the sample concentration could be discerned. Consequently, the Zimm plots in Figures S25-26 show that a subset of the data ($80 \leq \theta \leq 130°$) was used in each case to estimate $\overline{M}_W$ , the deviance at small angles usually being associated with dust or aggregates. Results are summarised in Table S5.

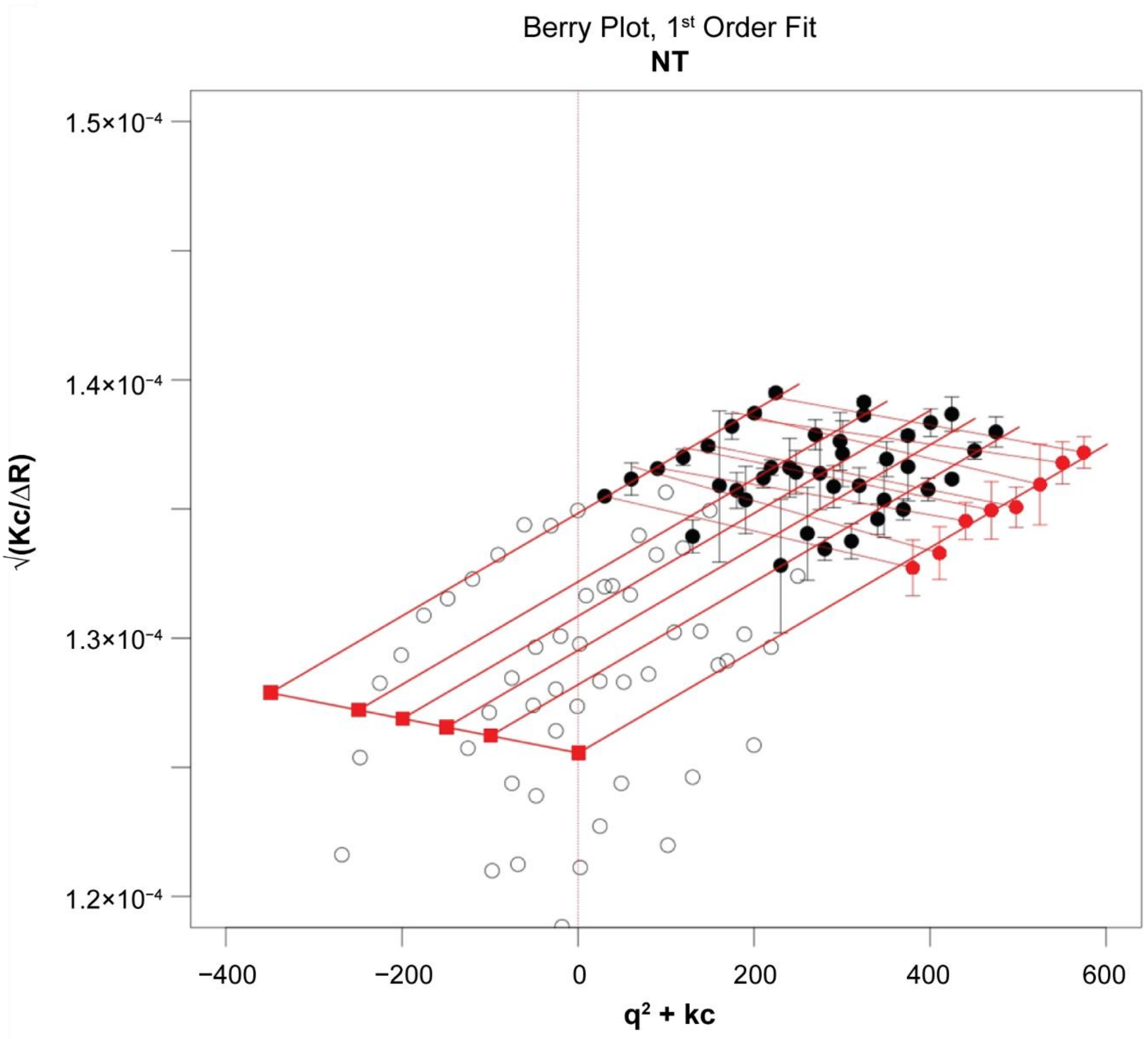

**Figure S25.** Zimm plot (Berry transformation) of spherical nanoparticles **NT**, which gives $M_w = 63.4 \pm 0.9 \times 10^6$ Da. The sample concentrations were 0.2, 0.3, 0.4, 0.5 and 0.7 mg mL$^{-1}$, respectively.

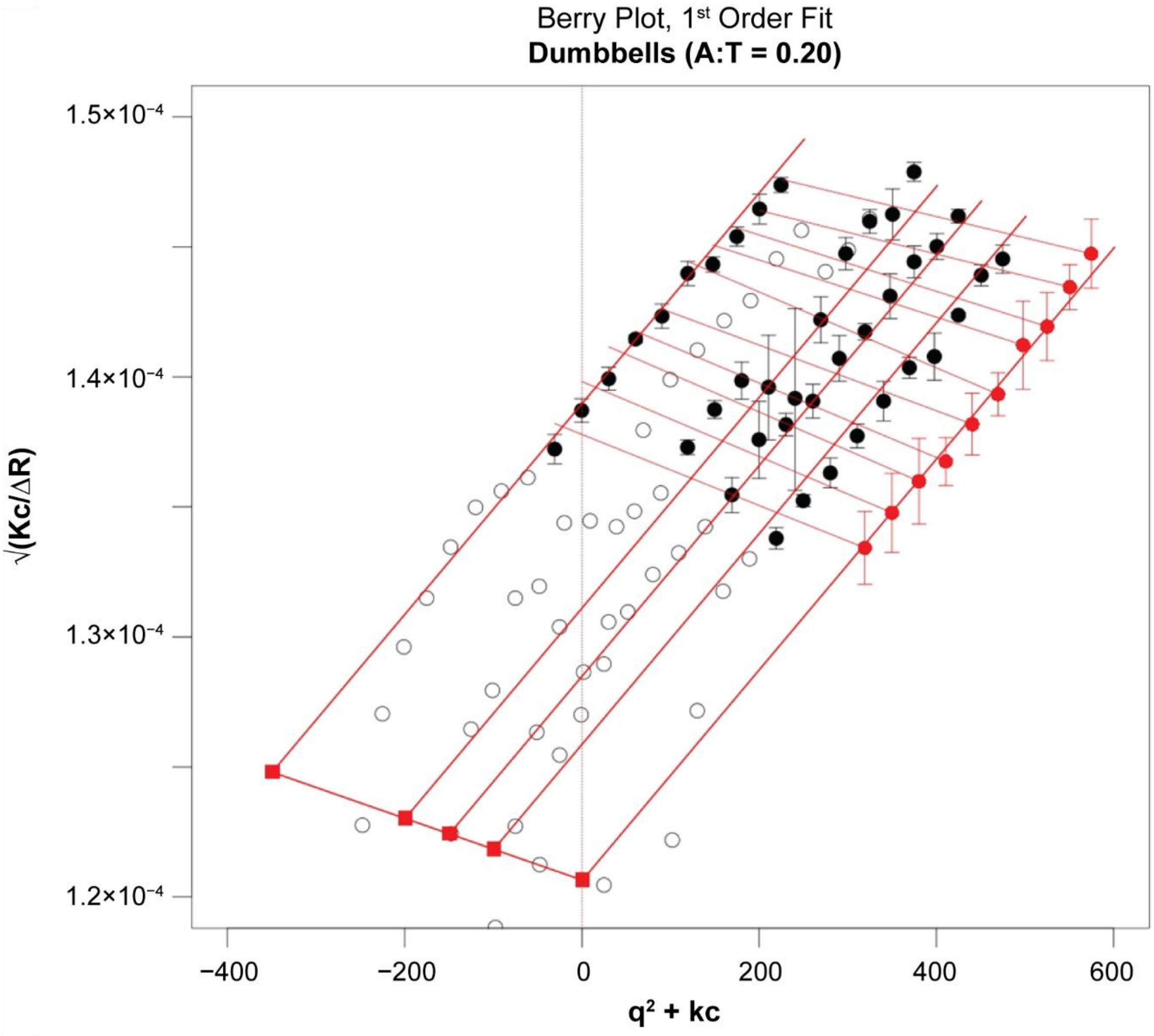

**Figure S26.** Zimm plot (Berry transformation) of dumbbells with an A:T molar ratio of 0.20, which gives $M_w = 68.7 \pm 0.9 \times 10^6$ Da. The sample concentrations were 0.2, 0.3, 0.4, 0.5 and 0.7 mg mL$^{-1}$, respectively.

**Table S5.** Summary of SLS characterization data for spherical nanoparticle **NT** and dumbbells with an A:T molar ratio of 0.20.

| Sample | $M_w$ / Da | $R_g$ / nm | $R_h$ / nm | $R_g/R_h$ |
|---|---|---|---|---|
| **NT** | $63.4 \pm 0.9 \times 10^6$ | 30.8 | 38.2 | 0.81 |
| **Dumbbells (A:T 0.20)** | $68.7 \pm 0.9 \times 10^6$ | 44.9 | 45.5 | 0.99 |

## S8.2 AFM Analyses of Nanoparticles

We imaged the seed nanoparticles (**NT**), dumbbells, and a sample of the worms using AFM (Figure S27). There was a consistent decrease in the height profile of individual particles across the series, which provided further evidence for a single particle transformation process (particle–particle fusion was expected to produce structures with similar height profiles).

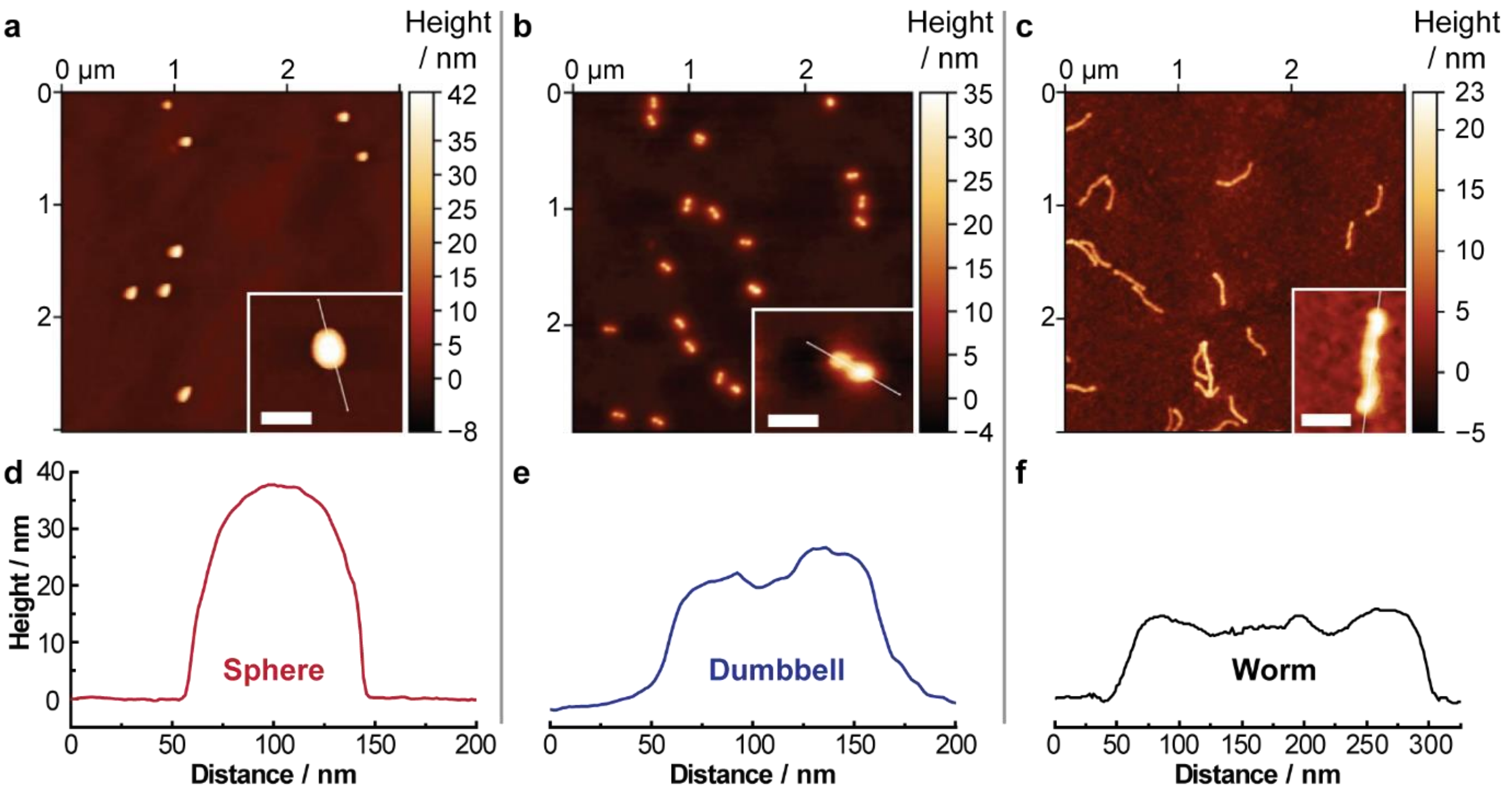

**Figure S27.** AFM images and the corresponding height profiles of (a, d) spherical nanoparticles **NT**, (b, e) dumbbell-like micelles at an A:T molar ratio of 0.20 and (c, f) worm-like micelles formed by further adding **PA** into dumbbells at a total A:T molar ratio of 0.33; scale bars = 100 nm.

# S9 FLUORESCENT TAGGING USING MORPHOLOGICAL TRANSFORMATION

## S9.1 Syntheses of Fluorescently-Labelled PA

Samples of **PA** tagged with a green (**PA$^{G}$**) or red (**PA$^{R}$**) dye were synthesised as shown in Scheme S3 by modifying the carboxylic acid end group of **PA** with amine-containing dye molecules. Successful dye incorporation was confirmed by $^{1}$H NMR spectroscopy (Figure S28).

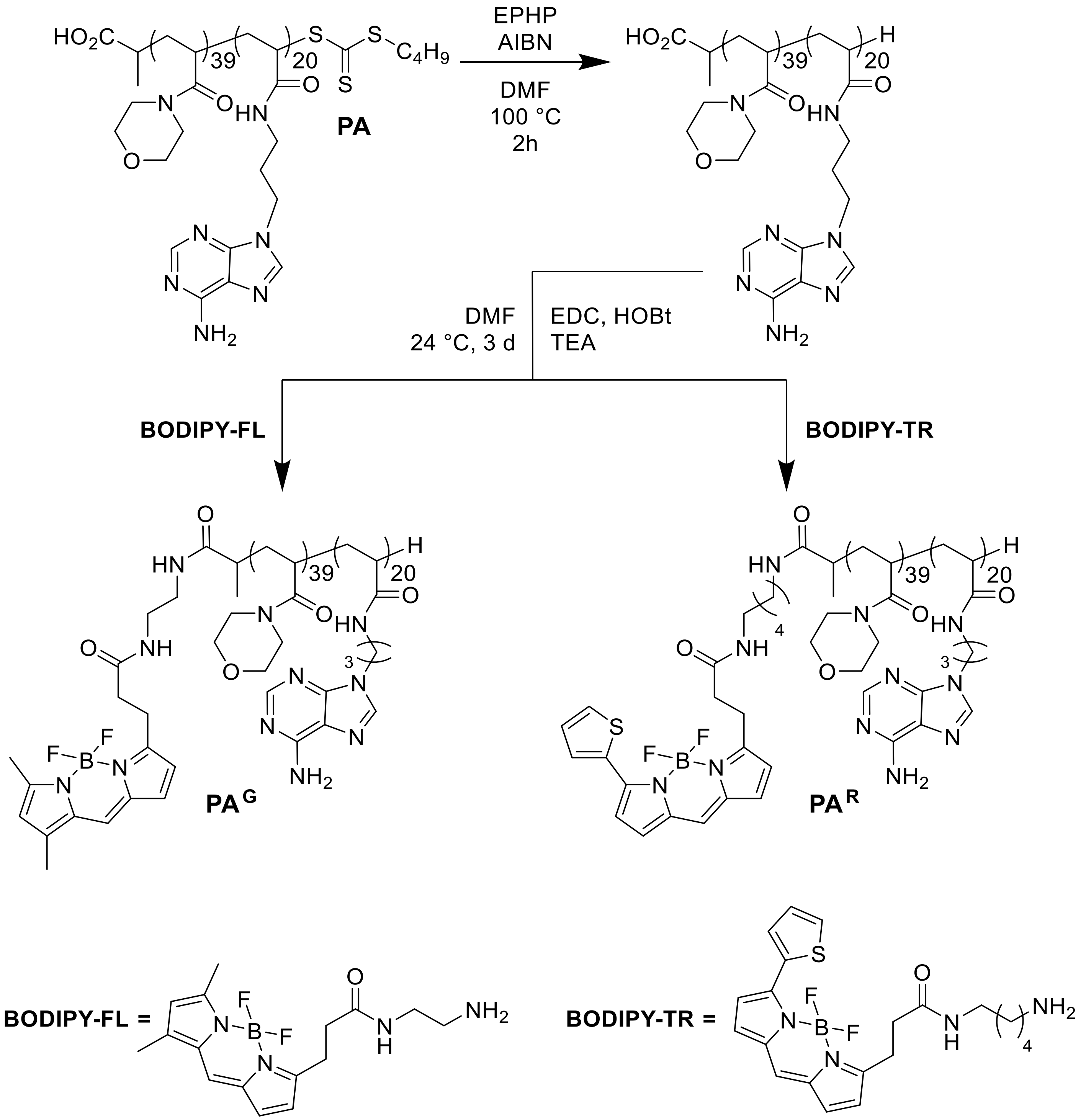

**Scheme S3.** Syntheses of **PA**[G] ((BODIPY-FL)-PNAM$_{39}$-*b*-PAAm$_{20}$) and **PA**[R] ((BODIPY-TR)-PNAM$_{39}$-*b*-PAAm$_{20}$).

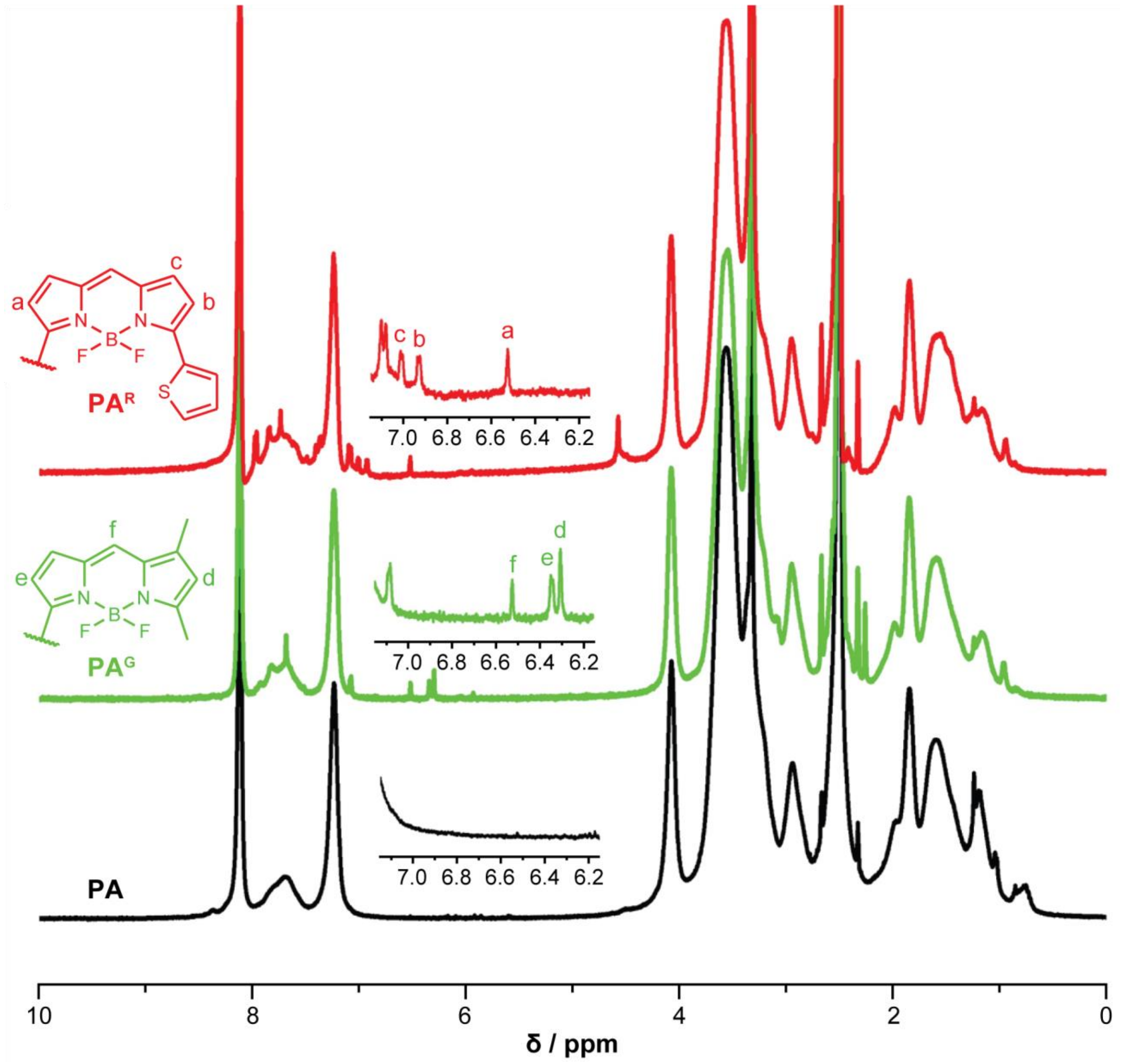

**Figure S28.** $^1$H NMR spectra of **PA$^G$** ((BODIPY-FL)-PNAM$_{39}$-*b*-PAAm$_{20}$), **PA$^R$** ((BODIPY-TR)-PNAM$_{39}$-*b*-PAAm$_{20}$) and **PA** (400 MHz, $d_6$-DMSO).

## S9.2 Stepwise Growth of Fluorescent Wormlike Nanoparticles

Fluorescent wormlike nanoparticles were fabricated as described in the main paper (Figure 4). Further control experiments are presented in Figure S29, which demonstrate that colocalisation of the dyes was only observed when worms were grown with sequential additions of **PA$^G$** and **PA$^R$**, and not when solutions of pure red and green worms were physically mixed.

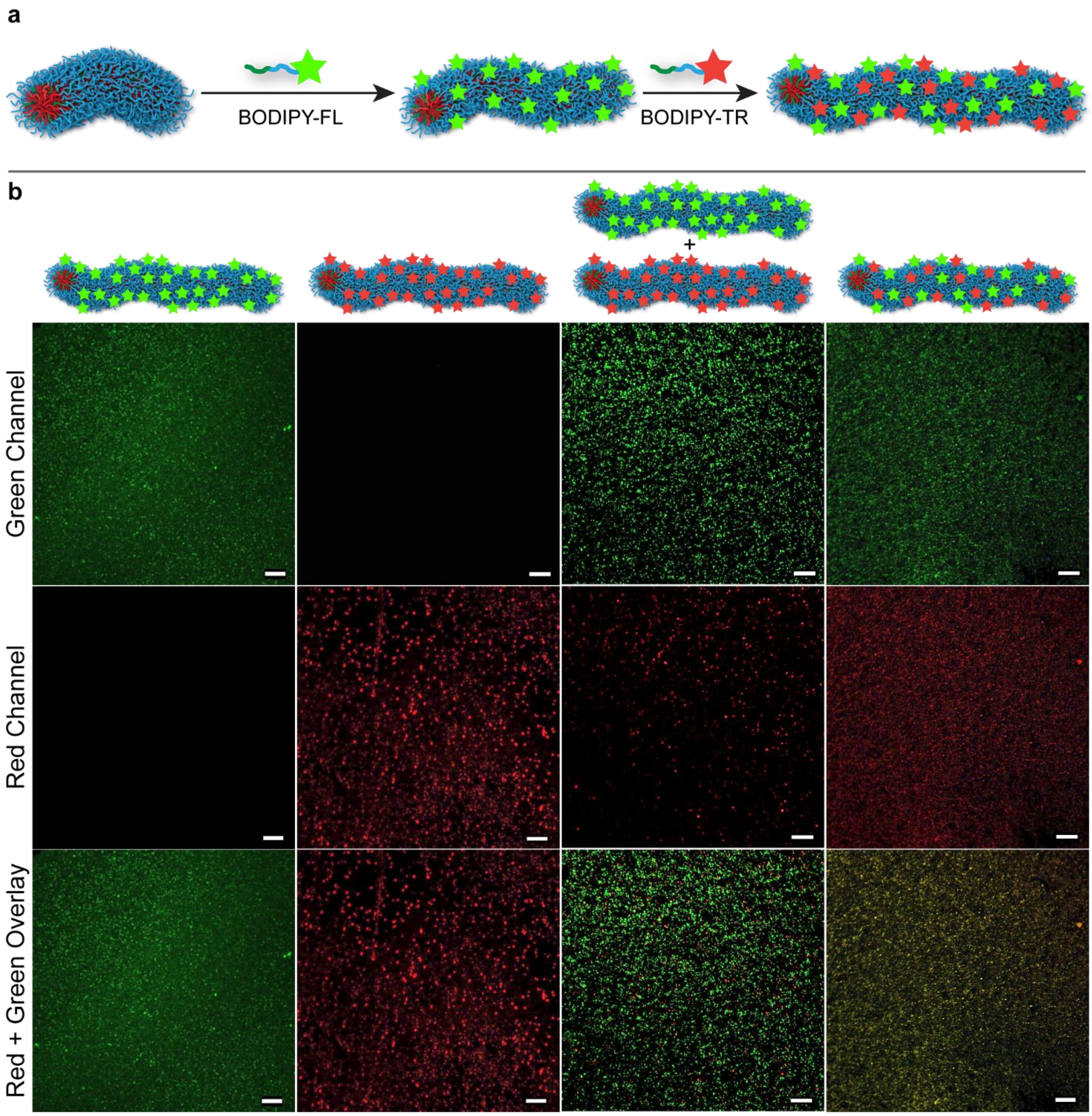

**Figure S29.** Further control experiments for the controlled fabrication of fluorescent nanoparticles mediated by complementary H-bonding interactions. (a) Schematic overview showing the sequential addition of **PA** tagged with BODIPY-FL (**PA**$^{G}$) to generate green fluorescent worms, followed by addition of **PA** tagged with BODIPY-TR (**PA**$^{R}$) to generate yellow fluorescent worms. (b) Confocal miscopy images of different possible combinations of dye additions. 1$^{st}$ column: 2 sequential additions of **PA**$^{G}$, showing only green particles. 2$^{nd}$ column: 2 sequential additions of **PA**$^{R}$, showing only red particles. 3$^{rd}$ column: physical mixture of the particles shown in columns 1 and 2, showing discrete red and green particles. 4$^{th}$ column: Sequential addition of **PA**$^{G}$

followed by **PA**$^{\mathbf{R}}$, showing almost complete overlay of the red and green fluorescence, confirming colocalization. Scale bars = 10 μm.

## S10 PHYSICAL MODEL FOR MORPH

This section contains the technical details related to the physical model for MORPH presented in the main text. The focus is on relevant timescales in the experimental nanoparticle system, comparison with other polymer systems that tend to relax to equilibrium spherical micelles, and on formulating a more complex mathematical model that takes into account both nanoparticle shape and volume, and which reduces to the equation for eccentricity presented in the main text.

### S10.1 Relevant Timescales

There are many timescales relevant to the formation and growth of the nanoparticles. In the main text, we emphasize the timescales $\tau_I$ for insertion of a polymer into the nanoparticle core and timescale $\tau_R$ for the rearrangement and relaxation of the nanoparticle core chains. We assume $\tau_I$ is inversely proportional to the concentration of the polymer (i.e., more polymer will lead to faster insertion). We assume that $\tau_R$ will be determined principally by the bulk properties of the core chains (i.e., the bulk modulus and viscosity) and remain more or less independent of the polymer concentration. These timescales are microscopic, because they arise from the properties of individual polymer chains and chain–chain interactions.

A different approach to the dynamics is a phenomenological model with timescales for the growth of surface area and volume. The surface-area growth proceeds via timescale $\tau_A$ (and, intuitively, surface-area growth should depend strongly on the insertion timescale $\tau_I$). The volume growth proceeds via a different timescale $\tau_V$, which depends on the relaxation time $\tau_R$.

### S10.2 Comparison with Equilibrium Phenomena

In addition to these timescales involved in the MORPH process, there are several timescales that describe equilibrium processes and which, in general, enter the physical description for an equilibrium micelle system. However, due to the non-equilibrium nature of the nanoparticles considered in this work, especially the glassiness of the nanoparticle core, these equilibrium timescales are too long to effect the morphological change from spheres to elongated nanoparticles that we observe.

In most diblock copolymer systems that form micelles, the dominant relaxation process for approaching equilibrium is single-chain extrusion (see, e.g., Ref. [7]). In this process, a micelle loses a single chain to the solution, and this chain diffuses and can rejoin a different micelle. This process leads to a broad equilibrium distribution of particle sizes. However, in the system we consider, single-chain extrusion is strongly suppressed by a combination of the glassiness of the nanoparticle core and the reversible H-bonding interactions between thymine and adenine (which are also not sufficiently strong for **PA** to pull the core chains out into the solution). As a result, polymers from the core do not individually leave the nanoparticle. Instead, any reformation into smaller nanoparticles must proceed via the budding and break-up of parts of the larger nanoparticles. This budding process is energetically costly and not observed during the shape-change step of the process. Instead, budding and break-up is only observed when these processes are strongly driven by the need to accommodate more polymers, i.e., at the end of the process in Figure 6 when the nanoparticle shell can no longer expand to accommodate more polymer insertion. Another potential disassembly process is a collective modulational instability of a cylinder, described in Refs. [8,9]. This process is expected to be as slow as the budding of individual smaller nanoparticles, and may be responsible for the cylinder-to-small-sphere transformation shown in the final stages of Figure 6.

Finally, surface tension of the core-solute interface is responsible for the (viscous) relaxation from anisotropic shapes into spherical nanoparticles. In equilibrium, this process is responsible for the generic spherical shapes of many micelle systems. This relaxation process depends on the relaxation rate of the core. In our system, the suppression of this process through glassy core dynamics is essential for stabilizing a worm-like nanoparticle shape.

### S10.3 Swelling Dynamics

In this subsection, we proceed to derive the quantitative relations between the microscopic timescales ($\tau_R$, $\tau_I$) and macroscopic timescales ($\tau_A$, $\tau_V$). We demonstrate the importance of having a thin shell for the MORPH mechanism: the shell thickness effectively rescales the relaxation rate. As a starting point, consider the equations for the growth of nanoparticle volume $V$ and surface area $A$:

$$\partial_t A = (\tau_A{}^{-1})A, \ \ \partial_t V = (\tau_V{}^{-1})aR^2$$

where $A \sim (1+\epsilon^2)R^2$, $\epsilon$ is the eccentricity of the spheroidal shape, $V \sim R^3$, R is the nanoparticle radius, and $a$ is the core thickness. We assume that the volume growth occurs primarily inside a shell of thickness $a$ much smaller than particle radius R (although the case $a \sim R$ can be considered for shells of thickness comparable to the nanoparticle size).

The timescale $\tau_A$ for area growth depends on the core surface tension: the larger the surface tension, the slower the area growth rate. In the absence of polymer insertion, the nanoparticle will prefer a spherical shape that minimizes the core-solution interfacial area and the area growth rate at fixed volume would be negative, driven entirely by surface tension. More generally, the surface tension of the core favours shapes that are more isotropic and can present a barrier to the development of anisotropy.

### S10.4 MORPH Dynamics

The phenomenological model that we present for MORPH dynamics results from putting the above ingredients together. The equation for the evolution of the area can be rewritten as $R^2\,\partial_t\epsilon^2 + 2\,R\,\partial_t R = (\tau_A{}^{-1})R^2$ to lowest order in $\epsilon^2$, whereas the equation for the volume will take the form $\partial_t \mathrm{R} = (\tau_V{}^{-1})a/3$. Substituting the equation for $\partial_t R$ into the equation for $\partial_t\epsilon^2$, we obtain the dynamics introduced in the main text:

$$\partial_t\epsilon^2 = [\tau_A{}^{-1} - \tau_V{}^{-1}a/(3R)]$$

where we identify $\tau_I{}^{-1} = \tau_A{}^{-1}$ and $\tau_R{}^{-1} = \tau_V{}^{-1}a/(3R)$. The nonlinear terms ignored here can have two effects: stabilizing a dumbbell shape, and (if odd in $\epsilon$) establishing a preference for elongated (prolate) over squished, pancake-like (oblate) shapes.

Glassiness of the core guarantees a slow relaxation time, so the transition from $\tau_I{}^{-1} < \tau_R{}^{-1}$ ($\epsilon = 0$) to $\tau_I{}^{-1} > \tau_R{}^{-1}$ ($\epsilon \neq 0$) can be realized. The relation $\tau_R{}^{-1} = \tau_V{}^{-1}a/(3R)$ shows the importance of having a thin shell (achieved through a short length of the added polymer core block): the ratio of shell thickness $a$ to nanoparticle radius $R$ rescales the effective relaxation rate that enters the equation for $\partial_t\epsilon^2$: the anisotropic regime is easier to probe with a thinner shell.

One simple generalisation of the model that may better reflect the complexity of the polymer system is that the polymer insertion and relaxation timescales may be coupled due to many-body effects. As a result, the eccentricity equation may take the more general form, $\partial_t\epsilon^2 = \tau_I{}^{-\alpha}\tau_R{}^{-\beta}\left[\tau_I{}^{-1+\alpha+\beta} - \tau_R{}^{-1+\alpha+\beta}\right]$ with exponents $\alpha$ and $\beta$ satisfying $\alpha + \beta < 1$. Although this form changes the dynamical scaling, this more general form leaves the phenomenology of this mechanism intact, i.e., similar development of anisotropic behaviour may be observed in systems with quite different rates of insertion and core relaxation.